\documentclass[12pt]{article}
\usepackage{epsfig,multicol,pstricks}
\date{\today}

\newcommand{\bmat}{\left(\begin{array}}
\newcommand{\emat}{\end{array}\right)}
\newcommand{\be}{\begin{equation}}
\newcommand{\ee}{\end{equation}}
\newcommand{\bea}{\begin{eqnarray}}
\newcommand{\eea}{\end{eqnarray}}
\newcommand{\nn}{\nonumber}
\newcommand{\eq}[1]{Eq.~(\ref{#1})}

\def\lsim{\raise0.3ex\hbox{$\;<$\kern-0.75em\raise-1.1ex\hbox{$\sim\;$}}}
\def\gsim{\raise0.3ex\hbox{$\;>$\kern-0.75em\raise-1.1ex\hbox{$\sim\;$}}}

\def\BR{\rm{BR}}
\def\bsgmm{b \rightarrow s ~ \gamma}
\def\nn{\nonumber}
\def\ep{\epsilon}

\def\la{\langle}
\def\ra{\rangle}

\def\PR{{Phys.~Rev.} }

\newcommand{\sft}{\mathrm{soft}}
\newcommand{\mss}{\mathrm{MSSM}}

\begin{document}
\begin{titlepage}
\pagestyle{empty}
\baselineskip=21pt
\rightline{UMN--TH--2629/07}
\rightline{FTPI--MINN--07/38}

\vspace{2cm}
\begin{center}
{\Large{\bf{
Constraints on Supersymmetric Flavour Models from $b\rightarrow s \gamma$}}}

\vspace{1cm}

Keith A. Olive$^1$ and L. Velasco-Sevilla$^{1,2}$\\
{\it 1. ~William I. Fine Theoretical Physics Institute, \\
University of Minnesota, Minneapolis, MN 55455, USA}\\
{\it 2. ~The Abdus Salam International Center for Theoretical Physics, \\
Strada Costiera 11, 34014 Trieste, Italy}

\vspace{2cm}

{\bf Abstract}
\end{center}
We consider the effects of departures from minimal flavour violations (MFV) in the context of CMSSM-like theories. Second and third generation off-diagonal elements in the Yukawa, sfermion, and trilinear mass matrices are taken to be non-zero at the GUT scale. These are run down together with MSSM parameters to the electroweak scale. We apply constraints from fermion masses and CKM matrix elements to limit the range of the new free parameters of the model.  We determine the effect of the departure from MFV on the branching ratio of $\bsgmm$. We find that only when the expansion parameter in the down-squark sector is relatively large  there is a noticeable effect, which tends to relax the lower limit from $\bsgmm$ on the universal gaugino mass. We also find that the expansion parameter associated with the slepton sector needs to be smaller than the corresponding parameter in the down-squark sector in order to be compliant with the bound imposed by the branching ratio of $\tau\rightarrow \mu \gamma$.
% BR$(\tau\rightarrow \mu \gamma)$.

\end{titlepage}
%\newpage
\section{Introduction}

Flavour symmetries (FS), \cite{gflsym}-\cite{su3refs}, are constructed using experimental data from fermion masses and mixings, with or without imposing a supersymmetric framework. 
In theories with supersymmetric FS,  there is no general principle which forbids off-diagonal terms in the soft squared mass or the trilinear mass matrices at the electroweak (EW) scale. 
Indeed supersymmetric FS are constructed so as to be independent of the specific mechanism for breaking supersymmetry. In addition, it is well known that in the minimal supersymmetric standard model (MSSM), dangerous supersymmetric contributions to flavour changing processes can be avoided assuming flavour diagonal  sfermion soft squared mass and trilinear mass matrices at the EW scale.
However, there is no intrinsic reason why off-diagonal terms should be rotated away using the same unitary matrices which diagonalize the Yukawa matrices. In what follows, FS will always refer to supersymmetric flavour symmetries.

%In a supersymmetric theory, one can refer to minimal flavour violation (MFV) as those flavour transitions which 
%occur only in
A supersymmetric theory can be considered to be minimal flavour violating (MFV) if flavour transitions 
occur only in the charged-current sector and such that can be entirely determined by the CKM angles \cite{Ciuchini:1998xy}. In this sense, it is natural to assume that FS can be postulated assuming these MFV conditions. 
If FS are embedded in a supersymmetric grand unified theory (GUT) \cite{gflsym,Ross:2004qn},  the Yukawa matrices have a well defined structure which contains off-diagonal terms at $M_{\rm{GUT}}$ and which evolve to the EW scale, reproducing the observed fermion masses and mixings.
However, every possible assumption for the supersymmetric flavour structure at $M_{\rm{GUT}}$ can 
not be compatible with this version of MFV. Clearly,  the running of general sfermion mass squared and trilinear mass matrices generate angles that cannot be absorbed by the CKM elements. 
As a consequence, it is often assumed that one can impose vanishing off-diagonal terms at the GUT scale in the soft squared mass and trilinear mass matrices \cite{ex_mvf_gut}. In addition, when the trilinear terms are diagonal, the off-diagonal terms in the squark mass matrices generated radiatively at the EW scale can be rotated to the MFV form defined in \cite{D'Ambrosio:2002ex} with the same unitary matrices diagonalizing the Yukawa matrices. 
In FS with supergravity \cite{Ross:2004qn,Dudas:1996fe,Ross:2002mr}, the scalar fields breaking such symmetries (flavons) generate off-diagonal terms at the GUT scale in the supersymmetric sector that are correlated to those of the fermion sector in a model dependent way. The available model dependent freedom allows one to construct safe theories at the EW scale.

Rather than determining the possible structure of the Yukawa matrices at the GUT scale (which is the basis for constructing FS) using only the fermion masses and mixings of the SM at the EW scale and hoping that the FCNC bounds can be satisfied, here we propose a bottom up approach for determining the parameters, both in the fermion sector and in the sfermion sector, of possible flavour symmetries taking into account (i) fermion masses with their corresponding supersymmetric corrections and mixings at the EW scale and (ii) measurements or bounds from FCNC experiments.

We will begin by first discussing the flavour violating processes that we will use in our subsequent analysis in section 2. We will focus primarily on the branching ratio for $b \rightarrow s \gamma$, given the current improvement in measuring its experimental value \cite{bsgexp,hfag}. %and its calculation in the SM at NNLO \cite{bsgtheonew}.
We discuss its computation in models without MFV in section 2 as well. In section 3, we briefly discuss general features of the
fermion mass matrices and the importance of considering their supersymmetric corrections in order to properly determine their profile at the GUT scale. We also point out that these theories naturally go beyond the MVF conditions at the EW scale.  

In section 4, we employ conditions based on supergravity and FS to determine the form of the soft squared mass and trilinear mass matrices that we consider.  The approach we take here makes use of an underlying supergravity theory  with an effective K\"ahler potential from which one can construct the potential and through minimization, calculate the soft squared masses and trilinear couplings. It turns out that in supersymmetric theories with non-Abelian FS, the K\"ahler potential \cite{Ross:2004qn} has fewer model dependent parameters than other sectors of the theory and hence can be effectively used to impose the size of the off-diagonal terms of soft squared mass and trilinear mass matrices at $M_{\rm{GUT}}$. 
Using the powerful tools of supergravity, one can then determine the structure in the supersymmetric sector at $M_{\rm{GUT}}$ and test its consequences at  the EW scale. The form of the Yukawa matrices and soft terms are developed in section 5.

In Section 6 we present the results of the analysis. We first focus on explaining the impact of considering non-diagonal Yukawa matrices at the GUT scale on the running of off-diagonal soft squared masses and trilinear couplings. 
Even when off-diagonal terms are taken to be zero at the GUT scale, due to the dependence of the beta functions of these terms, they will not be zero at the EW scale, but we show that they are well below current experimental bounds. We then determine the profile of the Yukawa matrices at the GUT scale when considering their supersymmetric corrections at the EW scale.
We do not include GUT-specific effects and assume that all GUT-scale physics has decoupled below the GUT scale.

We then go on to determine the sensitivity of $\BR(\bsgmm)$ on the parameters used to define the Yukawa couplings at the GUT scale. Although FCNC bounds serve to restrict the parameters determining the Yukawa matrices, soft squared matrices and trilinear couplings we find a region where the contribution to  $b \rightarrow s \gamma$ is enhanced with respect to the constrained minimal supersymmetric standard model (CMSSM) with MFV and brings its value closer to the experimental value, thus relaxing the lower bound imposed on the universal gaugino mass at large $\tan \beta$. We show that this is possible while at the same time keeping flavour violating parameters in the lepton sector below the bounds imposed by $\BR(\tau\rightarrow \mu\gamma)$. Finally, we mention the sensitivity of the determination of the MSSM spectra on these new parameters.

%
%%%%%%%%%%%%%%%%%%%%%%%%%%%%%%%%%%%%%%%%%%%%%%%%%%%%%%%%%%%%%%%%%%%%%%%%%%%%%%%%%%%%%%%%%%%%%%%%%%%%%%%%%%%%%%%%%%%%%%%%
%
\section{Supersymmetric flavour and CP violation problems in FS}
%
%%%%%%%%%%%%%%%%%%%%%%%%%%%%%%%%%%%%%%%%%%%%%%%%%%%%%%%%%%%%%%%%%%%%%%%%%%%%%%%%%%%%%%%%%%%%%%%%%%%%%%%%%%%%%%%%%%%%%%%%
%
It is well known that the supersymmetric soft Lagrangian, $\cal{L}^{\mss}_{\sft}$
\bea
\cal{L}^{\mss}_{\sft} &=& -\frac{1}{2}\left(M_1\tilde{B}\tilde{B} + M_2 \tilde{W}\tilde{W} + M_3 \tilde{g}\tilde{g}\right) + h.c.\nn\\
                      & & -\left(\tilde{\bar{Q}} a_u H_u\tilde{u} + \tilde{\bar{Q}} a_d H_d\tilde{d} + \tilde{\bar{L}} a_e H_d\tilde{e} \right) + h.c.\nn\\
                      & & - \tilde{Q} M^2_{\tilde{Q}} \tilde{Q}^{\dagger}
- \tilde{L} M^2_{\tilde{L}} \tilde{L}^{\dagger}    
- \tilde{u} M^2_{\tilde{u}} \tilde{u}^{\dagger}
- \tilde{d} M^2_{\tilde{d}} \tilde{d}^{\dagger}
- \tilde{e} M^2_{\tilde{e}} \tilde{e}^{\dagger} \nn\\
                      & & - M^2_{H_u} H^*_u H_u  - M^2_{H_d} H^*_d H_d 
- (B \mu H_u H_d+ c.c.)          
\label{eq:sftLg}
\eea
introduces many new parameters that are not present in the SM. There are 106 mass terms, 
phases and mixings in the Lagrangian of the minimal version of the supersymmetric SM (MSSM), 
which cannot be rotated away by redefining the phases and flavour basis for the quark and lepton supermultiplets, and which have no counterpart in the SM. 

However, most of the parameters involve flavour mixing or CP violation of the type which is severely restricted by experimental data. In the slepton sector, the most severe constraint is found 
when $M^2_{\tilde{e}}$ is not diagonal leading to the decay $\mu\rightarrow e \ \gamma$ 
which could take place, for example, via a one-loop diagram involving a virtual bino and a wino. 
The current bound on $\BR(\mu\rightarrow e \gamma)\lsim 10^{-11}$ \cite{RPP} constrains all of  
the parameters involved in the decay and in particular restricts 
specific terms in the squared masses of the slepton sector. 
The current bound on $\BR(\tau\rightarrow \mu \gamma)$ ($\lsim 7 \times 10^{-8}$) 
\cite{RPP} can also impose important constraints in the slepton sector.

Another example comes from $K^0\leftrightarrow \overline{K}^0$ mixing where 
the effective Hamiltonian gets an important contribution from %$M^2_{\tilde{s}\tilde{d}}$ 
 ${M^2_{\tilde d}}_{12}$ via a box diagram involving gluinos. 
Winos and binos can also mediate these kind of diagrams. 
If the squark and gaugino masses are order $1$ TeV or less, it is possible to
put bounds on the parameters $\Delta m_K$ and $\epsilon_K$ appearing
in the neutral kaon system effective Hamiltonian which restricts the amount
of down-strange squark mixing and CP violating complex phases that
one can tolerate among the soft parameters. Constraints related to the
second and third generations come from
the $D^o, \overline{D}^o$ and $B^o, \overline{B}^o$ neutral meson
systems, and the decay $b \rightarrow s \gamma$.  After the Higgs 
fields get vevs, the $a_u$, $a_d$ and $a_e$ matrices defined in eq. (\ref{eq:sftLg}) 
contribute to off-diagonal squark and slepton 
mass squared terms and hence
their form is also constrained by FCNC limits. There are other significant constraints on CP-violating phases from limits on the electric dipole moments of the neutron and electron and Hg \cite{edm}.

All dangerous FCNC and CP violation effects in the MSSM can be avoided
if one assumes that in the soft MSSM Lagrangian, \eq{eq:sftLg}, all
the mass matrices  are diagonal. If the sfermion mass matrices are diagonal, 
\bea
M^2_{\tilde{Q}}= m^2_{\tilde{Q}}{\mathbf{1}},\
M^2_{\tilde{u}}= m^2_{\tilde{u}}{\mathbf{1}},\
M^2_{\tilde{d}}= m^2_{\tilde{d}}{\mathbf{1}},\
M^2_{\tilde{L}}= m^2_{\tilde{L}}{\mathbf{1}},\
M^2_{\tilde{e}}= m^2_{\tilde{e}}{\mathbf{1}}
\label{eq:propto1sms}
\eea
then all squark and slepton mixings vanish. Squarks and sleptons with the same electroweak quantum numbers are degenerate in mass and can be rotated into each other. There are  also supersymmetric contributions to FCNC modulo mixing due to $a_u$, $a_d$ and $a_e$. Hence a further assumption usually implemented is that the trilinear terms are proportional to the Yukawa matrices
\bea
a_u=A_u Y_u,\quad a_d=A_d Y_d,\quad a_e=A_e Y_e,
\label{eq:at_proptoyktrl}
\eea
which ensures that only the squarks and sleptons of the third family can have large scalar cubic couplings.

Family symmetries attempt to reproduce mixings and masses in the quark and lepton sector usually by breaking the family symmetry spontaneously with scalar fields, {\it flavons}, and coupling different powers of them to quarks and leptons to reproduce the appropriate size of Yukawa couplings and hence explain the different hierarchies in fermion masses. However once these FS are implemented, if the theory is supersymmetric then not only will fermion mass terms be generated but  sfermion mass terms as well and these will induce FCNC and CP violating effects. These effects then can be used to test different family symmetries. Unfortunately, unless the mechanism by which 
supersymmetry is broken is known,  
one is forced to make more assumptions, as in the case of the constrained version of the MSSM
(CMSSM) discussed below.

%%%%%%%%%%%%%%%%%%%%%%%%%%%%%%%%%%%%%%%%%%%%%%%%%%%%%%%%%%%%%%%%%
%
%\subsection{${\mathbf{b}}\rightarrow {\mathbf{s \gamma}}$ in models with  non-minimal flavour violation \label{eq:bsg_gfm_rs}}
%
%%%%%%%%%%%%%%%%%%%%%%%%%%%%%%%%%%%%%%%%%%%%%%%%%%%%%%%%%%%%%%%%%
%
%

In this paper, we will focus our attention to the constraints imposed by the rare decays of $\bsgmm$ on departures from MFV as the parameters that we consider in this analysis are more sensitive to this process than to other observables such as BR$(B_s\rightarrow \mu^+ \mu^-)$  and $\Delta M_{B_s}$. As a consequence,  we will need to consider a generic method for computing $\BR(\bsgmm)$.  QCD LO effects have been computed for the running of the chromo-magnetic Wilson coefficients of the effective Hamiltonian describing the decay $\BR(\bsgmm)$, from the EW scale to the decay scale, $\mu_b$ \cite{Borzumati:1999qt}. These coefficients arise as a result of the virtual exchange of gluinos, neutralinos, charged Higgs bosons and charginos.

The MFV QCD NLO contributions have been computed for the exchanges of charged Higgs boson and
charginos in \cite{Ciuchini:1997xe,Ciuchini:1998xy,Bobeth:1999ww,Degrassi:2000qf}.
In \cite{Degrassi:2006eh}, the MFV MSSM NLO QCD contributions have been computed.
These contributions complete the NLO calculation if the MFV condition is imposed at the electroweak scale. In addition, some beyond leading order effects (BLO), when the MFV is not assumed, have been calculated in \cite{Foster:2005wb}. 

Thus in a general framework, 
the only complete calculations of BR$(\bsgmm)$ are available at LO. However, the
present value of $\BR(\bsgmm)$ in the SM at NNLO is calculated \cite{bsgtheonew}
\bea
{\rm BR}(\bsgmm)= (3.15\pm0.23)\times 10^{-4}
\label{sm_brbsg}
\eea
and the present experimental value estimated by the Heavy Flavour
Average Group (HFAG) is \cite{bsgexp,hfag}
\bea
{\rm BR}(\bsgmm)= (3.55\pm0.24^{+0.09}_{-0.10})\times 10^{-4}.
\label{exp_brbsg}
\eea

Here, we will not need to consider general departures from MFV at the EW 
scale. Instead, we are restricting the soft terms at the GUT scale. The
expansion  parameter of the $u$ sector, $\ep_u$, is taken to be much smaller
than the corresponding parameter in the down sector,  $\ep_d$.  In addition,
we start with the Yukawa matrix, $Y_u$, 
diagonal at the GUT scale, and  the departure from MFV will be sensitive mainly to
gluinos. We provide the procedure that we use to obtain ${\rm{BR}}(\bsgmm)$ in the framework 
described below in Appendix {\ref{sec:bsg_bmfv}}.

%%%%%%%%%%%%%%%%%%%%%%%%%%%%%%%%%%%%%%%%%%%%%%%%%%%%%%%%%%%%%%%%%%%%%%%%%%%%%%%%%%%%%%%%%%%%%%%%%%%%%%%%%%%%%%%%%%%%%
%
\section{Fermion Masses and MFV conditions}
%
%%%%%%%%%%%%%%%%%%%%%%%%%%%%%%%%%%%%%%%%%%%%%%%%%%%%%%%%%%%%%%%%%%%%%%%%%%%%%%%%%%%%%%%%%%%%%%%%%%%%%%%%%%%%%%%%%%%%%
%
When flavour symmetries are constructed, the quark sector is usually fixed by 
precision fits of the CKM elements. The lepton
sector is similarly determined by the UMNS elements although only two
mixing angles out of  the three mixing angles and three phases
(one CKM-like and two Majorana) are known. Thus, the determination of the lepton mixing matrix
is not precise as in the case of the quark sector. Hence it is more common to construct
flavour symmetries by fixing elements of Yukawa matrices in the quark
sector and then adjusting parameters in the lepton sector by relating
these to the quark sector and of course using the available information
from leptons. Since only the left handed states enter into the
electroweak mixings, only the Yukawa couplings along and above (or
below, depending on the convention) the diagonal can be determined, so
it is necessary to make an assumption of the Yukawa couplings below
(or above) the diagonal in the Yukawa matrix. It is natural then to classify these
choices as symmetric or non-symmetric Yukawa matrices. For the
non-symmetric matrices, however there is less precision in determining the elements because the number of parameters is larger than in the symmetric case. The form of these matrices for the symmetric case is \cite{King:2004tx}:
\bea
Y_{d~ sym}\propto\left[
\begin{array}{ccc}
\leq \ep_d^4 & \ep_d^3       & \leq\ep_d^3\\
\ep_d^3      & \leq \ep_d^2  & \ep_d^2\\
\leq\ep_d^3  & \ep_d^2       &  1
\end{array}
\right],\quad
Y_{u ~sym}\propto\left[
\begin{array}{ccc}
\leq \ep_u^4 & \ep_u^3       & \leq\ep_u^3\\
\ep_u^3      & \leq \ep_u^2  & \ep_u^2\\
\leq\ep_u^3  & \ep_u^2       &  1
\end{array}
\right],
\label{eq:yuksymm}
\eea
while for the non-symmetric \cite{King:2004tx} case we have:
\bea
Y_d\propto\left[
\begin{array}{ccc}
\leq \ep_d^4 & \ep_d^3       & \leq \ep_d^3 \\
\leq \ep_d^3 & \leq \ep_d^ 2 & \ep_d^2      \\
\leq \ep_d   & \leq 1        & 1             
\end{array}
\right],\quad
Y_u\propto\left[
\begin{array}{ccc}
\leq \ep_u^4 & \ep_u^3       & \leq \ep_u^3 \\
\leq \ep_u^3 & \leq \ep_u^ 2 & \ep_u^2      \\
\leq \ep_u   & \leq 1        & 1             
\end{array}
\right],
\eea
with $\ep_d=O(\lambda_c)$ and $
\ep_u=O(\lambda^2_c)\label{eq:yukmat_sym}$, where $\lambda_c \approx
  0.23$ is
  the Wolfenstein parameter $V_{us}=\lambda_c$.
The coefficients of Yukawa elements are taken to be $O(1)$. 
In general, we are free to independently choose either $Y$ or $Y_{sym}$ for the up-type and 
down-type sectors. As the non-symmetric matrices have more parameters, we will restrict our attention here to the symmetric cases which will be the most restrictive.

The extension of FS to the supersymmetric sector is somewhat
arbitrary, independently of the many complications inherent in the MSSM.
The problem rests in the lack of an underlying theory which relates the
parameters of a FS to the supersymmetric parameters. 
From a bottom up approach we can tackle
this problem by using the structure of the running of the soft
parameters in the MSSM. 

The global symmetry of the gauge sector of the SM is given by
%
%\bea
$U(3)^5=SU(3)_Q\times SU(3)_U \times SU(3)_D\times ..., ~$%\nn
%\eea
and is broken only by the Yukawa couplings
\bea
Y_d\rightarrow \overline{3}_Q\times 3_d,\quad
Y_u\rightarrow \overline{3}_Q\times 3_u,\quad
Y_e\rightarrow \overline{3}_L\times 3_e.
\eea
The MFV hypothesis \cite{Ciuchini:1998xy} basically constrains this breaking to the fermion
sector and imposes \eq{eq:propto1sms}  and \eq{eq:at_proptoyktrl} at some low scale to ensure small FCNCs.

Given the dependence of the beta functions of the MSSM soft terms \cite{Martin:1993zk} on $Y^\dagger_f Y_f$ and $Y_fY^\dagger_f$, their running  will alter the conditions \eq{eq:propto1sms} and \eq{eq:at_proptoyktrl} and hence the MFV scale is set at the EW scale.
Thus, within the MSSM, the MFV hypothesis implies a strong restriction on the scale at which FCNCs are evaluated (e.g. $\mu_b$):
\bea
 \tilde {Q} M^2_{\tilde{Q}} \tilde{Q}^{\dagger}\quad &\propto&
 {\tilde Q} \left[ \sum x_n (Y_u Y_u^\dagger)^n\right] \tilde{Q}^{\dagger}  \sim {\tilde Q} \left[x_o I + x_1
   Y_uY_u^\dagger\right] \tilde{Q}^{\dagger}
\label{eq:mQ_mfv}
\eea
As a consequence we obtain the same CKM factors for the soft masses as in the SM and only the flavour-independent magnitude of FCNC amplitudes are modified, for example:
\bea
A(b\rightarrow s\gamma)\propto \left[ (V_{\rm CKM})^*_{ts} (V_{\rm
    CKM})_{tb} \right], \quad \Delta M_{B_{d,s}} \propto\left[
  (V_{\rm CKM})_{tb} (V_{\rm CKM})_{td,s}\right]^2%\nn.
\eea
Of course, the motivation for the MFV conditions is the strong suppression of BSM effects in flavour parameters (e.g. SM CKM fits). However, one must check carefully that the form of  $\tilde{Q} M^2_{\tilde{Q}} \tilde{Q}^{\dagger}$ in fact does not violate that assumption in the sense that it could give a large deviation from the MFV conditions.  Once the running of Yukawa matrices is included this can never be the case.

If we put the MFV conditions at the GUT scale for arbitrary initial conditions, this can result into catastrophic FCNCs at the EW scale.  Since the main goal of FS is to restrict the form of Yukawa matrices, $Y_f$, we then know  the form of the terms
\bea
Y^{\dagger}_f Y_f, \quad Y_fY^{\dagger}_f, \quad a^{\dagger}_f a_f\quad
{\rm{and}} \quad a_f a^{\dagger}_f
\label{eq:prod_yfdag_yf}
\eea
at the GUT scale.
We can restrict then the initial conditions of off-diagonal terms of the soft parameters by the off-diagonal elements of the terms \eq{eq:prod_yfdag_yf}. The specific way of correlating this can be obtained with a FS governed by a non-minimal supergravity theory.

The MFV conditions in supersymmetry have been extended to include terms such as  $x_2 Y_dY^\dagger_d$  $+x_3Y_dY^\dagger_d Y_uY^\dagger_u$   $+x_4Y_uY^\dagger_u  Y_dY^\dagger_d$ in \eq{eq:mQ_mfv} and  a similar 
structure to other supersymmetric terms \cite{D'Ambrosio:2002ex}. In this case, gluino contributions to FCNC can compete with those of chargino and charged Higgs contributions. We present examples below of such cases when $A_0$ is diagonal  at the GUT scale.
%
%
%%%%%%%%%%%%%%%%%%%%%%%%%%%%%%%%%%%%%%%%%%%%%%%%%%%%%%%%%%%%%%%%%%%%%%%%%%%%%%%%%%%%%%%%%%%%%%%%%%%%%%%%%%%%%%%%%%%%%%%%%
%
%
\section{Flavour symmetries with an underlying supergravity theory \label{sec:fsug}}

\subsection{Trilinear terms}

The expansion parameters appearing in the Yukawa couplings, $\ep_d$ and $\ep_u$, in flavour symmetries arise due to scalar fields, flavons, acquiring a vev, $\la \theta \ra$,  and hence breaking the flavour symmetry at some scale $M$:
\bea
(Y_{f})_{ij} \propto \left(\frac{\la\theta \ra}{M}\right)^{\alpha^f_{ij}},
\label{eq:yukflav}
\eea
For example in $SU(3)$ flavour symmetries, the Yukawa matrices have the form
\bea
Y_f/(Y_f)_{33}=\left(
\begin{array}{ccc}
0        & c^f_{12}\ep^3_f e^{i\varphi^f_{12}} & c^f_{13}\ep^3_f e^{i\varphi^f_{13}}  \\
c^f_{21}\ep^3_f e^{i\varphi^f_{21}} & c^f_{22}\ep^2_f e^{i\varphi^f_{22}} & c^f_{23} e^{i\varphi^f_{23}}\ep^2_f\\
c^f_{31}\ep^3_f e^{i\varphi^f_{31}} & c^f_{32}e^{i\varphi^f_{32}} \ep^2_f & 1 \\
\end{array}
\right),\label{eq:yuksu3_pr}
\eea
Note that we will be assuming a symmetric form for the Yukawa matrices
so that $c_{ij} = c_{ji}$ and we drop the subscript $sym$. As noted earlier, there are only 3 physical CP violating phases and therefore
there are relations between the phases $\varphi^f_{ij}$ in
  \eq{eq:yuksu3_pr}. This equation
% \eq{eq:yuksu3_pr} 
represents the most general parameterization of the Yukawa matrices for quarks and charged leptons with an $SU(3)$ FS
 \cite{Ross:2004qn,su3refs}. In any supersymmetric FS, the flavon fields also couple to the sfermions through the trilinear terms
\bea
(a_f)_{ij} H_f Q_iq^c_j.
\eea
The generic form of the trilinear coupling matrices in models of supersymmetry breaking such as supergravity mediation, gauge mediation or anomaly mediation, is of the form
\bea
(a_f)_{ij}=(Y_f)_{ij}(A^f_{0})_{ij}.
\label{eq:trilsft}
\eea
In minimal supergravity $A_0$ is a constant and hence the proportionally in \eq{eq:at_proptoyktrl} is achieved.
Once a family symmetry is considered \cite{Ross:2004qn}, there are additional contributions to the trilinear couplings given by derivatives of the Yukawa couplings with respect to the flavon fields \cite{Ross:2004qn,Ross:2002mr},
and  the trilinear terms can then be written as 
\bea
(a_f)_{ij}=(Y_f)_{ij}((A^f_{0})_{ij}+k^f_{ij}), 
\label{eq:trilsft+fys}
\eea
where the $k^f_{ij}$ are the coefficients produced when taking the derivatives with respect to the flavon fields times the gravitino mass,  $k^f_{ij} \propto \alpha^f_{ij}m_{3/2}$.

\subsection{Soft squared masses}

The soft squared mass terms in general scenarios of soft supersymmetry breaking remain diagonal, although the diagonal terms can be different from each other. However in theories with underlying supergravity and family symmetries, the off-diagonal terms involve powers of the flavon fields. Of course the diagonal terms are free of such contributions since the terms in the Lagrangian involving the soft squared masses, \eq{eq:sftLg}, need to respect Hermiticity.

Once the flavour symmetry is specified, the K\"ahler potential can be trivially written as
\bea
K&=&\sum_{\psi}\psi^i\psi^{\dagger \bar{j}}K_{i\bar{j}}(\psi),\quad \psi=u_R,\ d_R,\ e_R,\ \nu_R,\ Q_L,\ L_L,\\
K_{i\bar{j}}(\psi)&=&\delta_{i\bar{j}}\left[c(\psi)+d(X_p,\psi) X^{p}X^\dagger_{p}\right]+  \frac{\theta^a_i\theta^{\dagger}_{a \bar{j}}}{M^2(\theta_a)}\left[c(\theta_a, \psi)+ d(\theta_a, X_p, \psi) X^p X^{\dagger}_p\right].
\label{eq:Kahler}
\eea
Here $\theta_{k}$ are the flavon fields and $X_p$ are the fields breaking supersymmetry.
The coefficients $c$ and $d$  are a-priori un-related coefficients that can be different for each different flavon and $X$ field. In abelian flavour symmetries, they are different for each fermion species but in non-Abelian symmetries the coefficients of one species to another are often related. For example in $SU(3)$ flavour models, \cite{Ross:2004qn,su3refs,King:2004tx}, we have structures like
\bea
M^2 =m^2_{3/2}\mathbf{1}-m^2_0
\left(
\begin{array}{ccc}
r_1  & 0   & 0 \\
0    &r_2 &O\left(\frac{\langle\theta\rangle^2}{M^2}\right) \\
0    &  O\left(\frac{\langle\theta\rangle^2}{M^2}\right)  & r_3
\end{array}
\right),
\eea
where it is important to stress that due to the vacuum alignment
  of the flavon fields, \cite{Ross:2004qn,su3refs} there are no $D$
  terms contributing to the soft-squared masses, which often
present dangerous contributions to FCNC \cite{Babu:2003zz}.
In what follows, we adopt the following form for the soft-squared masses
\bea
M^2_{\tilde{f}}=
\left(
\begin{array}{ccc}
{M^2_{\tilde{f}}}_{11}  & 0                            & 0                  \\
0                  & {M^2_{\tilde{f}}}_{22}            & {M^2_{\tilde{f}}}_ {23} \\
0                  & {M^{2\dagger}_{\tilde{f}}}_{23}  & {M^2_{\tilde{f}}}_ {33}  \\
\end{array}
\right),\quad f=Q,u,d,L,e.
\eea
%

%%%%%%%%%%%%%%%%%%%%%%%%%%%%%%%%%%%%%%%%%%%%%%%%%%%%%%%%%%%%%%%%%%%%%%%
%
\section{Determination of the flavour structure at the GUT scale \label{sec:det_f_GUT}}
%
%%%%%%%%%%%%%%%%%%%%%%%%%%%%%%%%%%%%%%%%%%%%%%%%%%%%%%%%%%%%%%%%%%%%%%%
%
%
\subsection{General assumptions}
Having specified the assumptions that we make based on an underlying supergravity structure, we 
next describe how these structures are realized at the electroweak scale and how we might implement experimental constraints on the structure at the GUT scale.
In a complete theory of FS, the full superpotential and K\"ahler potential would be specified, including all the coefficients appearing in \eq{eq:yukflav}, \eq{eq:trilsft+fys} and \eq{eq:Kahler}. One can then impose those initial conditions at the GUT scale and run those parameters down to the EW scale.
Lacking a complete theory we can at best begin with
 the general form of the Yukawa matrices, \eq{eq:yukmat_sym}, and the supergravity assumptions described in the previous section at the GUT scale and apply the constraints from FCNC, mainly $\BR(b\rightarrow s \gamma)$ to limit the possible values of the parameters of the flavour structure for Yukawa matrices and soft terms at the GUT scale. To this end, we implement an iterative procedure, like the one followed in analyses of the constrained MSSM (CMSSM) \cite{cmssm,cmssmwmap}.
 
In the CMSSM,  gaugino masses are assumed to be unified at the GUT scale with value $m_{1/2}$, as are the soft scalar masses with value $m_0$ and trilinear couplings with value $A_0$. Typically, the flavour structure for the scalar masses and trilinear terms is assumed to be diagonal at both the GUT and EW scales.  That is, the running of the off-diagonal terms is also neglected. With this set of boundary conditions at the GUT scale, we can use the radiative electroweak 
symmetry breaking conditions by specifying the ratio of the two Higgs vacuum expectation values, $\tan \beta$, and the mass, $M_Z$, to predict the values of the Higgs mixing mass parameter, $\mu$ and the bilinear coupling, $B$. The sign of $\mu$ remains free. 

The CMSSM is subject to a number of phenomenological constraints. These include the LEP limits on the chargino mass: $m_{\chi^\pm} > 104$~GeV~\cite{LEPsusy}, on the selectron mass: $m_{\tilde e} > 99$~GeV~ \cite{LEPSUSYWG_0101} and on the Higgs mass: $m_h >114$~GeV~\cite{LEPHiggs}. The former two constrain $m_{1/2}$ and $m_0$ directly via the sparticle masses, and the latter indirectly via the sensitivity of radiative corrections to the Higgs mass to the sparticle masses, principally $m_{\tilde t, \tilde b}$. The Higgs limit  imposes important constraints principally on $m_{1/2}$ particularly at low $\tan \beta$. Another constraint is the requirement that the branching ratio for $b \rightarrow
s \gamma$ is consistent with the experimental measurements \cite{bsgexp,hfag}. 
These measurements agree at the 1-$\sigma$ level with the Standard Model, and therefore provide bounds on MSSM particles \cite{Degrassi:2000qf},  such as the chargino and charged Higgs masses, in particular. Typically, the $b\rightarrow s\gamma$ constraint is more important for $\mu < 0$, but it is also relevant for
$\mu > 0$,  particularly when $\tan\beta$ is large. The constraint imposed by
measurements of $b\rightarrow s\gamma$ also excludes small
values of $m_{1/2}$. To apply the constraint from $b\rightarrow s\gamma$
to the supersymmetric parameter space, we convolve the experimental result
given in Eq. \ref{exp_brbsg} with the theoretical calculated result and uncertainty.
We add linearly the
uncertainty generated from the variation of the input
matching scale to the SM error given in Eq \ref{sm_brbsg}. 
We also add linearly the systematic uncertainties in the experimental result 
given in Eq. \ref{exp_brbsg}
which could allow central values of $b\rightarrow s\gamma$ significantly
outside the range inferred from the statistical error in Eq. \ref{exp_brbsg}. 
For example, in the CMSSM, the 95\% CL lower limit to $m_{1/2}$ at high $\tan \beta$
occurs when the calculated central value of the branching ratio of $b\rightarrow s\gamma$ is
as low as $1.9 \times 10^{-4}$ \cite{ehoww}.
Finally, there are
regions of the $(m_{1/2}, m_0)$ plane that are favoured by
the BNL measurement \cite{newBNL} of $g_\mu - 2$ at the 2-$\sigma$ level, corresponding to 
a deviation  from the Standard Model 
calculation~\cite{Davier} using $e^+ e^-$ data.  

If R-parity is conserved, the lightest supersymmetric particle becomes a cold dark matter
candidate \cite{EHNOS} and
 therefore the model becomes subject to cosmological
constraints on the relic density of dark matter (usually the lightest neutralino) \cite{cmssmwmap}.
In the context of the $\Lambda$CDM model, the WMAP only results indicate \cite{wmap}
\bea
\Omega_{CDM} h^2 = 0.1042^{+0.0081}_{-0.0080}
\label{wmap}
\eea
or a 2$\sigma$ range of 0.0882 -- 0.1204 for $\Omega_{CDM} h^2$.
This range of densities forces one into relatively narrow regions in a $m_{1/2},m_0$ plane
when $\tan \beta$ and $A_0$ are fixed.

In this paper, we will restrict our attention to models with
fixed $\tan \beta =50$ with $\mu > 0$. This is done to enhance somewhat the importance of
$b \to s \gamma$ constraints. 
We take $m_t = 171.4$ GeV and $m_b(m_b) = 4.25$ GeV. 
In the CMSSM with $\tan \beta = 50$ and $A_0 = 0$,
the Higgs mass constraint requires $m_{1/2} \gsim 350$ GeV
and is similar to the constraint from $b \to s \gamma$ for this choice of $\tan \beta$ and $A_0$.
At larger values of $m_{1/2}$, the value of $m_0$ is constrained to lie in a relatively 
narrow range to ensure that the co-annihilation of neutralinos and staus are
effective in reducing the relic density to the WMAP range.  At very large values of 
$m_{1/2}$, s-channel annihilations of neutralinos through heavy Higgs scalars
and pseudo-scalars control the relic density and larger values of $m_0$ are allowed.
The parameter choices used in the results presented in the next section have been made
such that the the relic density is within the WMAP range in the CMSSM.
We also note that to explain the $g_\mu - 2$ result at the 2-$\sigma$ level,
places an upper bound on $m_{1/2}$ of approximately 900 GeV.
We have not considered here parameter choices with very large values of $m_0$
typically associated with the focus point as this part of the plane is not very sensitive to 
FC processes.

\subsection{Form of the Yukawa matrices and soft terms}
Our goal in this paper is to test the effect of including off-diagonal terms in the scalar and 
trilinear matrices. We will restrict our attention to the mixing between the second and third
generations only (mixing with the first generation will be of higher order in the expansion
parameters $\ep_u$ or $\ep_d$ as discussed in the previous section).
At the EW scale, we set up off-diagonal entries of $Y_d$ using the CKM mixings. We also assume the same mixings for the charged lepton sector. At the GUT scale we choose the following form of the Yukawa matrices, which contain 10 parameters:
\bea
Y_u=\left(
\begin{array}{ccc}
0 & 0                & 0\\
0 & c^u_{22} \ep^2_u & 0\\
0 & 0                & c^u_{33}
\end{array}
\right),\quad 
Y_d=\left(
\begin{array}{ccc}
0 & 0                & 0\\
0 & c^d_{22} \ep^2_d & c^d_{23} \ep^2_d\\
0 & c^d_{23} \ep^2_d & c^d_{33}
\end{array}
\right), \quad
Y_e=\left(
\begin{array}{ccc}
0 & 0                & 0\\
0 & c^e_{22} \ep^2_d & c^e_{23} \ep^2_d\\
0 & c^e_{23} \ep^2_d & c^e_{33}
\end{array}\label{eq:yuk_o_init}
\right).
\eea
In the up sector, we have the masses of the top and charm quarks to determine 
two of the three parameters in $Y_u$, thus we have just one free parameter which we choose 
to be $\ep_u$. The $d$ and $e$ sectors are linked because we are
choosing to generate the mixing in the lepton sector using the
$V_{\rm{CKM}}$ matrix:
\bea
Y_d=V^*_{\rm{CKM}} \hat Y_d V^\dagger_{\rm{CKM}},\quad
Y_e=V^*_{\rm{CKM}} \hat Y_e V^\dagger_{\rm{CKM}},
\label{eq:rel_yd_ye}
\eea
where $\hat Y_f$ are the diagonal matrices of the bare Yukawa couplings at the EW scale. 
In the $e$ and $d$ sectors combined, there are seven parameters.
We can use the masses of $\mu$, $\tau$, and the masses of the strange and bottom quarks
to eliminate four parameters. An additional parameter is eliminated by the $V_{\rm{CKM}}$ matrix,
leaving us with two free parameters which we choose to be $\ep_d$ and $c^d_{22}$.
Thus choosing
\bea
\ep_u,\quad \ep_d,\quad {\rm{and}}\quad c^d_{22}
\eea
 as free parameters at the GUT scale allows us to determine  the remaining coefficients 
\bea
c^u_{22},\quad c^u_{33},\quad c^d_{23},\quad c^d_{33}, \quad c^e_{22},\quad c^e_{23}, \quad {\rm{and}} \quad c^e_{33}.
\eea
The fits are done at the appropriate low energy scale and run up and down iteratively 
to obtain a consistent solution given our set of input parameter values.

Although we are choosing $\ep_u$ and $\ep_d$ as free parameters, their range is restricted by
requiring convergence of the running of the RGEs. Outside this range, Yukawa couplings tend
to diverge while running down to the EW scale. This is reminiscent of the restriction to $\tan \beta$
in the CMSSM. 
The allowed range can be estimated from the running of the mass ratios $m_s/m_b$ and $m_c/m_t$, which are directly correlated to the values $\ep_d$ and $\ep_u$ respectively by Eq. (38). These ratios increase from the EW to the GUT scale in the MSSM. We use  an analytic form 
\cite{Olechowski:1990bh} for the ratios of Yukawa couplings derived from one-loop  RGEs for quark mixings and  solved in the presence of the heaviest Yukawa generation. In this approach the 
evolution parameter, $\chi$  
is simply related to the evolution of the mass ratios $m_s/m_b$ and $m_c/m_t$ by
\bea
\chi=\left(\frac{M_G}{M_Z}\right)^{-y^2_t/16/\pi^2}\approx 0.8,\quad
\frac{(m_s/m_b)(M_G)}{(m_s/m_b)(M_Z)}=\chi,\quad
\frac{(m_c/m_t)(M_G)}{(m_c/m_t)(M_Z)}=\chi^3.
\eea
Then at $M_{\rm{GUT}}$, $\ep_d \approx \sqrt{m_s/m_b} \approx 0.1$ and $\ep_u \approx \sqrt{m_c/m_t} \approx 0.05$  and hence the range for $\ep_d$ and $\ep_u$ should be restricted to be close to $0.1$ and $0.05$ respectively.

With the above parameterization of Yukawa matrices then we can parameterize the soft masses as follows:
\bea
M^2_{\tilde{Q}} & = & M^2_{\tilde{u}_R}=\left(
\begin{array}{ccc}
1 &  0               & 0     \\
0 &  x^u_{22}        & x^u_{23}\ep^2_{Su} \\
0 &  x^u_{23}\ep^2_{Su} & x^u_{33}        
\end{array}
\right)m^2_0,\quad
M^2_{\tilde{d}_R}\ = \ \left(
\begin{array}{ccc}
1 &  0     &  0  \\
0 &  x^d_{22}     & x^d_{23}\ep^2_{Sd} \\
0 &  x^d_{23}\ep^2_{Sd} & x^d_{33}  
\end{array}
\right)m^2_0\nn\\
M^2_{\tilde{L}} & = & \left(
\begin{array}{ccc}
1 &  0     &  0  \\
0 &  x^\nu_{22}        & x^\nu_{23}\ep^2_{Su} \\
0 &  x^\nu_{23}\ep^2_{Su} & x^\nu_{33}  
\end{array}
\right)m^2_0,\quad \quad \quad \quad
M^2_{\tilde{e}_R} \ = \ \left(
\begin{array}{ccc}
1 &  0     &  0  \\
0 &  x^e_{22}     & x^e_{23}\ep^2_{Se} \\
0 &  x^e_{23}\ep^2_{Se} & x^e_{33}  
\end{array}
\right)m^2_0\nn\\
a_u &=& \left(
\begin{array}{ccc}
0  &  0      &  0 \\
0  &  z^u_{22}\ep^2_{Su}  &  0 \\
0  &  0      &  z^u_{33} 
\end{array}
\right)A_0,\quad \quad \quad \quad \quad \quad
a_d \ = \ \left(
\begin{array}{ccc}
0  &  0         &  0 \\
0  &  z^d_{22} \ep^2_{Sd}  &  z^d_{23} \ep^2_{Sd} \\
0  &  z^d_{23} \ep^2_{Sd}  &  z^d_{33} 
\end{array}\right)A_0,\nn\\
a_e & = & \left(
\begin{array}{ccc}
0  &  0         &  0 \\
0  &  z^e_{22} \ep^2_{Se}  &  z^e_{23} \ep^2_{Se} \\
0  &  z^e_{23} \ep^2_{Se}  &  z^e_{33} 
\end{array}\right)A_0
\label{eq:sofmasses_offneq0}
\eea

In the spirit of the CMSSM, we can simply choose all of the $x^f_{ij} = 1$,
$\ep_{Sf} = \ep_f$, and
set the trilinear couplings to be aligned with the Yukawa matrices,
such that $z^f_{ij}=c^f_{ij}$. We will make this simplification in sections 6.1 - 6.3.

Later, in section 6.4, we will use the generality obtained when  $\ep_{Sd}, \ep_{Se}$, and  
$\ep_{Su}$ are assumed to be unrelated to  $\ep_{d}$ and  $\ep_{u}$ to
set constraints on these parameters from $\bsgmm$.  We can test
departures from the CMSSM by effectively taking coefficients
$x^f_{ij}$ different from unity, but that is clearly beyond the scope
of the present study. The matrices are then simplified to:

\bea
M^2_{\tilde{Q}} & = & M^2_{\tilde{u}_R}=\left(
\begin{array}{ccc}
1 &  0               & 0     \\
0 &  1               & \ep^2_{Su} \\
0 &  \ep^2_{Su}      & 1
\end{array}
\right)m^2_0,\quad
M^2_{\tilde{d}_R}\ = \ \left(
\begin{array}{ccc}
1 &  0     &  0  \\
0 &  1     & \ep^2_{Sd} \\
0 &  \ep^2_{Sd} & 1  
\end{array}
\right)m^2_0\nn\\
M^2_{\tilde{L}} & = & \left(
\begin{array}{ccc}
1 &  0     &  0  \\
0 &  1     & \ep^2_{Su} \\
0 &  \ep^2_{Su} & 1  
\end{array}
\right)m^2_0,\quad \quad \quad \quad
M^2_{\tilde{e}_R} \ = \ \left(
\begin{array}{ccc}
1 &  0     &  0  \\
0 &  1     & \ep^2_{Se} \\
0 &  \ep^2_{Se} & 1  
\end{array}
\right)m^2_0\nn\\
a_u &=& \left(
\begin{array}{ccc}
0  &  0         &  0 \\
0  &  c^u_{22}\ep^2_{Su}  &  0 \\
0  &  0      &  c^u_{33} 
\end{array}
\right)A_0,\quad \quad \quad \quad \quad \quad
a_d \ = \ \left(
\begin{array}{ccc}
0  &  0         &  0 \\
0  &  c^d_{22}\ep^2_{Sd}  &  c^d_{23}\ep^2_{Sd}  \\
0  &  c^d_{23}\ep^2_{Sd}  &  c^d_{33} 
\end{array}\right)A_0,\nn\\
a_e & = & \left(
\begin{array}{ccc}
0  &  0        &  0 \\
0  &  c^e_{22} \ep^2_{Se} &  c^e_{23} \ep^2_{Se}\\
0  &  c^e_{23} \ep^2_{Se} &  c^e_{33} 
\end{array}\right)A_0
\label{eq:sofmasses_offneq0_ef_diff_efS}
\eea
Note that for this case the trilinear couplings ($\sim z^f_{ij}\ep^p_{f}$) will not be
exactly aligned with the Yukawa couplings but will still be proportional  to them and
effectively we are taking $z^f_{ij}=c^f_{ij}\ep^2_{Sd}/\ep^2_{d}$ for
$(i,j)\neq(3,3)$ and $z^f_{33}=c^f_{33}$.

%%%%%%%%%%%%%%%%%%%%%%%%%%%%%%%%%%%%%%%%%%%%%%%%%%%%%%%%%%%%%%%%%%%%%%%%%%%
%
\section{Results}
%
%%%%%%%%%%%%%%%%%%%%%%%%%%%%%%%%%%%%%%%%%%%%%%%%%%%%%%%%%%%%%%%%%%%%%%%%%%%
%

To get a feel for the range of parameter values of the coefficients $c^f_{ij}$ and $\ep_u$ and $\ep_d$ we first consider a
specific point in the susy parameter space with $m_{1/2} = 520$ GeV, $m_0 = 330$ GeV, $A_0 = 0$ , and $\tan \beta = 50$.   
We also consider a case with $A_0 = -m_{1/2}$  and we restrict our attention to $\mu > 0$. 
In the CMSSM, this point corresponds to one lying on the WMAP co-annihilation strip
\cite{cmssmwmap} so the that the relic density lies within the WMAP preferred range.
The Higgs mass is suitably large ($\gsim 116$ GeV), the susy contribution to $b \to s \gamma$ is small, and the contribution to $(g - 2)_\mu$ accounts for the discrepancy between theory and experiment.

The goal of the analysis is to constrain the allowed parameter space both from flavour changing neutral currents and from the parameters determining the appropriate values of fermion masses at electroweak scale. As noted earlier, our free parameters are $c^d_{22}$ and $(\ep_u,\ep_d)$. 
We have set $c^d_{22} = 0.7$.
The other $c^f_{ij}$ are then fixed by the low energy fermion masses and mixings. 
Since we expect the parameters $\ep_u$ and $\ep_d$ to describe the
structure of the Yukawa matrices, we do not want to rely on fine tuned
coefficients $c^f_{ij}$. Indeed, $V_{cb}$ would preclude a value of $c^d_{23}$ as small as 0.1
if $\ep_d$ is also taken of order 0.1.

%%%%%%%%%%%%%%%%
%
%
\subsection{Running of off-diagonal soft parameters\label{sec:runn_off_soft_p}}

Since we work in the context of the CMSSM, it is important to first establish the degree
of running of our off-diagonal terms between the GUT and EW scales. 
The 1-loop RGE's for $(M^2_{\tilde Q})_{ij}$, $(M^2_{\tilde d_R})_{ij}$ and  $a_d$  are given by\footnote{
Note that we employ a sign convention for $a$ which differs by a relative sign from that of \cite{Martin:1993zk}.}
\cite{Martin:1993zk}
\bea
\label{eq:beta1l_soft}
\beta^{(1)}_{M^2_{\tilde Q}} &=&(M^2_{\tilde Q} +2 M^2_{H_u}) Y^{\dagger}_u Y_u +
(M^2_{\tilde Q} + 2M^2_{H_d}) Y^{\dagger}_d Y_d + \left[ Y^{\dagger}_u Y_u + Y^{\dagger}_d Y_d  \right]  M^2_{\tilde Q}\nn\\
 &+& 2 Y^\dagger_u M^2_u Y_u + 2 Y^\dagger_d M^2_d Y_d  +2 a^\dagger_d a_d\nn\\
 &-& \frac{32}{3} g^2_3 M^2_3 - 6 g^2_2 M^2_2 -\frac{2}{9} g^2_1 M^2_1 + 
\frac{1}{5} g^2_1 {\mathcal{S}}\nn\\
\beta^{(1)}_{M^2_{\tilde d}} &=&  (2 M^2_{\tilde d} + 4 M^2_{H_d}) Y_d Y^{\dagger}_d + 4 Y_d M^2_{\tilde Q} Y^\dagger_{d} + 2 Y_d Y^\dagger_d M^2_{\tilde d}  \nn\\
&+& 4 a_d a^{\dagger}_{d} -\frac{32}{3} g^2_3 |M_3|^2 -\frac{8}{9} g^2_1 |M_1|^2 + \frac{2}{3} g^2_1 {\mathcal{S}}\nn\\
\beta^{(1)}_{a_{d}} &=& a_d \left\{ {\rm{Tr}} (3 Y_d Y^\dagger_d +Y_e Y^\dagger_e) + 5 Y^\dagger_d Y_d  Y^\dagger_u Y_u  - \sum_{a=1,2,3} u_a g_a^2\right\}\nn\\
&+& Y_d \left\{  {\rm{Tr}} (6 a_d Y^\dagger_d + 2 a_e Y^\dagger_e) + 4 Y^\dagger_d a_d +2 Y^\dagger_u a_u - 2\sum_{a=1,2,3} u_a g_a^2 M_a \right\}
\eea
where $u_a = 7/9, 3, 16/3$ are numerical coefficients, $g_a$ are the gauge couplings and $M_a$ are the masses of the gauginos. We have not shown (though we do include them in our calculations) the beta functions for $M^2_{\tilde u}$,  $M^2_{\tilde L}$, and $M^2_{\tilde e}$.
$\beta^{(1)}_{M^2_{\tilde u}} $  is similar to $\beta^{(1)}_{M^2_{\tilde d}}$   with the interchange $u \leftrightarrow d$ and
different contributions from the gauginos. The running in the $\tilde
u$ sector is in general small due to our assumption that the expansion
parameter in the up sector is small in comparison with that of the
down sector. The running of $M^2_{\tilde L}$ 
($M^2_{\tilde e}$) is similar to that of  $M^2_{\tilde Q}$ 
($M^2_{\tilde d}$) due to their common dependence on $\ep_u$ ($\ep_d$). 
Here, we will describe
the running of  $(M^2_{\tilde Q})_{ij}$, $(M^2_{\tilde d_R})_{ij}$ and
$a_d$, but the running for $(M^2_{\tilde L})_{ij}$, $(M^2_{\tilde e_R})_{ij}$ and
$a_e$, has a similar behaviour.

 $Y_d$ is not diagonal at the GUT scale, however,  $Y_u$ is, and  although 
at two loops the evolution of $Y_u$ produces off-diagonal elements we
neglect them here, due to the smallness of $\ep_u$. Therefore, the
most important contributions to the beta functions of the off-diagonal terms of the 
the soft squared masses 
in \eq{eq:beta1l_soft} are found in the terms 
\bea 
(M^2_{\tilde Q} Y^\dagger_f Y_f)_{ij},\quad  (Y^\dagger_f Y_f M^2_{\tilde Q})_{ij},\quad  (Y^\dagger_f M^2_{\tilde f} Y_f)_{ij}, \quad (a^\dagger_d a_d)_{ij}
\label{eq:rel_soft}
\eea 
For the off-diagonal trilinear terms,  important contributions arise from the terms:
\bea
(a_d Y^\dagger_d Y_d)_{ij},\quad (Y_d Y^\dagger_d a_d)_{ij},\quad (Y_d {\rm{Tr}} (a_d Y^\dagger_d))_{ij},\quad (Y_d {\rm{Tr}} (a_e Y^\dagger_e))_{ij}.
\eea
Of course we need only consider the case for $f=d$ and $(i,j)=(2,3)$. Expressing $Y_d$ in terms of $\ep_d$, the contribution to the beta functions of the soft squared masses is
\bea
(M^2_{\tilde Q} Y^\dagger_d Y_d)_{23}  &=& \left[ c^d_{23} c^d_{33} \ep^2_d M^2_{\tilde Q ~ 22} +  {c^d_{33}}^2 M^2_{\tilde Q ~ 23} \right]   \rightarrow c^d_{23} c^d_{33} M^2_{\tilde Q ~ 22} \ep^2_d = c^d_{23} c^d_{33} \ep^2_d m^2_0 \nn\\
(Y^\dagger_d Y_d M^2_{\tilde Q})_{23}  &=& \left[ ({c^d_{22}}^2 + {c^d_{23}}^2) \ep^2_d M^2_{\tilde Q ~ 23}  + c^d_{23} c^d_{33} M^2_{\tilde Q ~ 33}\right]  \ep^2_d \rightarrow c^d_{23} c^d_{33} M^2_{\tilde Q ~ 33} \ep^2_d = c^d_{23} c^d_{33} \ep^2_d m^2_0  \nn\\
(Y^\dagger_d M^2_{\tilde f} Y_d )_{23} &=& \left[ c^d_{22}c^d_{23}  M^2_{\tilde d ~ 22}  +  {c^d_{23}}^2  M^2_{\tilde d ~ 32}  \right] \ep^4_d  + \left[ c^d_{22}c^d_{33}  M^2_{\tilde d ~ 23}  +  c^d_{23}c^d_{33}  M^2_{\tilde d ~ 33}  \right] \ep^2_d \nn\\
&& \rightarrow c^d_{22}c^d_{23} M^2_{\tilde d ~ 22} \ep^4_d +   c^d_{23}c^d_{33}  M^2_{\tilde d ~ 33}  \ep^2_d \simeq  c^d_{23}c^d_{33} \ep^2_d m^2_0\nn\\
(a^\dagger_d a_d)_{23} &=&   z^d_{23} z^d_{33}\ep^2_d A_0^2\rightarrow 0
\label{eq:lead_rel_soft}
\eea
where we have kept only like terms to lowest order in $\ep_d$. The expressions on the right-hand side of the arrows are the remaining non-zero contributions when the off-diagonal terms in 
eqs. (\ref{eq:propto1sms}) and (\ref{eq:at_proptoyktrl}) are set to zero. 
From \eq{eq:sofmasses_offneq0} we can see that since $M^2_{\tilde Q ~ 22}$  $>> M^2_{\tilde Q ~ 23}$, even when the off-diagonal terms of \eq{eq:rel_soft} are set to zero, the leading terms $\sim M^2_{\tilde Q ~ ii}\ep^2_d$,  of \eq{eq:lead_rel_soft} remain, except for the term $a^\dagger_d a_d$, which vanishes.
For the contribution to the beta functions of the trilinear couplings we have
\bea
(a_d Y^\dagger_d Y_d)_{23} &=& \sum_l [z^{d}_{2l}\ep^2_d] A_0  (Y^{\dagger}_d Y_d)_{l3}\rightarrow z^{d}_{22} c^d_{23} c^d_{33} \ep^4_d A_0\nn\\
(Y_d Y^\dagger_d a_d)_{23} &=&  z^{d}_{23}\ep^2_d A_0  (Y^{\dagger}_d Y_d)_{22} + z^{d}_{33} A_0  (Y^{\dagger}_d Y_d)_{23}
  \rightarrow  z^d_{33} c^d_{23} c^d_{33} \ep^2_d A_0  \nn\\
\left[Y_d {\rm{Tr}} (a_d Y^\dagger_d) \right]_{23} &=& c^d_{22} c^d_{23} \ep^6_d z^d_{22} A_0 + 
2 {c^d_{23}}^2 \ep^6_d z^d_{23} A_0 + c^d_{23} c^d_{33} \ep^2_d z^d_{33} A_0 \nn\\
&& \rightarrow \left[c^d_{22} c^d_{23} \ep^6_d z^d_{22} + c^d_{23} c^d_{33} \ep^2_d z^d_{33}\right] A_0 \nn\\
\left[Y_d {\rm{Tr}} (a_e Y^\dagger_e)\right]_{23}  &=& c^e_{22} c^d_{23} \ep^6_d z^e_{22} A_0 + 
2 c^d_{23} c^e_{23} \ep^6_d z^e_{23} A_0 + c^d_{23} c^e_{33} \ep^2_d z^e_{33} A_0 \nn\\
&& \rightarrow \left[c^e_{22} c^d_{23} \ep^6_d z^e_{22} + c^d_{23} c^e_{33} \ep^2_d z^e_{33}\right] A_0 
\label{eq:cont_bf_a_f}
\eea
In \eq{eq:cont_bf_a_f} we see again that when the off-diagonal terms of the trilinear couplings $a_{d,e}$ are set to zero the leading terms of the original expression do not vanish. 

\begin{figure}[ht!]
\begin{center}
\includegraphics[width=.45\textwidth]{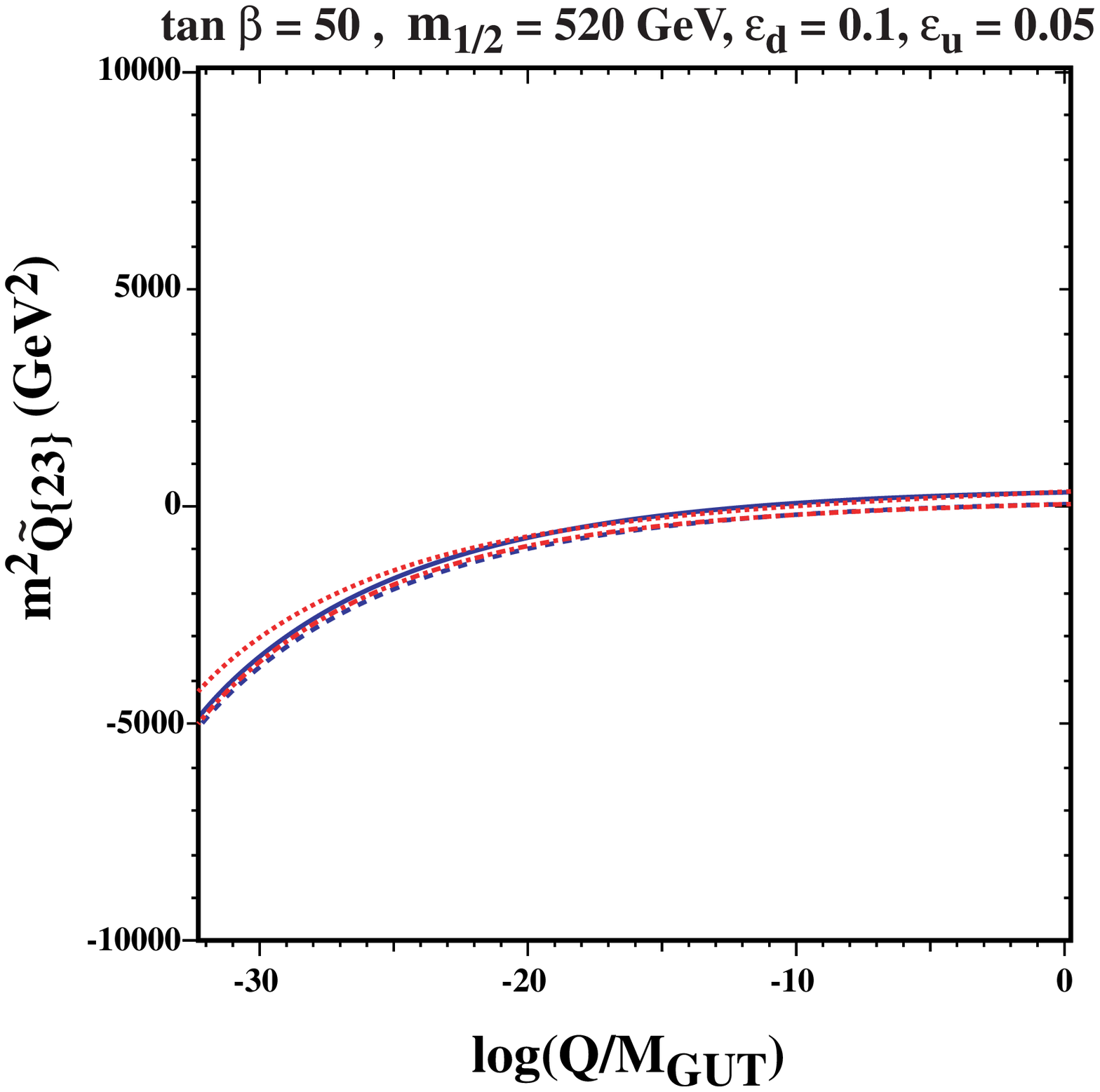}
\includegraphics[width=.45\textwidth]{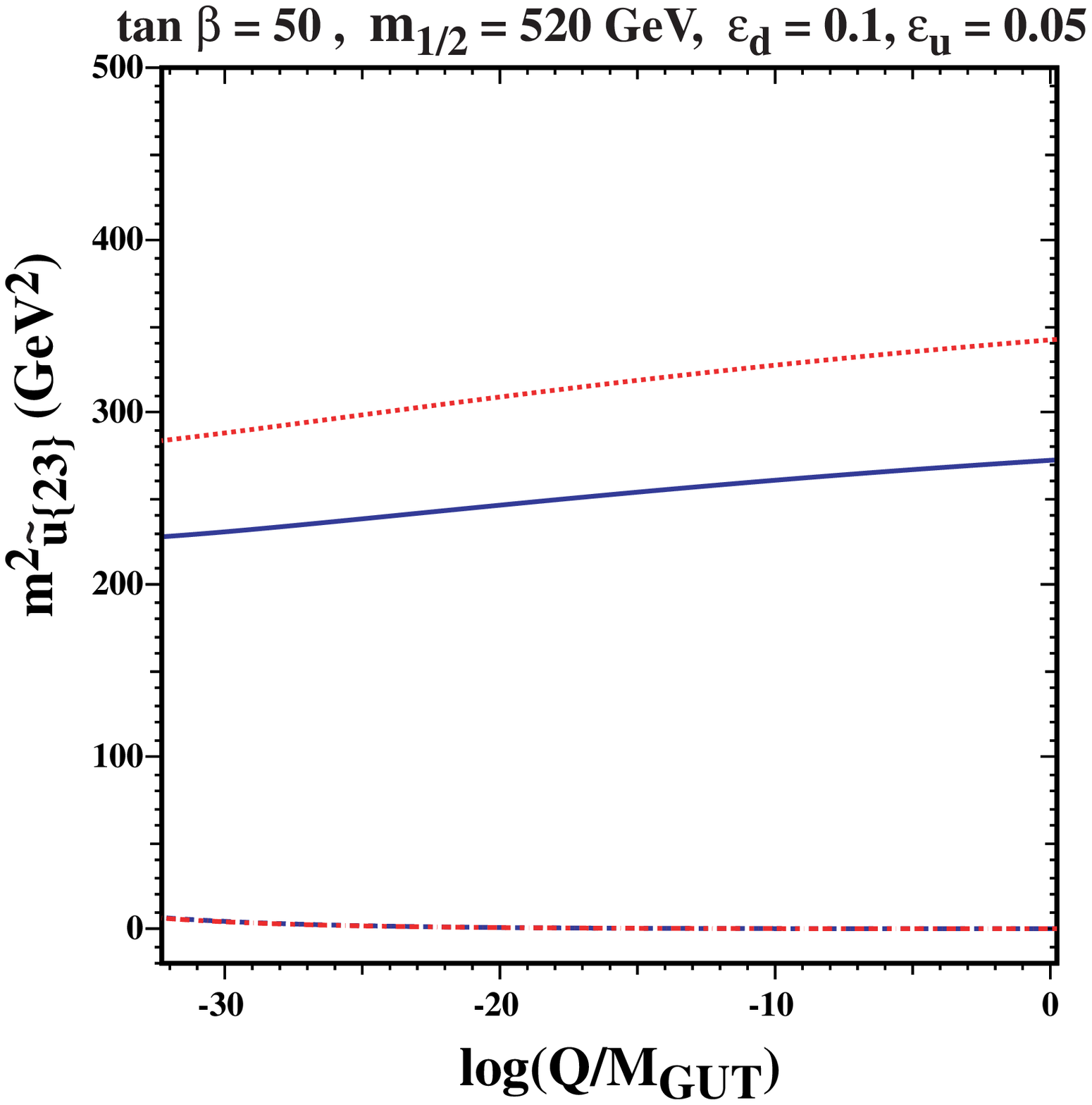}
\end{center}
\begin{center}
\includegraphics[width=.45\textwidth]{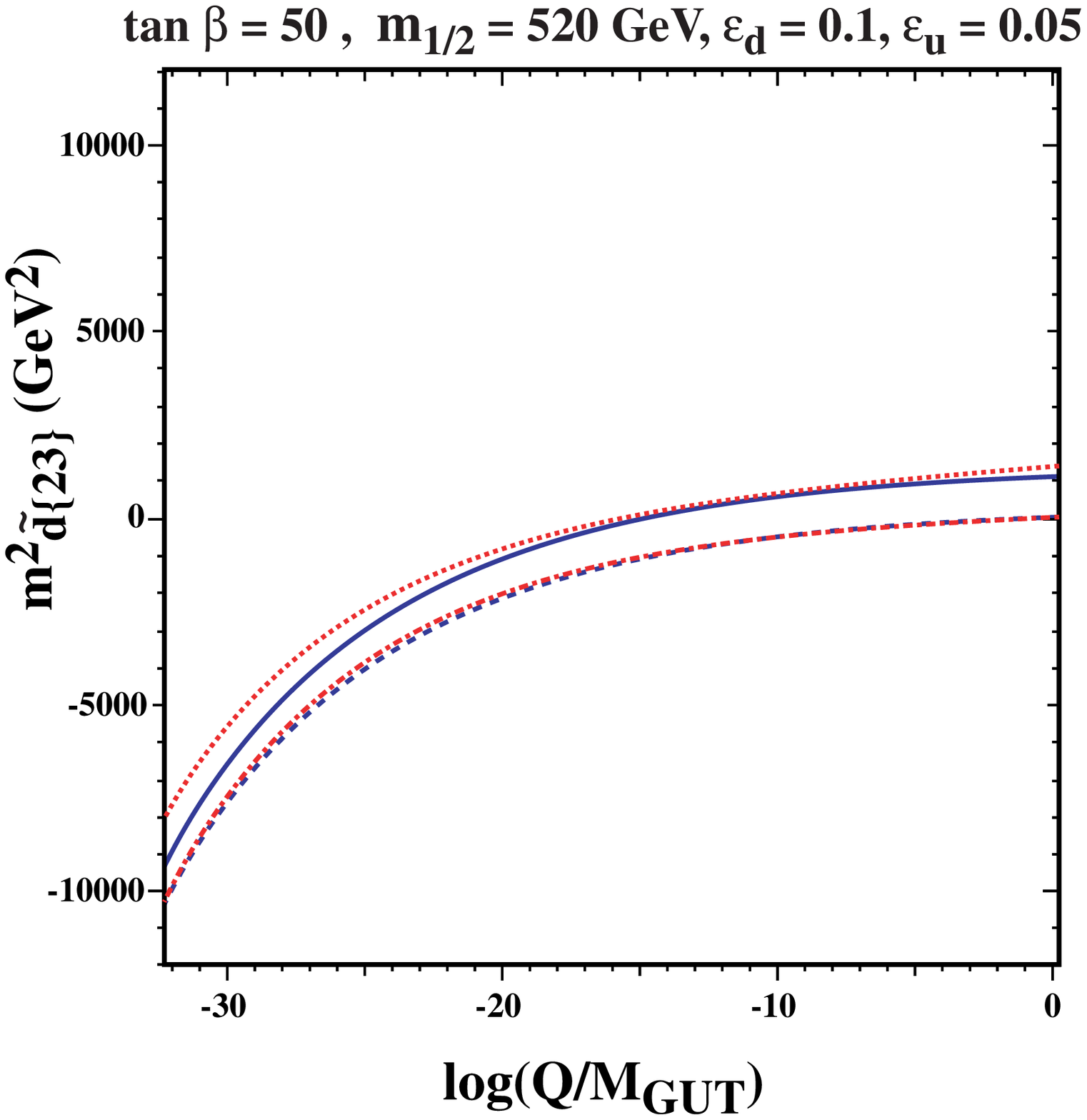}
\includegraphics[width=.45\textwidth]{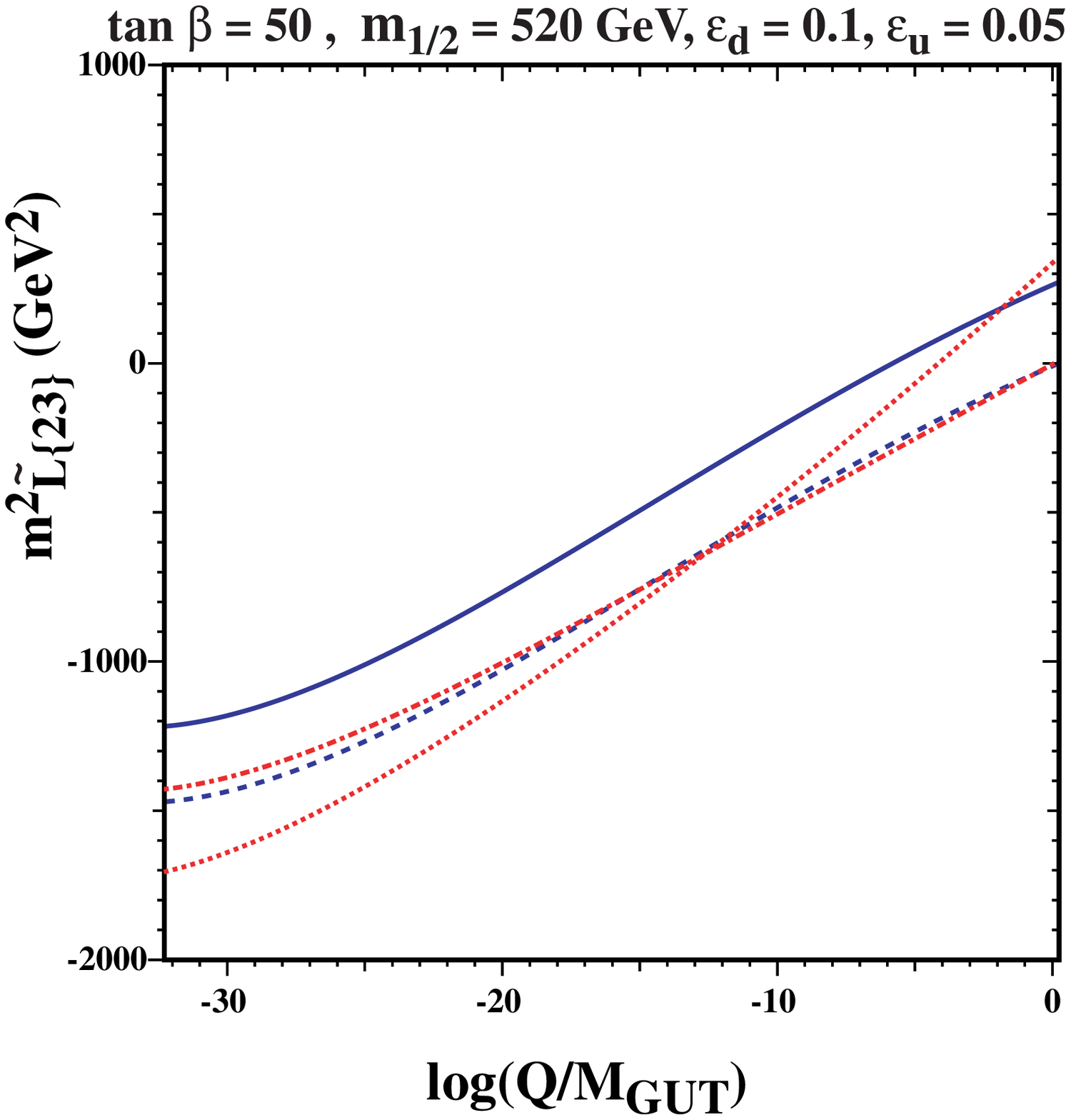}
\end{center}
\caption{\small The running of the parameters
of the soft squared masses a) $ M^2_{\tilde Q ~ 23}$, b) $ M^2_{\tilde u ~ 23}$, c)
$ M^2_{\tilde d ~ 23}$, and d) $ M^2_{\tilde L ~ 23}$ when off-diagonal terms are turned off 
(dashed blue)  and on (solid blue) at the GUT scale with $A_0 = 0$. 
Here $m_{1/2}=520$ GeV, $m_0=330$ GeV, $\ep_d=0.1$ and $\ep_u = 0.05$. 
Also shown is the running when $A_0 = -m_{1/2}$  and $m_0 = 370$ GeV.
The dot-dashed (red) curve corresponds to the case where off-diagonal terms are turned off
at the GUT scale and the dotted (red) curve corresponds to the case where they are 
turned on. }
\label{fig:ed_eu_softsqrd_comp_0.08_520_330_0}
\end{figure}

In Figure
\ref{fig:ed_eu_softsqrd_comp_0.08_520_330_0}, we show the running of the 
off-diagonal soft masses.
We begin by discussing the case for $A_0$ = 0.
Here, we have chosen $\ep_d = 0.1$ and $\ep_u = 0.05$ as well as $c^d_{22} = 0.7$
which will be used throughout this section.
The running of the soft masses in this case 
is shown by the dashed (when the off-diagonal terms of
the soft squared masses are set to zero at the GUT scale) 
and solid (when they are non-zero) blue curves.
In the first panel of Fig. \ref{fig:ed_eu_softsqrd_comp_0.08_520_330_0},
we show the running of $ M^2_{\tilde Q ~ 23}$ as a function of the renormalization scale $Q$.
When the off-diagonal terms are turned off, the running starts at $ M^2_{\tilde Q ~ 23} = 0$
as seen by the dashed (blue) curve.
The dominant contributions to the running of are suppressed by $\ep^2_d$ 
and proportional to $c^d_{23} c^d_{33}$. 
As a consequence, the running of  $ M^2_{\tilde Q ~ 23}$ is relatively significant.
When the off-diagonal elements are turned on at the GUT scale, the initial value for  $ M^2_{\tilde Q ~ 23}$ is about (16.5 GeV)$^2$ (as seen more clearly in panel b for  $ M^2_{\tilde u ~ 23}$).
On this scale, the difference in the running is relatively minor as all of the leading contributions
to these expressions are the same. As a result the evolution of these terms does
not change drastically according to the different initial conditions at the GUT
scale. 

Also shown in Fig. \ref{fig:ed_eu_softsqrd_comp_0.08_520_330_0}a is the
analogous running when $A_0 \ne 0$.  Here we show the running for the 
specific choice $A_0 = - m_{1/2} = -520$ GeV.  Note that for this case,
we have adjusted $m_0$ to 370 GeV in order to maintain the correct
relic density. 
When $A_0 \ne 0$, the
evolution of soft squared masses will have 
contributions from the terms $a^\dagger_f a_f$ when the
off-diagonal terms start with a non-zero value at the GUT scale. 
As the dominant contribution to the running of $ M^2_{\tilde Q ~ 23}$ 
comes from the gaugino (mostly gluino) terms in \eq{eq:beta1l_soft}, 
$a^\dagger_f a_f$  tends to lower (slightly) the running as seen by the dash-dotted 
(when off-diagonal terms are turned off at the GUT scale) and dotted (when they are 
turned on) (red) curves. 

In contrast, in panel b) of Fig. \ref{fig:ed_eu_softsqrd_comp_0.08_520_330_0}, 
we show the running of  $ M^2_{\tilde u ~ 23}$.  The beta function for $ M^2_{\tilde u ~ 23}$
is determined by $Y_u$ or $a_u$ rather than $Y_d$ and as a consequence the running is minimal.
As expected from the beta functions, the running of  $ M^2_{\tilde d ~ 23}$ is also significant
as seen in panel c).  We also show the running of  $ M^2_{\tilde L ~ 23}$ in panel d).
As noted earlier, the beta function for  $ M^2_{\tilde L ~ 23}$ is similar to that of 
 $ M^2_{\tilde Q ~ 23}$ though the running is smaller due to the lack of gluino contributions. 

Although the running of $(M^2_d)_{23}$ is the most
significant of those shown in Fig. \ref {fig:ed_eu_softsqrd_comp_0.08_520_330_0},
in the computation of $\BR(\bsgmm)$, discussed below, this
effect is sub-dominant.  The reason is that 
the contributions from the charged Higgs boson and the charginos is much
more important than the contribution from gluinos. We can see from
 \eq{eq:coups_for_mag_wilcoeff} that only
the diagonalization matrix of $M^2_{\tilde u}$,  $K^u$, is relevant for the
computation of the magnetic Wilson coefficients where the charginos
and charged Higgs boson are the virtual particles. The diagonalization
matrix of $M^2_{\tilde d}$, $K^d$, is relevant only for the gluino
contributions. This is because they enter into the computation of
chromo-magnetic Wilson coefficients arising from the virtual exchange
of gluinos \cite{Borzumati:1999qt}.

\begin{figure}[ht!]
\begin{center}
\includegraphics[width=.45\textwidth]{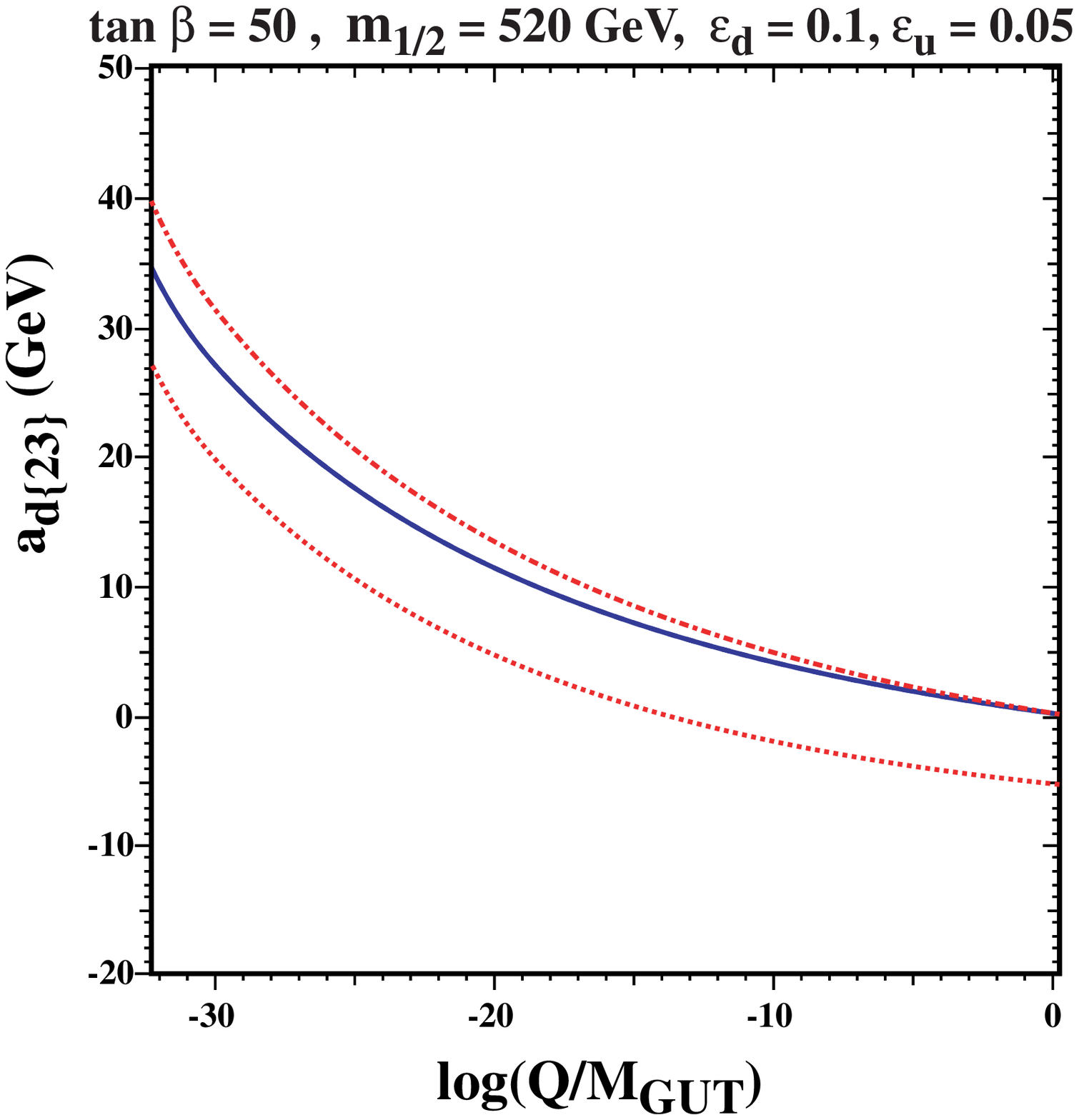}
\includegraphics[width=.45\textwidth]{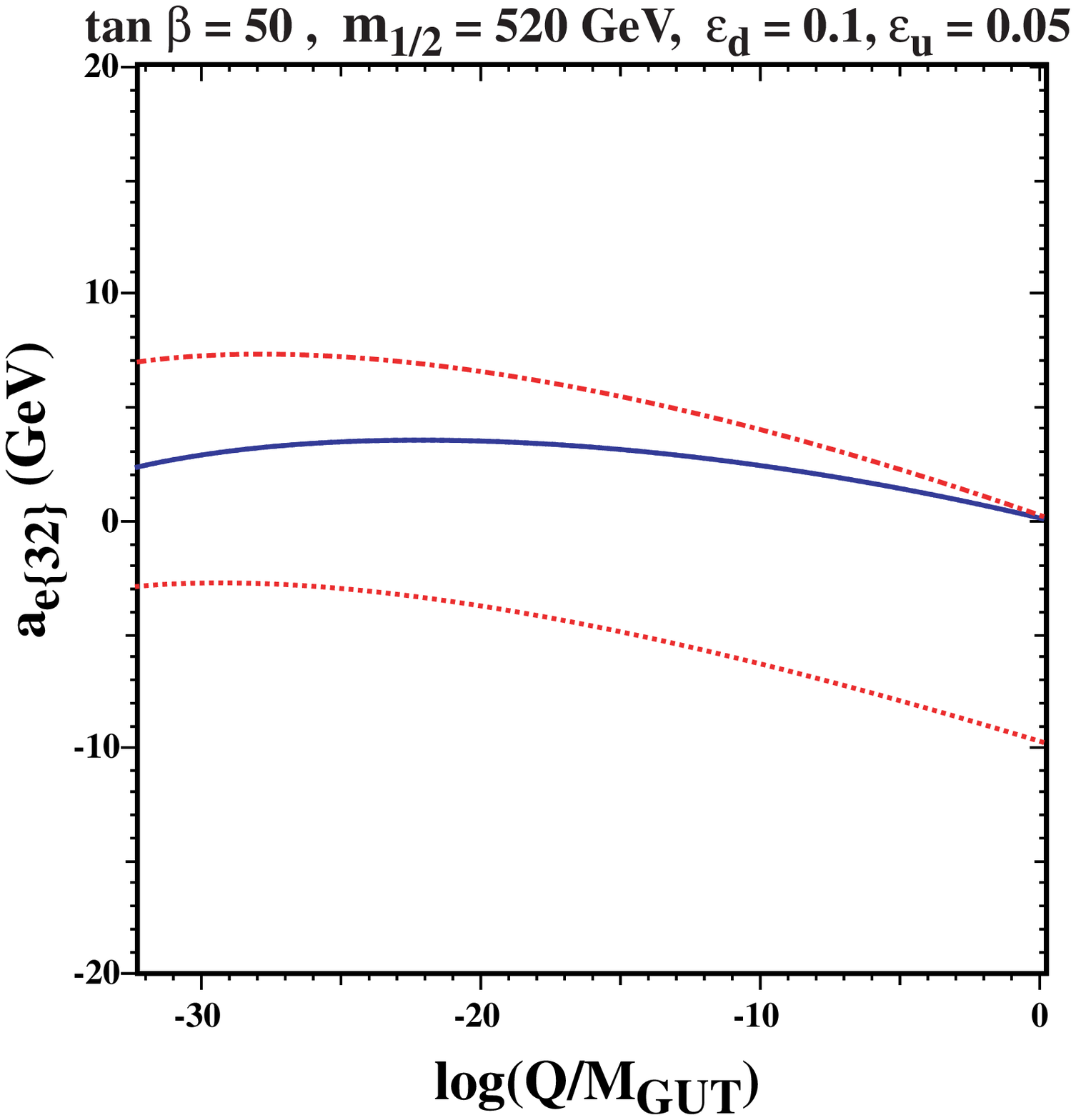}
\end{center}
\caption{\small As in Fig. \protect\ref{fig:ed_eu_softsqrd_comp_0.08_520_330_0}
for the running of a)  ${a_d}_{23}$ and b) ${a_e}_{32}$. }
\label{fig:ed_eu_aterms_comp_0.08_520_330_0}
\end{figure}

For $A_0=0$ the renormalization group equations generate a contribution
for $a_d$ of order $-Y_d \sum_a u_a M_a$  at one loop. In addition, when $A_0 = 0$,  the running of $a_d$ is independent of whether
off-diagonal terms of soft parameters are set to zero at the GUT scale. 
In Figure  \ref{fig:ed_eu_aterms_comp_0.08_520_330_0}, we show the running of ${a_d}_{23}$ 
in panel a)
and ${a_e}_{23}$ in panel b). The solid (blue) 
line corresponds to the case where $A_0 = 0$.
In both cases the running is very slight.

When $A_0 \ne 0$,  the first term of the one loop beta
function of $a_d$ will now be relevant. However since the  off-diagonal
terms $(a_f)_{ij}$ are suppressed with respect to the diagonal terms $(a_f)_{ii}$ as $(a_f)_{ij}/(a_f)_{ii}=\ep^2_d$, for $i,j=2,3$
and $f=d,e$, the difference will be small.  The running of ${a_d}_{23}$ 
and ${a_e}_{23}$ with $A_0 = -m_{1/2}$ is also shown in Fig. \ref{fig:ed_eu_aterms_comp_0.08_520_330_0}.

In order to determine the possible magnification of the flavour
violating parameters at the EW scale given the GUT scale initial conditions 
\eq{eq:sofmasses_offneq0}, it will be useful to define the parameters,
\bea
(\delta_{f,XY})_{ij}=\frac{(M^2_{f,XY})_{ij}}{\sqrt{(M^2_{f,XX})_{ii}
    (M^2_{f,YY})_{jj} }}
\label{eq:del_f_XY}
\eea
where $(M^2_{f,XY})_{ij}$, $X,Y=L,R$ are the running soft sfermion
parameters.
In order to give meaning to the flavour violating parameters, we must work
in the basis where fermion masses are diagonal.
The details of this rotation are given in Appendix \ref{app:inter} (see: \eq{eq:eff_rotatedmassm}).
In the definition of the $\delta$'s, all soft masses will be assumed to be in the 
rotated basis.
 
\begin{figure}[ht!]
\begin{center}
\includegraphics[width=.335\textwidth]{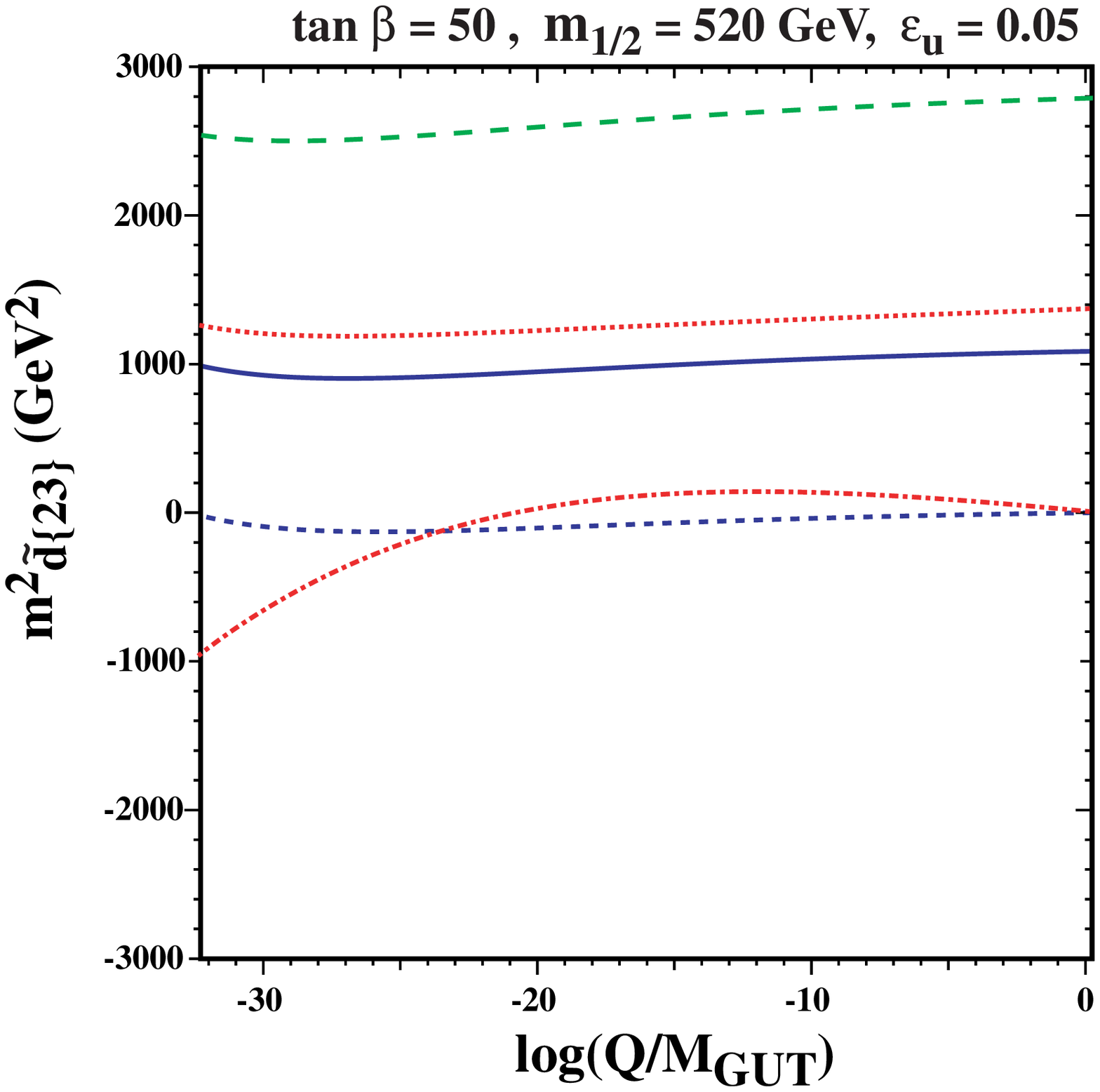}
\includegraphics[width=.32\textwidth]{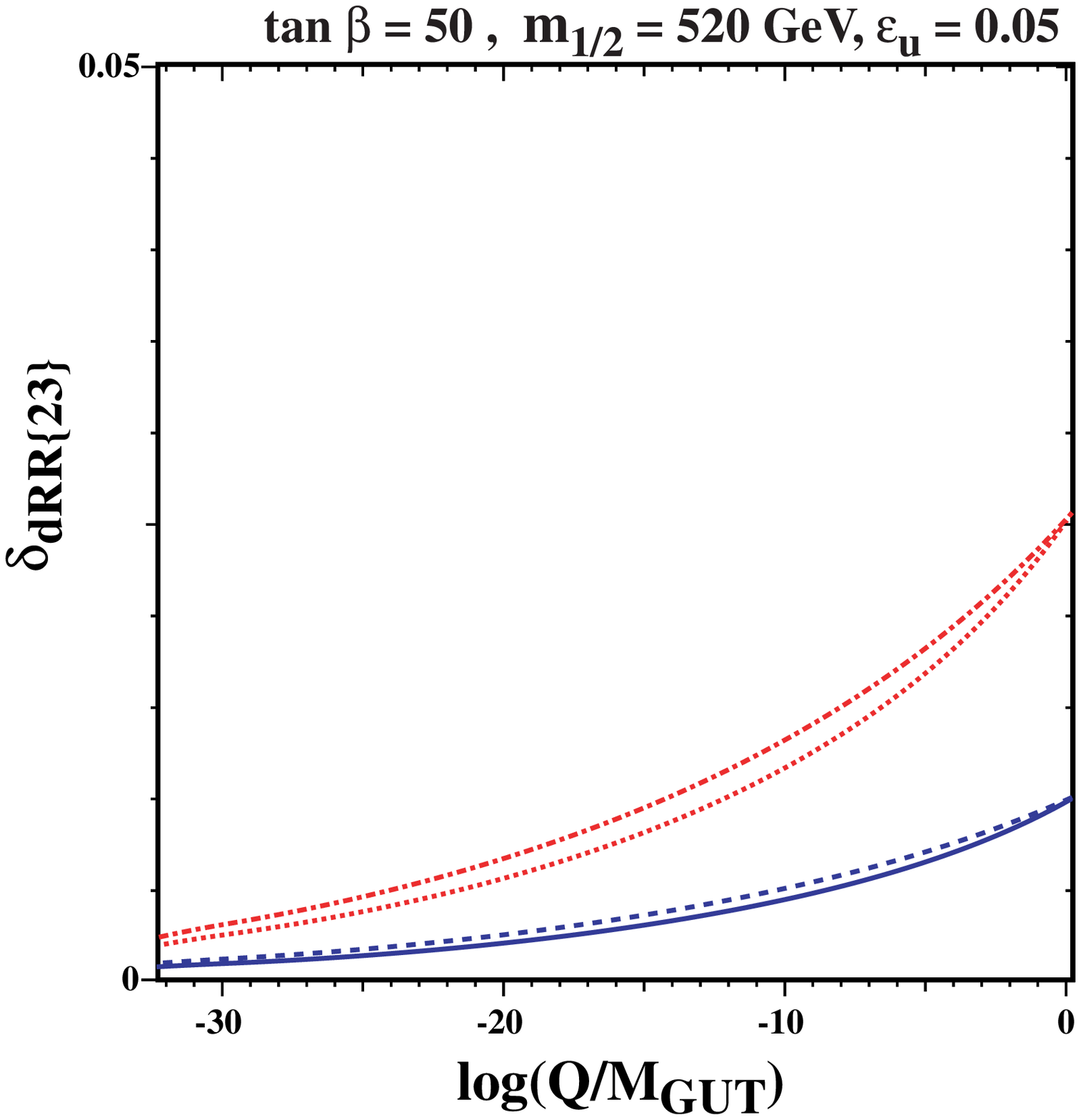}
\end{center}
\begin{center}
\includegraphics[width=.32\textwidth]{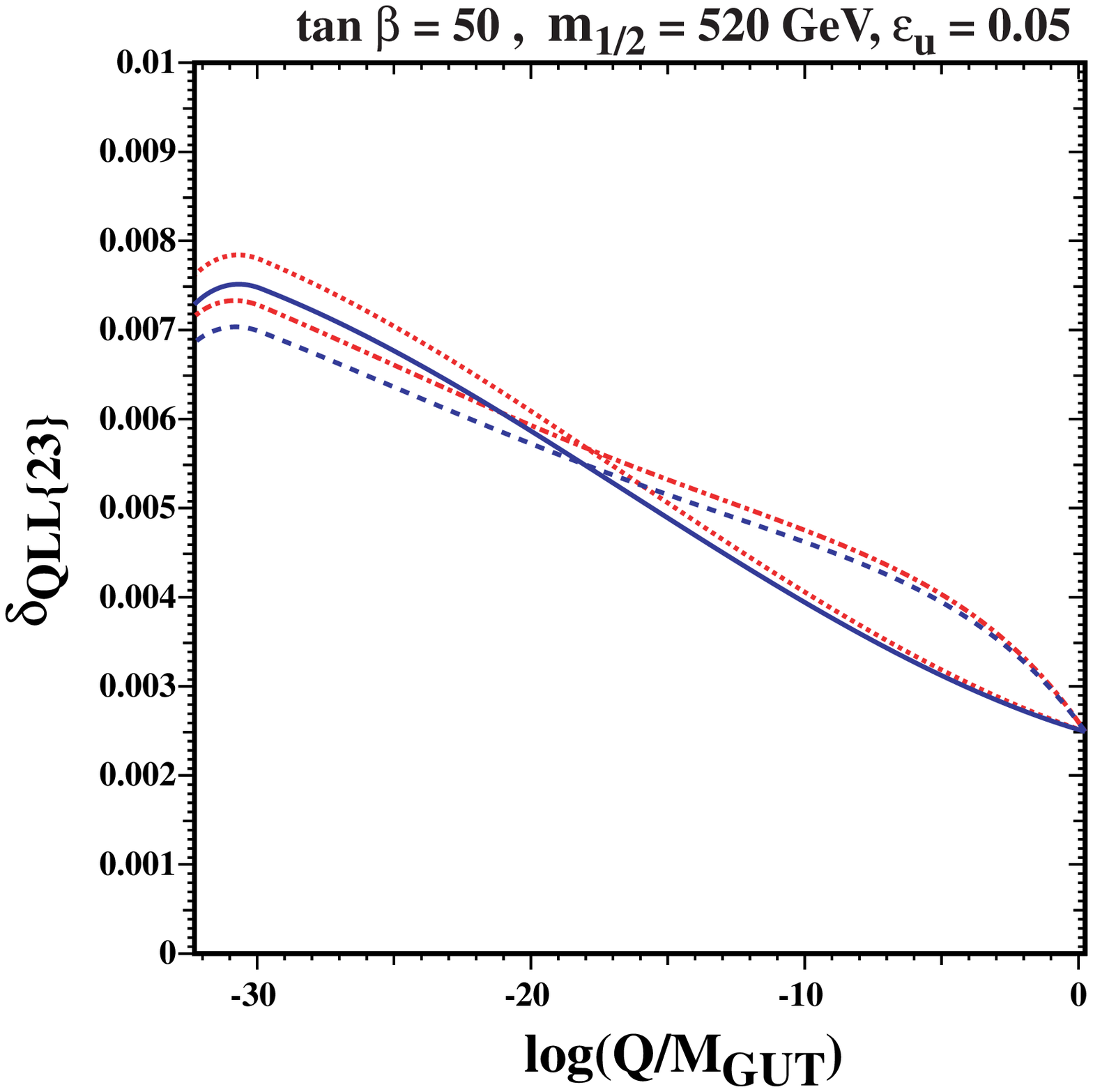}
\includegraphics[width=.32\textwidth]{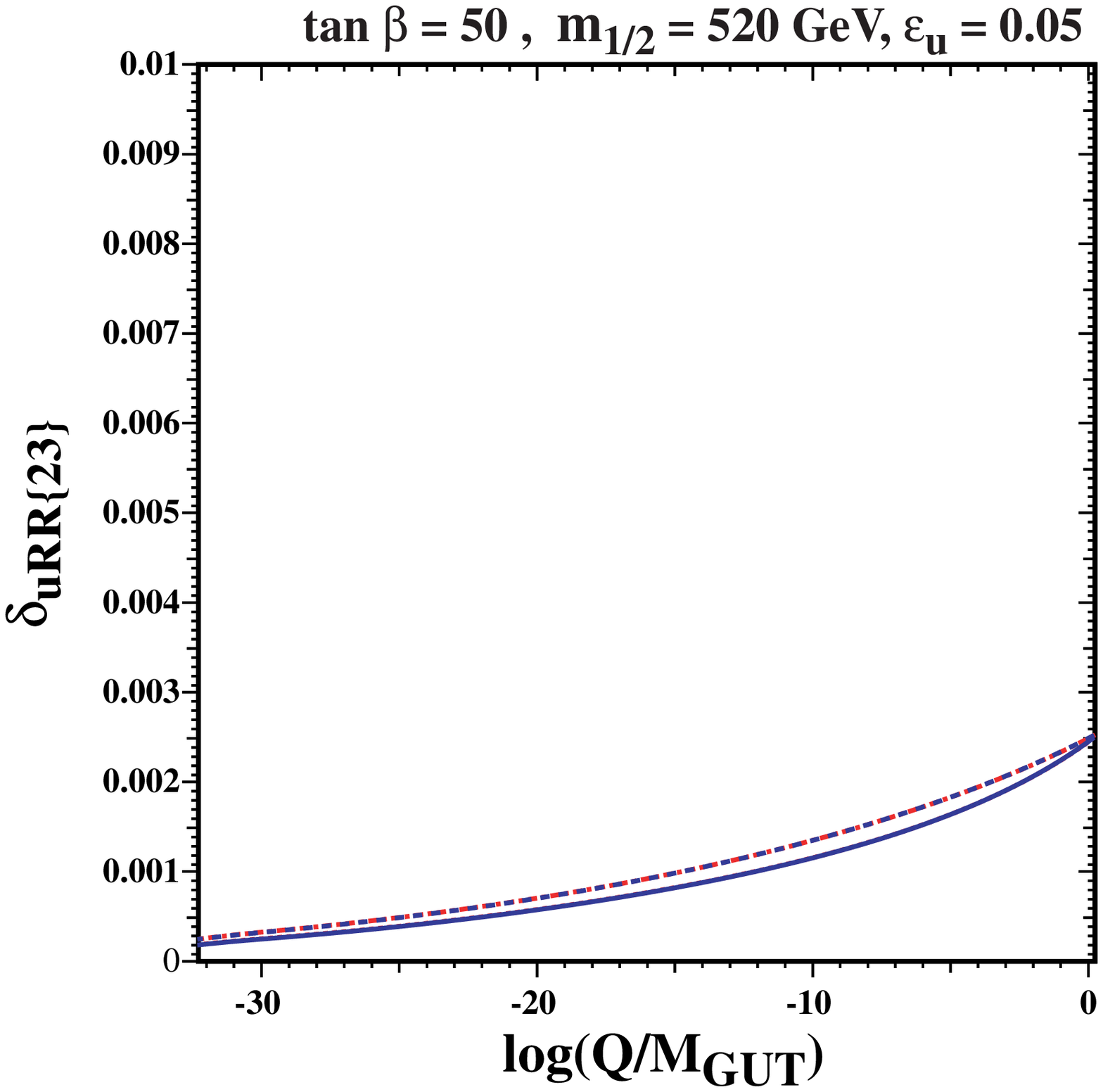}
\end{center}
\begin{center}
\includegraphics[width=.32\textwidth]{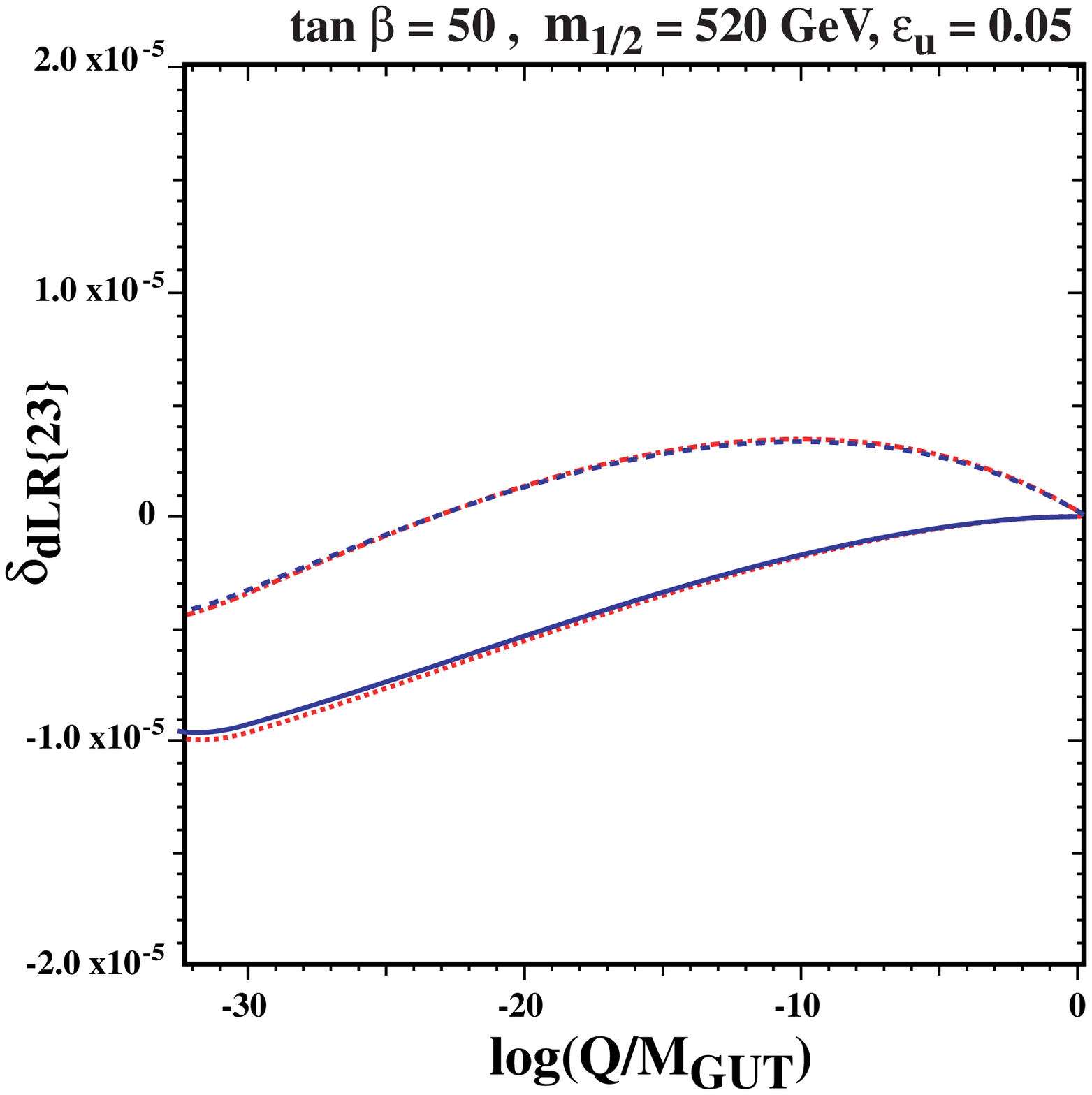}
\includegraphics[width=.32\textwidth]{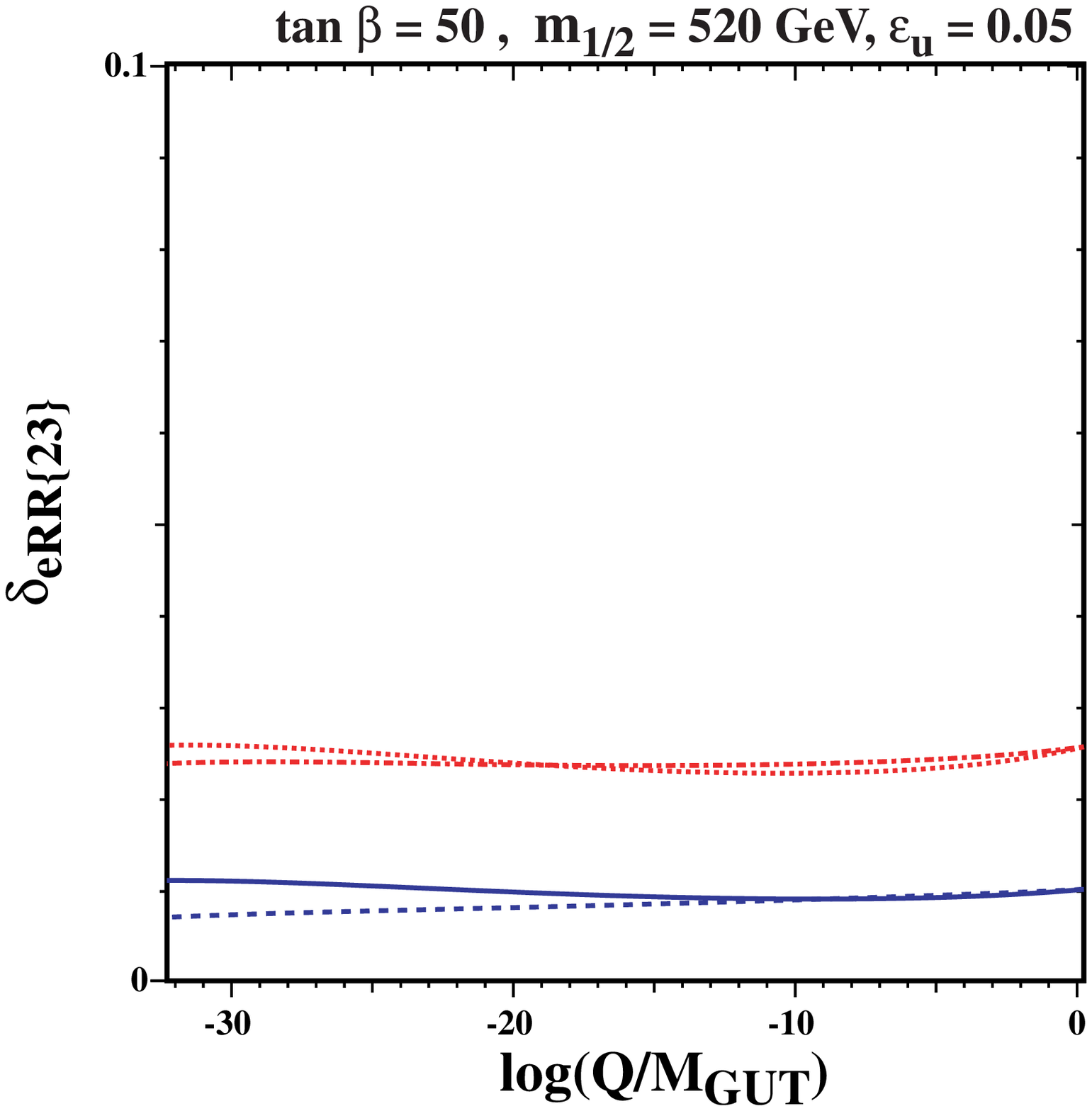}
\end{center}
\caption{\small In panel a), the running of $M^2_{\tilde d ~ 23}$ in the rotated basis 
as in Fig. \protect\ref{fig:ed_eu_softsqrd_comp_0.08_520_330_0}.
Also shown by the (green) long-dashed curve is the case $\ep_d = 0.16$ with $A_0 = 0$
and off-diagonal terms turned on at the GUT scale.
In the remaining panels, the running of squark flavour violating parameters $(\delta_{q,XX})_{23}$ 
(b-d), $(\delta_{d,XY})_{23}$ (e), and $(\delta_{e,XX})_{23}$  (f)  for
  $(m_{1/2},m_0,A_0)=$ ~ $(520,330,0)$ GeV, with $\tan\beta=50$, $\ep_d = 0.1$ and $\ep_u = 0.05$ shown
  by the solid (blue) curves. 
  In panels b) -f), off-diagonal terms are on at the GUT scale. 
  Also shown by the dashed (blue) curves are the results for $A_0 = -m_{1/2}$ with $m_0 = 370$ GeV. The (red) dotted and dash-dotted curves show results for $\ep_d = 0.16$ with $A_0 = 0$ and $A_0 = -520$ GeV respectively.}
\label{fig:ed_eu_dlt_qXX_520_330_0}
\end{figure}

In Figure \ref{fig:ed_eu_dlt_qXX_520_330_0}a,
we show the running of $M^2_{\tilde d ~ 23}$ in the rotated basis. 
Since $M^2_{\tilde d}$ depends primarily  on $(Y_d^{\dagger}Y_d)_{23}\propto \ep^2_d$ 
%$Y_d^{\dagger}Y_d\propto \ep^2_d$ 
(if $\ep^2_u \ll \ep^2_d$) when
the transformation to the rotated basis is performed, we expect that
the value of $M^2_{\tilde d ~ 23}$ will be reduced due to its alignment with the
Yukawa matrix $Y_d$. Additionally, for  $A_0=0$, when off-diagonal terms are turned off, there is no
  contribution from the trilinear terms to the running of soft squared
  masses, as shown in \eq{eq:lead_rel_soft}. Thus, when going to the
  basis where Yukawa couplings are diagonal ($|\hat Y_f|^2=V^{fT}_R
  Y_f^{\dagger} Y_f V^{f*}_R$), the contributions of the form $Y_f^{\dagger} Y_{f}$
  and $Y_f Y_f^{\dagger}$ will also be rotated away to a large extent. We can see
  then in Fig. \ref{fig:ed_eu_dlt_qXX_520_330_0}a, that this is
  indeed the case for $A_0=0$, $\ep_d=0.1$  (blue-dashed line). When
  $A_0\neq 0$,  we see from \eq{eq:lead_rel_soft} that we have an
  extra contribution to the running of $M^2_{\tilde d}$ from the
  trilinear terms, which cannot be greatly reduced when going to the
  rotated basis. When off-diagonal terms are turned on, for the case
  of $A_0=0$ we expect a greater effect in the running of  $M^2_{\tilde
    d}$ than in any of the other soft squared masses, because  (along
  with $M^2_{\tilde e}$) this is the only matrix that is sensitive 
  to the parameter $\ep_d$ at the GUT scale. In  Fig.
  \ref{fig:ed_eu_dlt_qXX_520_330_0}a, the (green) long dashed curve
  corresponds to $\ep_d=0.16$ while the (blue) solid curve for
  $\ep_d=0.1$. Comparing the ratio $0.1^2/0.16^2=0.39$ we see that in
  the un-rotated basis (Fig.
  \ref{fig:ed_eu_softsqrd_comp_0.08_520_330_0}c)
the difference between the two initial conditions for  $M^2_{\tilde d}$
will be $39\%$ and is increased to about $50\%$ in the rotated
basis because the diagonalization matrix also contains the parameter $\ep^2_d$.

The sensitivity to $\ep_u$ in the running of $M^2_{\tilde d}$ is
  very small as a result of the difference between $\ep^2_d$ and
  $\ep^2_u$. The sensitivity to $c^d_{22}$ in the running of
  the soft squared masses comes from the terms of the form
  $Y_fY_f^{\dagger}$ and $a_{f}a_f^{\dagger}$, however as we will
  describe in the next section, the elements of  ${Y_f}_{ij}$ are
  adjusted to reproduce the correct fermion masses at the EW scale for a
  given choice of $c^d_{22}$ and $\ep_d$. For $c^d_{22}=0.7$ and
  $\ep_d=0.1$, one is able to reproduce the correct value of $m_s(\rm{EW})$. Similarly,  for
  $c^d_{22}=0.5$ and $\ep_d=0.12$ we are also able to obtain the correct value for
  $m_s(\rm{EW})$. Indeed, the product $c^d_{22}\ep_d^2$ at the GUT scale is 
  very similar. Therefore, the quantities  $Y_fY_f^{\dagger}$ are more or
  less fixed by this requirement for different choices of  $c^d_{22}$
  and $\ep_d^2$ and as a result  the running of $M^2_{\tilde d}$ will
 not be very sensitive to these changes.

In the remaining panels of Fig. \ref{fig:ed_eu_dlt_qXX_520_330_0},
we show the  flavour violating
parameters $(\delta_{q,XX})_{23}$, $(\delta_{d,XY})_{23}$, and $(\delta_{e,XX})_{23}$  
for our test point
$(m_{1/2},m_0,A_0)=$ ~ $(520,330,0)$ GeV and $(\ep_d,\ep_u)=(0.1,0.05)$.
The curves for these cases are all shown as solid (blue).
Off-diagonal terms at the GUT scale are turned on for all curves in panels b) - f).
For comparison, we show the case with non-zero $A_0$, ie.,
$(m_{1/2},m_0,A_0)=$ ~ $(520,370,-520)$ GeV. This case is shown by the dashed (blue)
curves.  In addition for both $A_0 = 0$ and $A_0 \ne 0$, we show the sensitivity of the running to
$\ep_d$.  The (red) dotted and dash-dotted curves correspond to these cases respectively,
where we have chosen $\ep_d = 0.16$.

%

%
%%%%%%%%%%%%%%%%%%%%%%%%%%%%%%%%%%%%%%%%%%%%%%%%%%%%%%%%%%%%%%%%%%%%%%%
%
\subsection{Determination of coefficients of Yukawa matrices}
%
%%%%%%%%%%%%%%%%%%%%%%%%%%%%%%%%%%%%%%%%%%%%%%%%%%%%%%%%%%%%%%%%%%%%%%%
%
%
In this subsection, we will determine the values and sensitivities of the quantities 
\bea
c^d_{23},\quad c^d_{33},\quad c^e_{22}, \quad c^e_{23},\quad c^e_{33},\quad c^u_{22}\quad {\rm and} \quad c^u_{33}
\eea
to the variation of $\ep_d$ and $\ep_u$ for a given value $c^d_{22}$.
For now, we continue to use our example based on the point
$(m_{1/2},m_0,A_0)= (520,330,0)~ {\rm{GeV}}$, $\tan\beta=50$.
As a result of the small differences in the running of the off-diagonal
terms of soft parameters when we choose different initial conditions
at the GUT scale, the Yukawa parameters are not really sensitive to this difference. 
In this section, we will only consider cases where the off-diagonal elements
are turned on at the GUT scale.
These parameters are, however, sensitive to variations in $\ep_d$,
$\ep_u$ and $c^d_{22}$. 

\begin{figure}[ht!]
\begin{center}
\includegraphics[width=.3\textwidth]{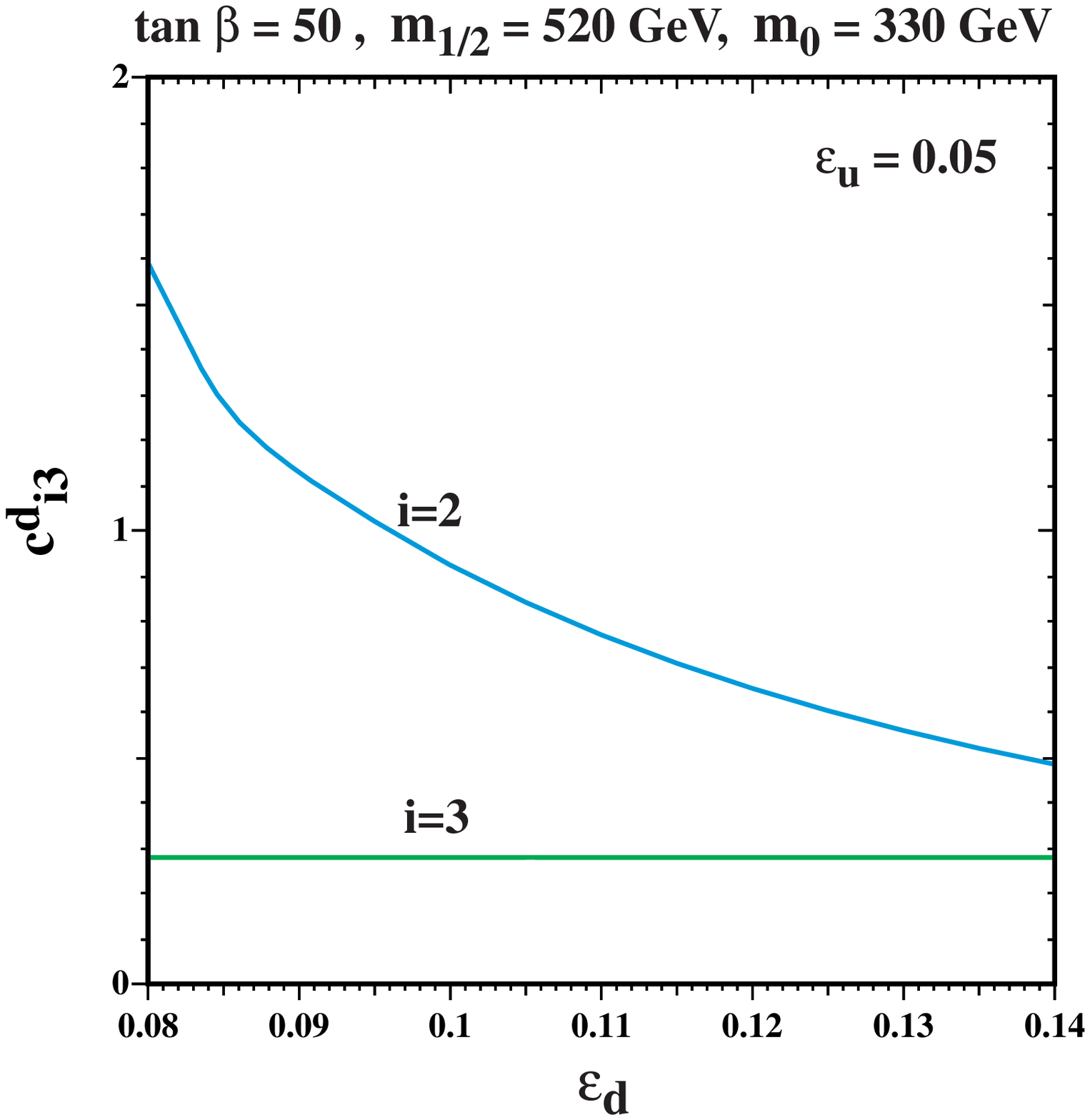}%{cd_yuk_cd22_0.7_eu_0.05_520_330_0.ps}
\includegraphics[width=.3\textwidth]{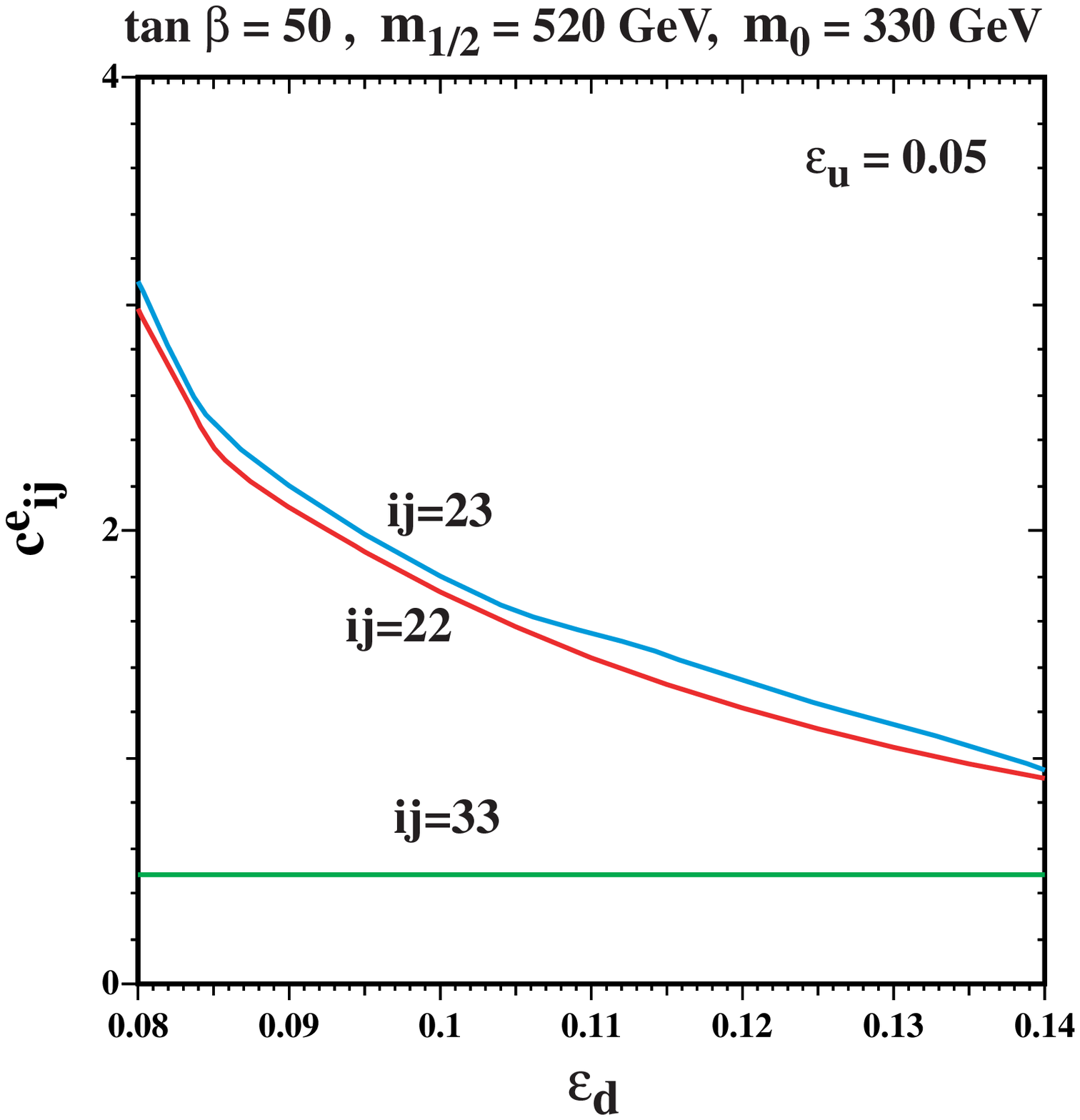}%{ce_yuk_cd22_0.7_eu_0.05_520_330_0.ps}
\includegraphics[width=.3\textwidth]{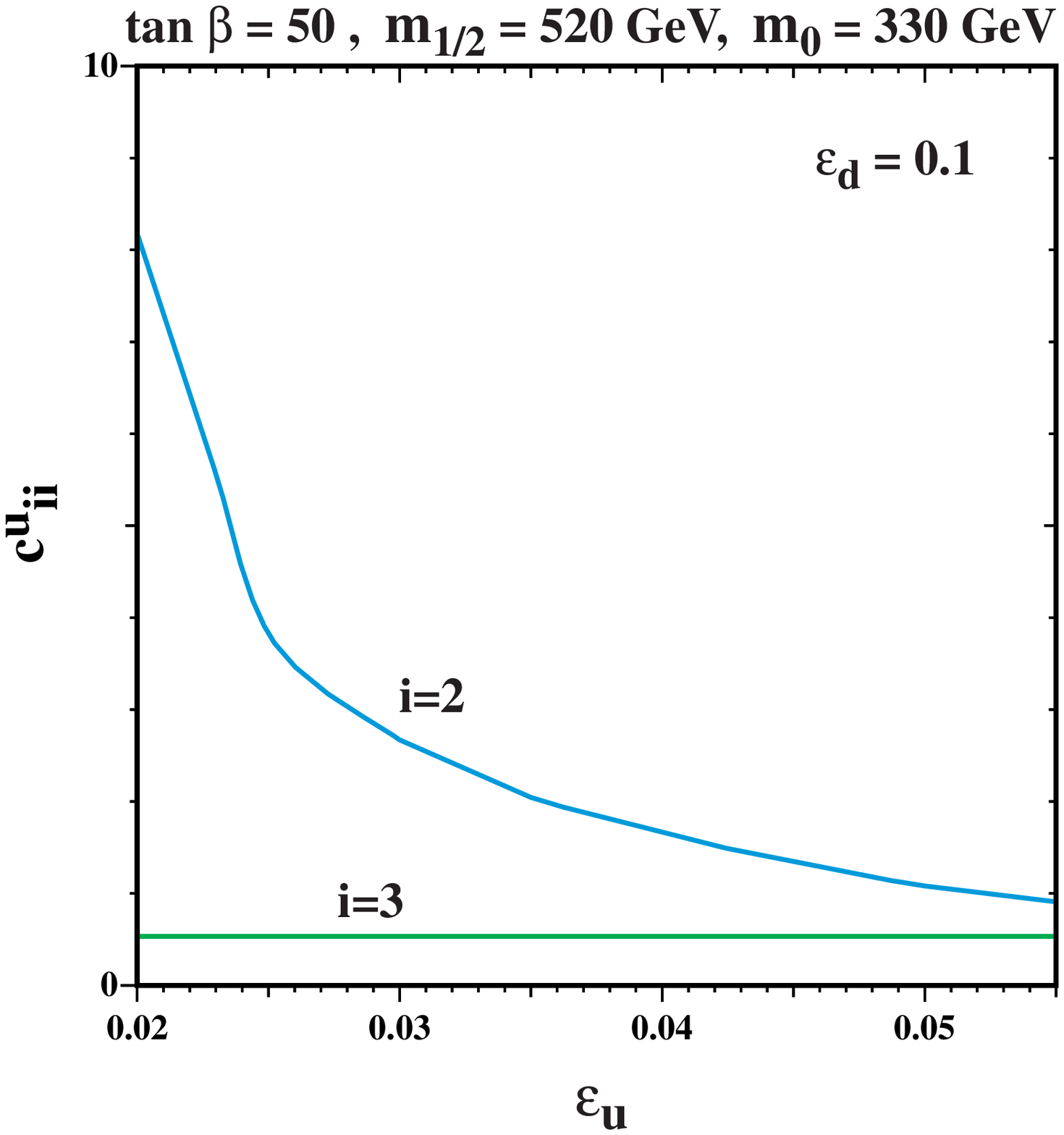}%{cu_yuk_cd22_0.7_ed_0.1_520_330_0.ps}
\end{center}
\caption{\small Yukawa parameters, $c^f_{ij}$ of the $d, e$ and $u$ sectors  
for the supersymmetric point  $(m_{1/2}, m_0,A_0)$ = $(520,330,0)$ GeV, $\tan\beta=50$ 
and $c^d_{22}=0.7$. In panel (a) we show $c^d_{i3}$ for $i = 2,3$ as labeled
as a function of $\ep_d$ for fixed $\ep_u = 0.05$.
In panel (b) we show $c^e_{ij}$ for $ij = 22,23,33$ as labeled
as a function of $\ep_d$ for fixed $\ep_u = 0.05$.
Finally in panel (c) we show $c^u_{ii}$ for $i = 2,3$ as labeled
as a function of $\ep_u$ for fixed $\ep_d = 0.1$. }
\label{fig:ed_eu_yuk_cfs_520_330_0_cd22_0.7}
\end{figure}

In Fig. \ref{fig:ed_eu_yuk_cfs_520_330_0_cd22_0.7},
we plot the relevant $c^f_{ij}$ versus the appropriate $\ep_f$  for $c^d_{22}=  0.7$.
There is little sensitivity to $\ep_d$ or $\ep_u$ for the 33 components of the Yukawa matrices
which are well fixed by the low energy fermion masses.
In contrast, the 22 and 23 components can be sensitive to $\ep_d$ and $\ep_u$ even
within their limited range of variation.

\begin{figure}[ht]
\begin{center}
\includegraphics[width=.45\textwidth]{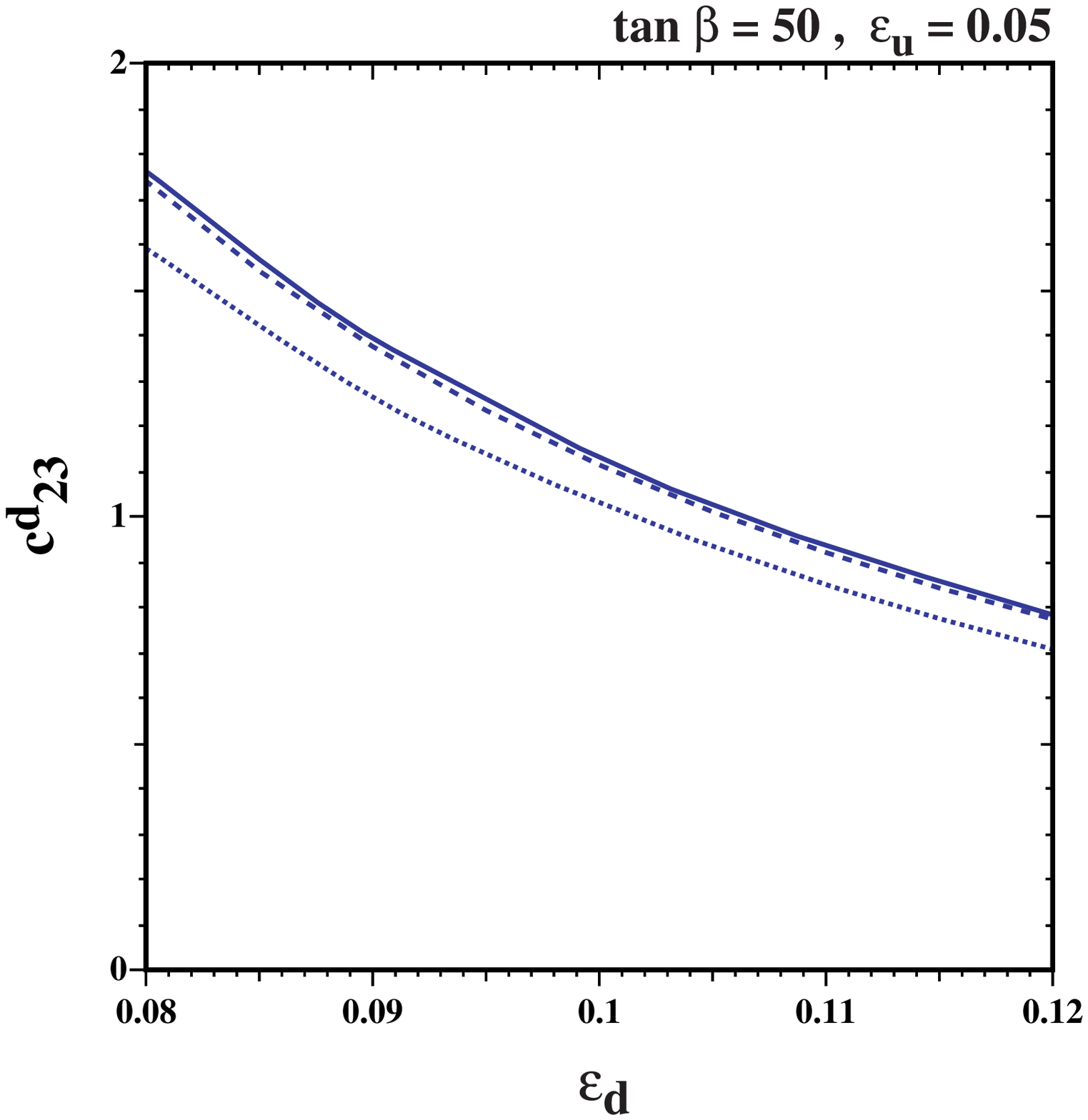} %{ed_eu_yuk_comp_b_zoom_cd23.ps}
\includegraphics[width=.45\textwidth]{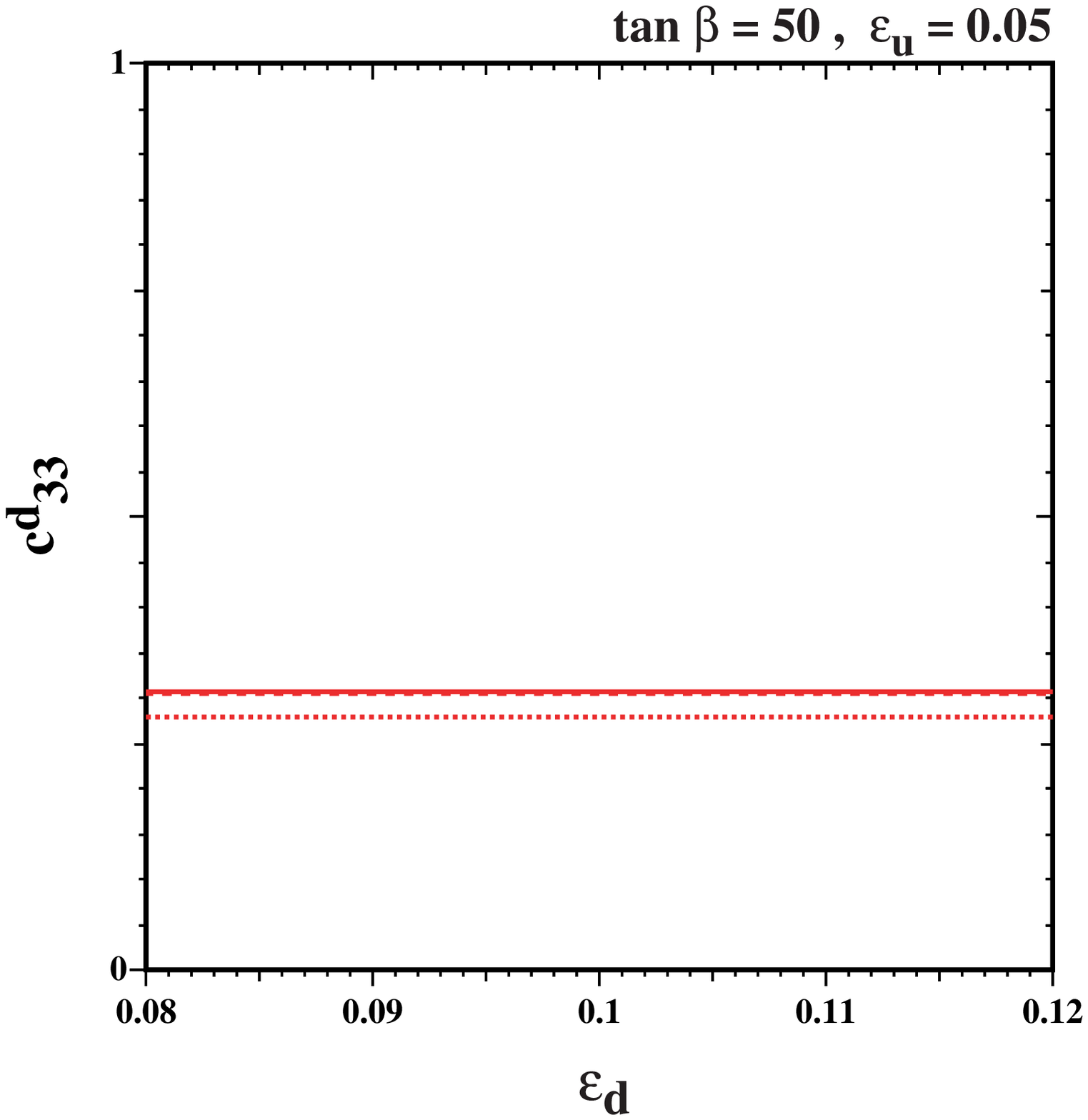} %{ed_eu_yuk_comp_b_zoom_cd33.ps}
\end{center}
\caption{\small As in Fig. \protect\ref{fig:ed_eu_yuk_cfs_520_330_0_cd22_0.7}, we plot
(a)  $c^d_{23}$ and (b) $c^d_{33}$  for three different choices of 
$(m_{1/2},m_0,A_0)$. The bottom curves (dotted) correspond to
$(m_{1/2},m_0,A_0)$  $=(520,330,0)$ GeV, the lines in the middle (dashed) to  
$(m_{1/2},m_0,A_0)$  $=(160,630,0)$ GeV and the lines at
the top (solid) to  $(m_{1/2},m_0,A_0)$  $=(1240,910,-1240)$ GeV.}
\label{fig:ed_eu_yuk_cd_comp}
\end{figure}

In Fig. \ref{fig:ed_eu_yuk_cd_comp}, we show the dependence of two of the Yukawa parameters,
$c^d_{23}$ and $c^d_{33}$, to our choice of the supersymmetric point as a function of $\ep_d$.
In addition to our original test point (shown here as the dotted curves), we consider
$(m_{1/2},m_0,A_0)$  $=(160,630,0)$ GeV shown by the dashed curves, and 
$(m_{1/2},m_0,A_0)$  $=(1240,910,-1240)$ GeV shown by the solid curves.
In all three cases, we have chosen $\tan \beta = 50$. 
In each case, the value of $m_0$ was chosen so as to obtain a relic density compatible with
WMAP. In the latter case, we chose $A_0 = -m_{1/2}$.   
The dependence of the parameters of $Y_e$ on the change of $(m_{1/2},m_0,A_0)$ is similar though
less pronounced and the dependence of the parameters of $Y_u$ is very small.
%

%%%%%%%%%%%%%%%%%%%%%%%%%%%%%%%%%%%%%%%%%%%%%%%%%%%%%%%%%%%%%%%%%%%%%%%%%%%%%%%%%%%%%%%%%%%

%

Given the parameterization of the Yukawa matrix $Y_d$ in \eq{eq:yuk_o_init},
the choice of $c^d_{22}$ is directly correlated to the
allowed value of $\ep_d$ given the value of $m_s/m_b({ M_{\rm GUT}})$:
\bea
\frac{m_s}{m_b}\left({M_{\rm GUT}}\right)\sim \frac{(Y_d)_{22}}{(Y_d)_{33}}
-\frac{(Y^{2}_d)_{22}+(Y^{2}_d)_{33}}{2 (Y^{2}_d)_{33}}= \frac{c^d_{22}}{c^d_{33}} \ep_d^2 - O(\ep^4_d),
\eea
which must be in the range $\sim (0.01,0.03)$ at $M_{GUT}$. Hence
for different choices of $c^d_{22}$, assumed to be $\mathcal{O}(1)$, we have the following allowed
ranges of $\ep_d$:
\bea
c^d_{22} = 0.5 ~ &\rightarrow & ~ \ep_d \simeq [0.10,0.12]\nn\\
c^d_{22}= 0.7 ~ &\rightarrow & ~ \ep_d \simeq [0.085,0.10]\nn\\
c^d_{22} = 1.0 ~ &\rightarrow & ~ \ep_d \simeq 0.08.
\eea
As a consequence the values of $(Y_d)_{ij}$ and  $(Y_e)_{ij}$ are different
for each case. In Table \ref{tbl:yuk_parscfij}, we show the allowed
ranges of the Yukawa parameters according to the allowed range for $\ep_d$ in each case.

\begin{table}[h!]
\begin{center}
\begin{tabular}{|l l l l l l |}
\hline
\multicolumn{6}{c}{Yukawa parameters $c^f_{ij}$}\\
\hline
$c^d_{22}$ & $c^d_{23}$   &  $c^d_{33}$ & $c^e_{22}$     & $c^e_{23}$ &  $c^e_{33}$ \\
$0.5$      & $[0.6,0.9]$  &  $0.3$      & $[1.2,1.7]$  & $[1.3,1.7]$    &  $0.5$\\
$0.7$      & $[0.9,1.3]$  &  $0.3$      & $[1.7,2.4]$    & $[1.9,2.4 ]$  &  $0.5$\\
$1.0$      & $1.6$        &  $0.3$      & $3.0$          & $3.1$  &  $0.5$\\
\hline
\end{tabular}
\end{center}
\caption{\small Values of the Yukawa parameters for different choices
  of $c^d_{22}$ for the supersymmetric point $(m_{1/2},m_0,A_0)= (520,330,0)~ {\rm{GeV}}$.}
\label{tbl:yuk_parscfij}
\end{table}

\begin{figure}[ht!]
\begin{center}
\includegraphics[width=0.32\textwidth]{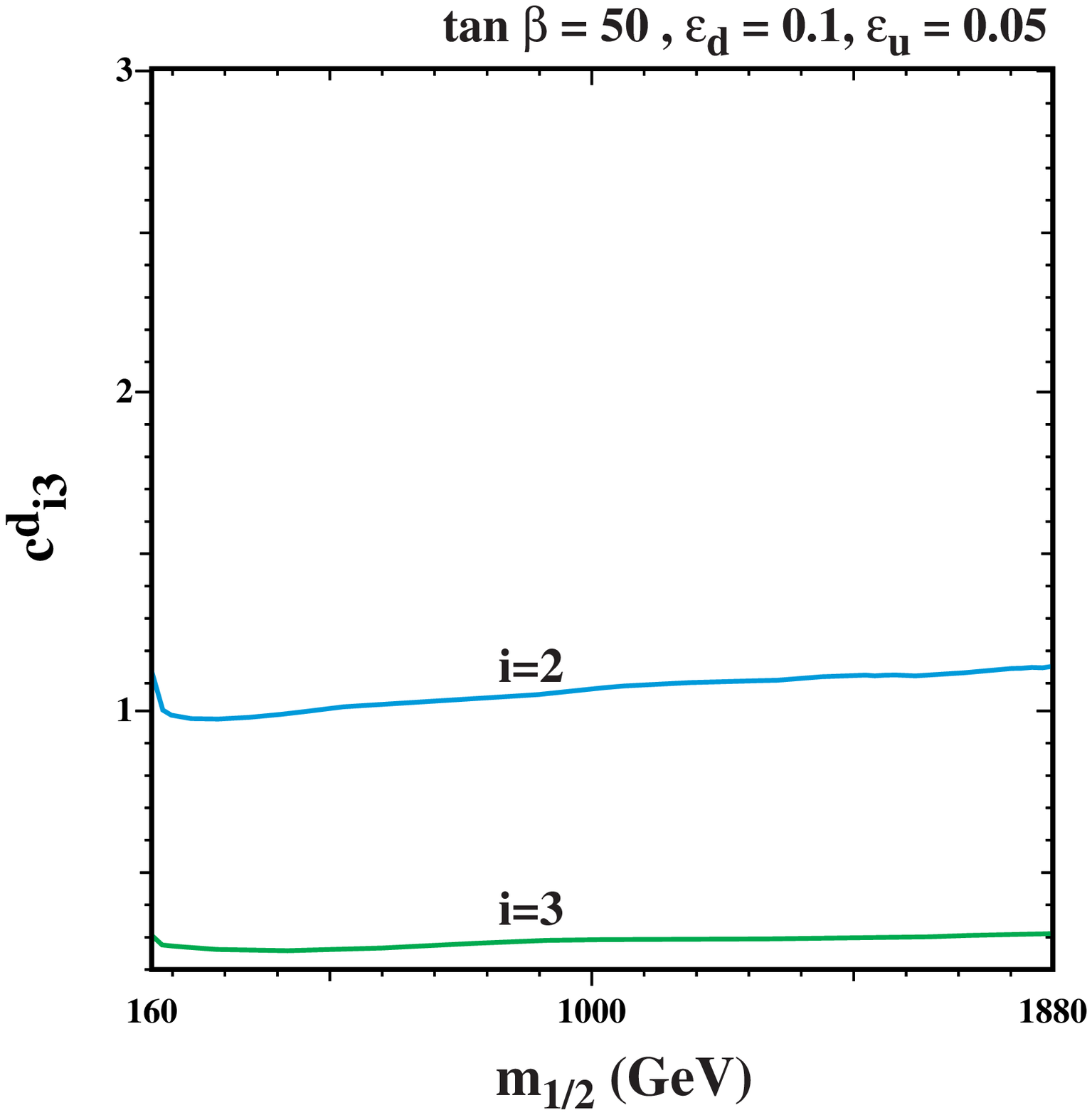}
\includegraphics[width=0.32\textwidth]{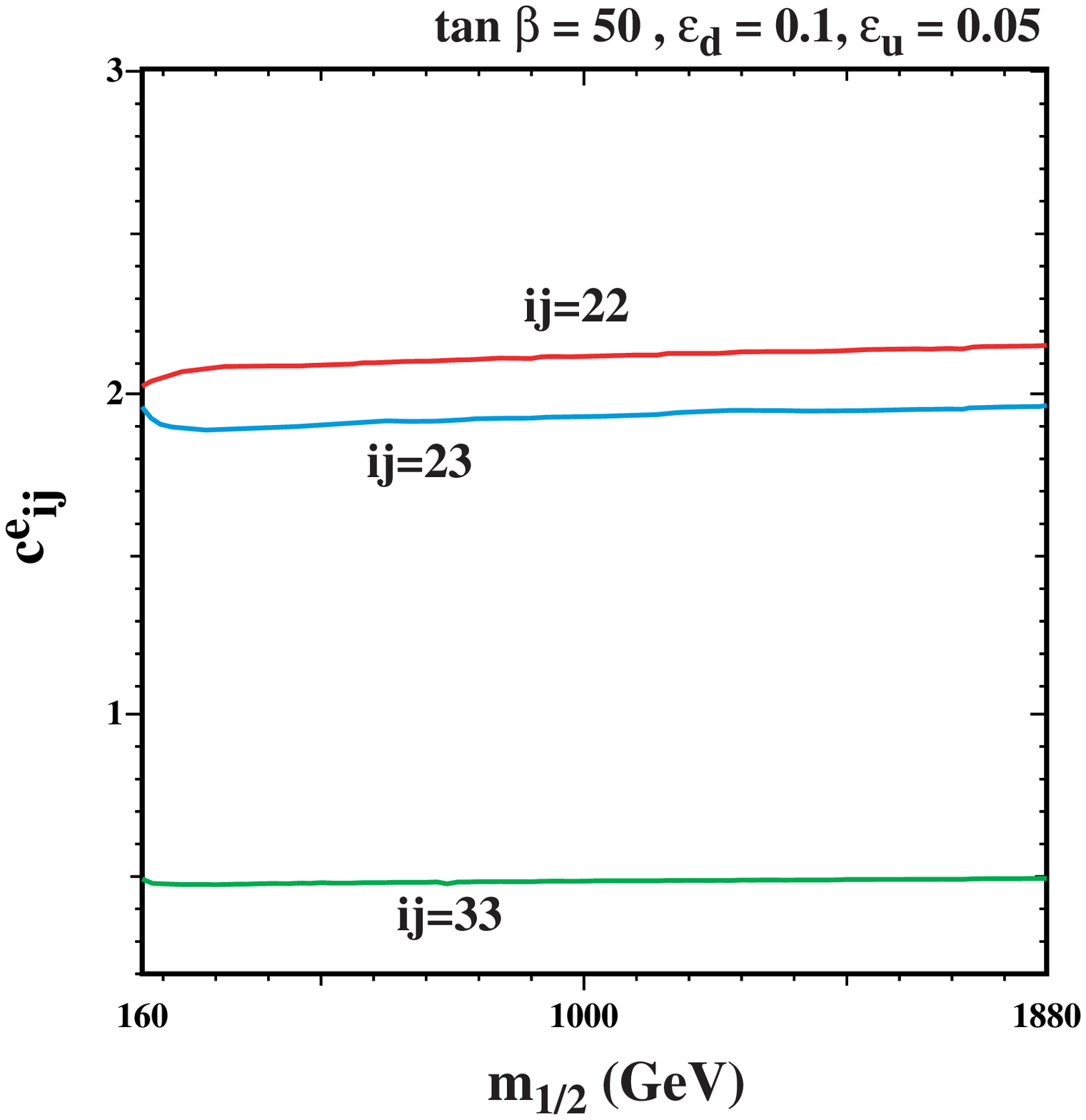}
\includegraphics[width=0.32\textwidth]{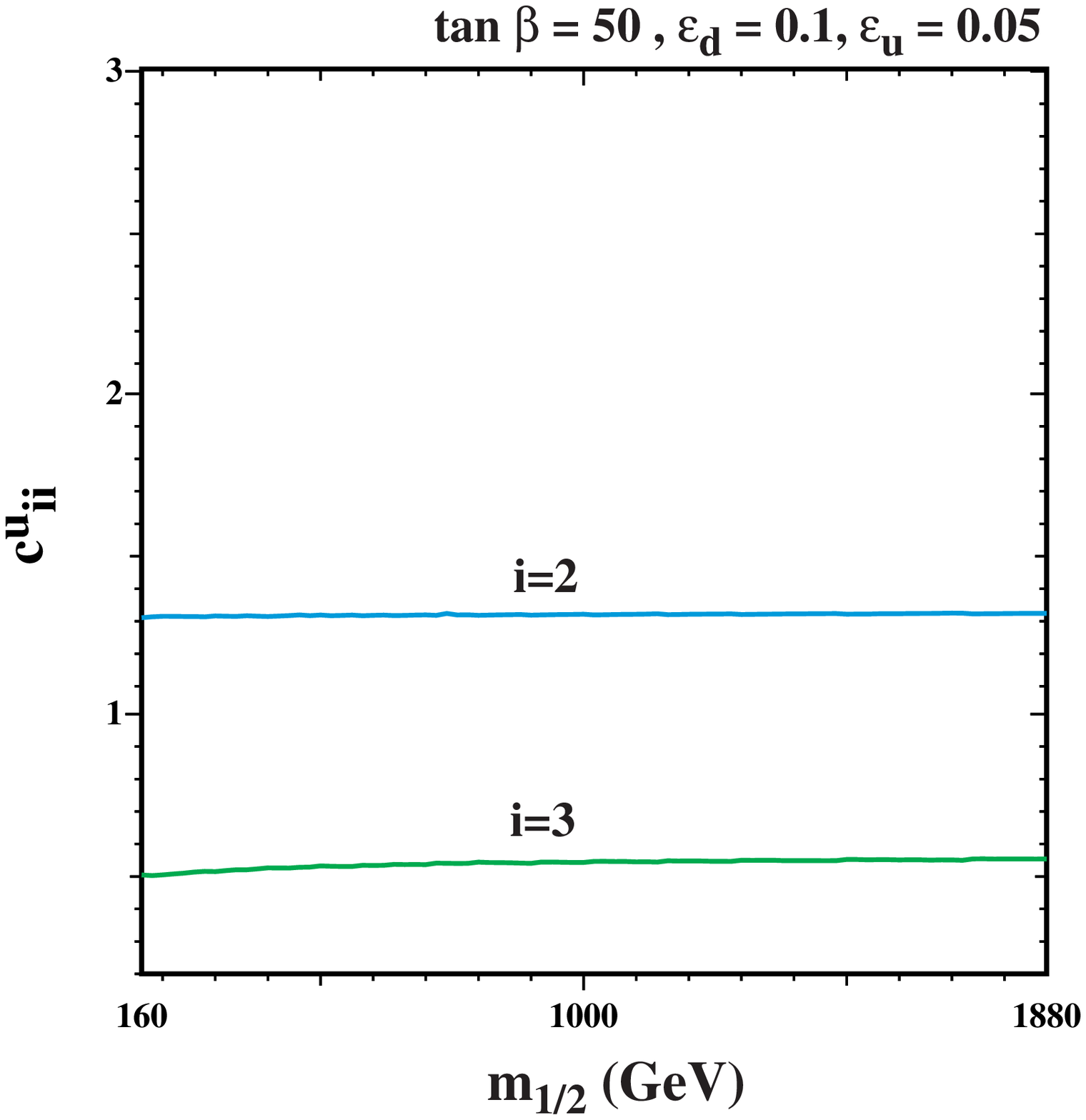}
\end{center}
\caption{\small The coefficients of the Yukawa matrices
plotted versus $m_{1/2}$ for $\ep_d=0.1$,
$\ep_u=0.05$ and $c^d_{22}=0.7$. 
In panel (a) we show $c^d_{i3}$ for $i = 2,3$ as labeled.
In panel (b) we show $c^e_{ij}$ for $ij = 22,23,33$ as labeled.
Finally in panel (c) we show $c^u_{ii}$ for $i = 2,3$ as labeled.}
\label{fig:spc_yukpars_ed_0.1_off1}
\end{figure}

To further compare how the determination of the
appropriate Yukawa parameters at the GUT scale depends on the chosen
supersymmetric parameters we show in Fig. \ref{fig:spc_yukpars_ed_0.1_off1},
the values of $c^f_{ij}$ as a function of the the gaugino mass, $m_{1/2}$. 
Here $A_0 = 0$ and $\tan \beta = 50$. For each value of $m_{1/2}$, $m_0$ 
is adjusted to insure the correct relic density of neutralinos.
As one can see the dependence on the supersymmetric model is rather minimal.

\begin{figure}[ht!]
\begin{center}
\includegraphics[width=.45\textwidth]{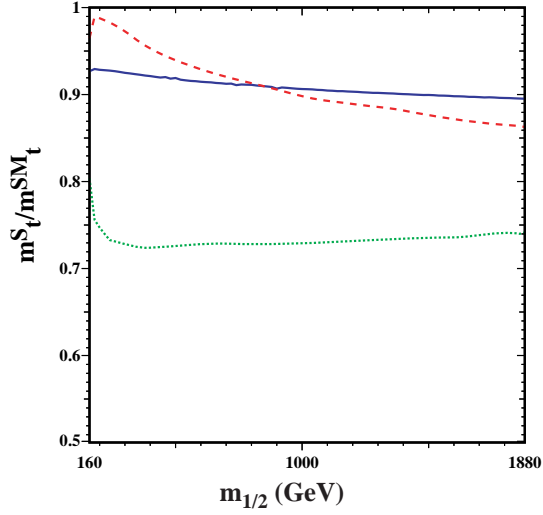} 
\end{center}
\caption{\small MSSM corrections at one loop to the masses of $m_t$
  (solid blue line),
  $m_b$ (dotted green line) and $m_s$ (dashed red line).}
\label{fig:mssm_mf_corr}
\end{figure}

Finally, we would like to emphasize  the importance of
taking into account the supersymmetric corrections to the fermion
masses for the proper determination of Yukawa parameters at the GUT
scale. From \cite{Pierce:1996zz}, we find that for large $\tan\beta$,
the supersymmetric corrections to the mass of the bottom can be as
large as $30\%$, as it is for the case of $\tan\beta=50$. 
We can express the relation between the masses of the fermions at a
certain scale $Q$, containing the supersymmetric corrections, and the
masses of fermions at the same scale as given by the SM through:
\bea
m_k^S(Q) = m^{SM}_k(Q) g^S_k(Q), \quad k=t,b,c,s.
\eea 
Here $m^{SM}_k(Q)$ are the SM values at $Q$, including the corrections
from gluons. A good analytical approximation to the factors $g^S_k(Q)$ is given
in \cite{Pierce:1996zz}. For example, for the top quark, this can be
expressed as $g^S_t(Q)=1+\Delta m_t/m_t$, where the most important
contributions to $\Delta m_t/m_t$ comes  from  top squark/gluino loops. %in this
                                %definition   $\Delta m_t/m_t$
                                %includes only susy corrections.
In Fig. \ref{fig:mssm_mf_corr}, we plot the numerical values of
$g^S_k(Q)$ for $k=t,b,s$, as a function of $m_{1/2}$ for $\tan\beta=50$ and for the
choice of Yukawa parameters $(c^d_{22},\ep_d)=(0.7,0.1)$.
These corrections include the corrections at one loop
from squark/gluino loops, charginos and neutralinos. We can see that,
in fact, for $m_b$ these corrections are about $30\%$, while for $m_s$, they
can be up to $14\%$. For $m_t$, the corrections are within $11\%$. The corrections for $m_c$,
though included are not shown as they are very small.

The supersymmetric corrections change the profile of Yukawa matrices at the GUT
scale. A SM analysis \cite{Ross:2004qn} 
determined  the
values for  $Y_d$ (for symmetric matrices) as follows:
\bea
\frac{(Y_d)_{22}}{(Y_d)_{33}}= 0.15^2=0.0225,\quad
\frac{(Y_d)_{23}}{(Y_d)_{33}}=1.3 (0.15^2)=0.0293.
\eea
If we take the values of the Yukawa parameters for $(c^d_{22},\ep_d)=$ $(0.7,0.1)$ from Table \ref{tbl:yuk_parscfij}, we have
\bea
\frac{(Y_d)_{22}}{(Y_d)_{33}}= 0.025,\quad
\frac{(Y_d)_{23}}{(Y_d)_{33}}=0.045.
\label{eq:ydk22_o_yd23}
\eea
Since $c^d_{22}$ is an input to our analysis, we are basically
adjusting the value of ${(Y_d)_{22}}/{(Y_d)_{33}}$, at the GUT scale, thus
there is no surprise in the similar values for this ratio in both
cases. What is interesting is the change in the off-diagonal parameter
${(Y_d)_{23}}/{(Y_d)_{33}}$, since this off-diagonal term is crucial when
constructing a specific flavour symmetry. 

It is important to emphasize the difference between the structure of
the Yukawa matrices at the GUT and EW scales.
Whilst at the GUT scale they can be written in terms of the expansion parameter
$\ep^2_d$ and the prefactors $c^f_{ij}$, at the EW scale this
is not longer the case. However if the prefactors  $c^f_{ij}$ are indeed to be
explained by a specific FS then the result at the EW can be expressed as
a series expansion in terms of $\ep_f$  and  the same coefficients.
For example,  we can express the Yukawa matrices
normalized to the elements $(Y_f)_{33}$ at the GUT scale.  The
numerical value using $c^d_{22}=0.7$, $\ep_d=0.1$ and  $\ep_u=0.05$ is
\bea
\frac{Y_d}{(Y_d)_{33}}&=&\left(
\begin{array}{ccc}
0 & 0                & 0\\
0 & \frac{c^d_{22}}{c^d_{33}} \ep^2_d & \frac{c^d_{23}}{c^d_{33}} \ep^2_d\\
0 & \frac{c^d_{23}}{c^d_{33}} \ep^2_d & 1
\end{array}
\right)=
\left(
\begin{array}{ccc}
0   &  0      & 0\\
0   &  0.025  & 0.033\\
0   &  0.033  & 1
\end{array}
\right)\nn\\
&\rightarrow& 
\left(
\begin{array}{ccc}
0   &  0      & 0\\
0   &  0.07  & 0.04\\
0   &  0.04  & 1
\end{array}
\right) =
\left(
\begin{array}{ccc}
0   &  0   & 0 \\
0   &  \left(\frac{c^d_{22}}{c^d_{33}}+4.49 \right) \ep^2_d &
\frac{c^d_{23}}{c^d_{33}} \ep^2_d + 7\ep^3_d \\
0   &  \frac{c^d_{23}}{c^d_{33}} \ep^2_d + 7\ep^3_d & 1
\end{array}
\right)
 \nn\\
\frac{Y_e}{(Y_e)_{33}}&=&\left(
\begin{array}{ccc}
0 & 0                & 0\\
0 & \frac{c^e_{22}}{c^e_{33}} \ep^2_d & \frac{c^e_{23}}{c^e_{33}} \ep^2_d\\
0 & \frac{c^e_{23}}{c^e_{33}} \ep^2_d & 1
\end{array}
\right)=
\left(
\begin{array}{ccc}
0   &  0      & 0\\
0   &  0.038  & 0.036\\
0   &  0.036  & 1
\end{array}
\right)\nn\\
&\rightarrow&
\left(
\begin{array}{ccc}
0   &  0      & 0\\
0   &  0.05  & 0.04\\
0   &  0.04  & 1
\end{array}
\right)=
\left(
\begin{array}{ccc}
0   &  0   & 0 \\
0   &  \left(\frac{c^e_{22}}{c^e_{33}}+1.25 \right) \ep^2_d &
\frac{c^d_{23}}{c^d_{33}} \ep^2_d + 3.9\ep^3_d \\
0   &  \frac{c^d_{23}}{c^d_{33}} \ep^2_d + 3.9\ep^3_d & 1
\end{array}
\right)\nn\\
\frac{Y_u}{(Y_u)_{33}}&=&
\left(
\begin{array}{ccc}
0 & 0                & 0\\
0 & \frac{c^u_{22}}{c^u_{33}} \ep^2_u & 0\\
0 & 0 & 1
\end{array}
\right)=
\left(
\begin{array}{ccc}
0 & 0      & 0 \\
0 & 0.005  & 0 \\
0 & 0      & 1
\end{array}
\right)\nn\\
&\rightarrow&
\left(\begin{array}{ccc}
0   &  0      & 0\\
0   &  0.01   & 0\\
0   &  0      & 1
\end{array}\right)
=
\left(
\begin{array}{ccc}
0 & 0      & 0 \\
0 & \left(\frac{c^u_{22}}{c^u_{33}}+2\right) \ep^2_u  & 0 \\
0 & 0      & 1
\end{array}
\right)
\label{eq:YkGUT_EW}
\eea
where the effect of running to the EW scale is given to the right of the arrows.

%
%
%%%%%%%%%%%%%%%%%%%%%%%%%%%%%%%%%%%%%%%%%%%%%%%%%%%%%%%%%%%%%%%%%%%%%%%%%%%%%%%%%%%%%%%%%%%%%%%%%%%%%%%%%%%
%
%

\subsection{Effects on $\mathbf{\bsgmm}$}
\begin{figure}[ht!]
\begin{center}
\includegraphics[width=0.45\textwidth]{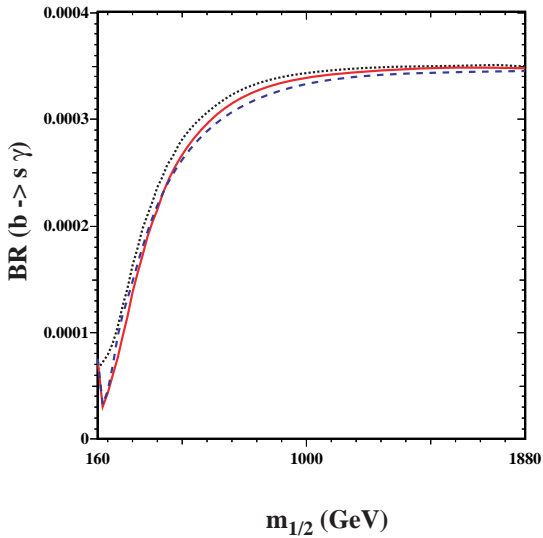} 
\end{center}
\caption{\small  The branching ratio for $\bsgmm$ for $A_0=0$ (solid red) and  $A_0= - m_{1/2}$ (dotted black) when $\ep_d=0.1$,  $\ep_d=0.05$ and $c^d_{22}=0.7$. When only the running of the 3rd generation is included, the analogous result is shown by the dashed (blue) curve.
  }
\label{fig:bsg_0.1_comp}
\end{figure}

We have seen that a good choice of Yukawa parameters was $c^d_{22}=0.7$
and $\ep_d=0.1$,  $\ep_u=0.05$, thus we present the analysis on the sensitivity of
$\bsgmm$ on $\ep_d$ centered on those parameter values. As we explain in
Appendix \ref{sec:bsg_bmfv} 
due to the hierarchy of $\ep^2_u \ll
\ep^2_d$, we expect that the main departures from MFV come from the
gluino contributions, entering through the magnetic Wilson
coefficients due to the diagonalisation of the soft
squared matrices in the $\tilde d$ sector, \eq{eq:sofmasses_offneq0}.
In Fig. \ref{fig:bsg_0.1_comp}
we show our calculation of  $\bsgmm$ as a function of $m_{1/2}$
when the running of the parameters
of the second family have not been considered (dashed blue curve). For comparison,
we also show the same result when the running of both the 2nd and 3rd generations
are included for 
$A_0=0$ (solid red curve) and when $A_0= - m_{1/2}$ (dotted black curve).
As before, for each value of $m_{1/2}$, $m_0$ is chosen so as to obtain the WMAP
value for the relic density of neutralinos.
Note that  the inclusion of the 2nd generation running 
relaxes slightly the bounds obtainable from a comparison of the computed
value of $\bsgmm$ with the experimental measurement at relatively low $m_{1/2} (\sim 400$ GeV).
As long as we fix the relation between $c^d_{22}$ and $\ep_d$ through the fermion masses,
there will be little sensitivity of $\bsgmm$ to $\ep_d$. 

The contribution of the charged Higgs to the Wilson coefficients, 
depends crucially on $x_{tH}=m^2_t/m^2_{H^-}$ where $m_{H^-}$ is the mass of the
charged Higgs. The determination of $m_{H^-}$ is quite sensitive to changes in the running of the
supersymmetric spectrum \cite{deBoer:1994he} and thus any change in the running of these
parameters will affect the final value of $m_{H^-}$:
\bea
m^2_{H^{-}}=m^2_A+M^2_W+\Delta^2_H,\quad  \Delta^2_H=\frac{-3g^2_2}{32
\pi^2}\frac{m^4_t \mu^2}{\sin^4\beta M^2_W}\frac{f(M^2_{\tilde u_1,
  \tilde u_2})}{M^2_{\tilde u_1}-M^2_{\tilde u_2}},
\eea
where $m_A$ is the pseudo-scalar Higgs mass and $f(M^2_{\tilde u_1,
  \tilde u_2})$ is a function of the lightest sups: $\tilde u_i$
for $i=1,2$. The running of $\mu$ is quite sensitive to the changes in
the running of the susy spectra and as a result  $m^2_{H^{-}}$ is
also sensitive.
 Thus, when we
include the running of the second family, the added small changes to the
supersymmetric spectrum alters the final value of $m_{H^-}$.

We have tested the effect of the gluino contribution by computing the ratio, 
$\delta_{\tilde g} = \frac{\hat K^2_{7\rm{eff}}}{{\hat K^2_{7\rm{eff} \tilde g}}}$, where these quantities have been defined in \eq{eq:pertb_conts_bsg} and \eq{eq:Kgluin} in Appendix \ref{sec:bsg_bmfv}. 
The ratio is very close  to 1 for $m_{1/2} > 300$ GeV. At lower
gaugino masses, the contribution from gluinos increases but the ratio
only decreases by about 1\% at $m_{1/2} = 160$ GeV.
When $m_{1/2}$ is below
$300$ GeV, the stops become quite light and a NLO calculation is not
possible using the approximation of 
\cite{Degrassi:2000qf}. %{\it Also found in the program by Gambino}. 
At lower $m_{1/2}$,
a LO calculation has to be used also for charginos, charged Higgs and
neutralinos. In fact when $m_{1/2}\leq 280$ GeV this is the case. Thus
all the results for $\bsgmm$ constrained to $m_{1/2}\leq 280$ GeV are
calculated at LO.

\subsection{Constraints on the parameter $\ep_S$ \label{sec:const_efS} }

As we have pointed out in Section \ref{sec:fsug}, once a family
symmetry is introduced, additional contributions to the trilinear
couplings are given by the derivatives of the Yukawa couplings with
respect to the flavon fields, $\sim Y_{ij}^f
F^{\alpha}\partial_{\alpha}$, and then in a realistic flavour
symmetry $(a_f)_{ij}$ must be of the form of \eq{eq:trilsft+fys} with
$k^f_{ij}\neq 0$. Given the limitation on the number of parameters
that we can use to scan the possibilities of the general form of
\eq{eq:sofmasses_offneq0}, with coefficients $x^f_{ij}\neq 1$ and
$z^f_{ij}\neq c^f_{ij}$, we choose to analyze the matrices of the form given in
\eq{eq:sofmasses_offneq0_ef_diff_efS}, and hence effectively allowing 
$\ep_{Sf} > \ep_f$. To remain consistent with the approximations
made earlier, we maintain the restriction $\ep_{Su} < \ep_{d}$. 

\begin{figure}[ht!]
\begin{center}
\includegraphics[width=.45\textwidth]{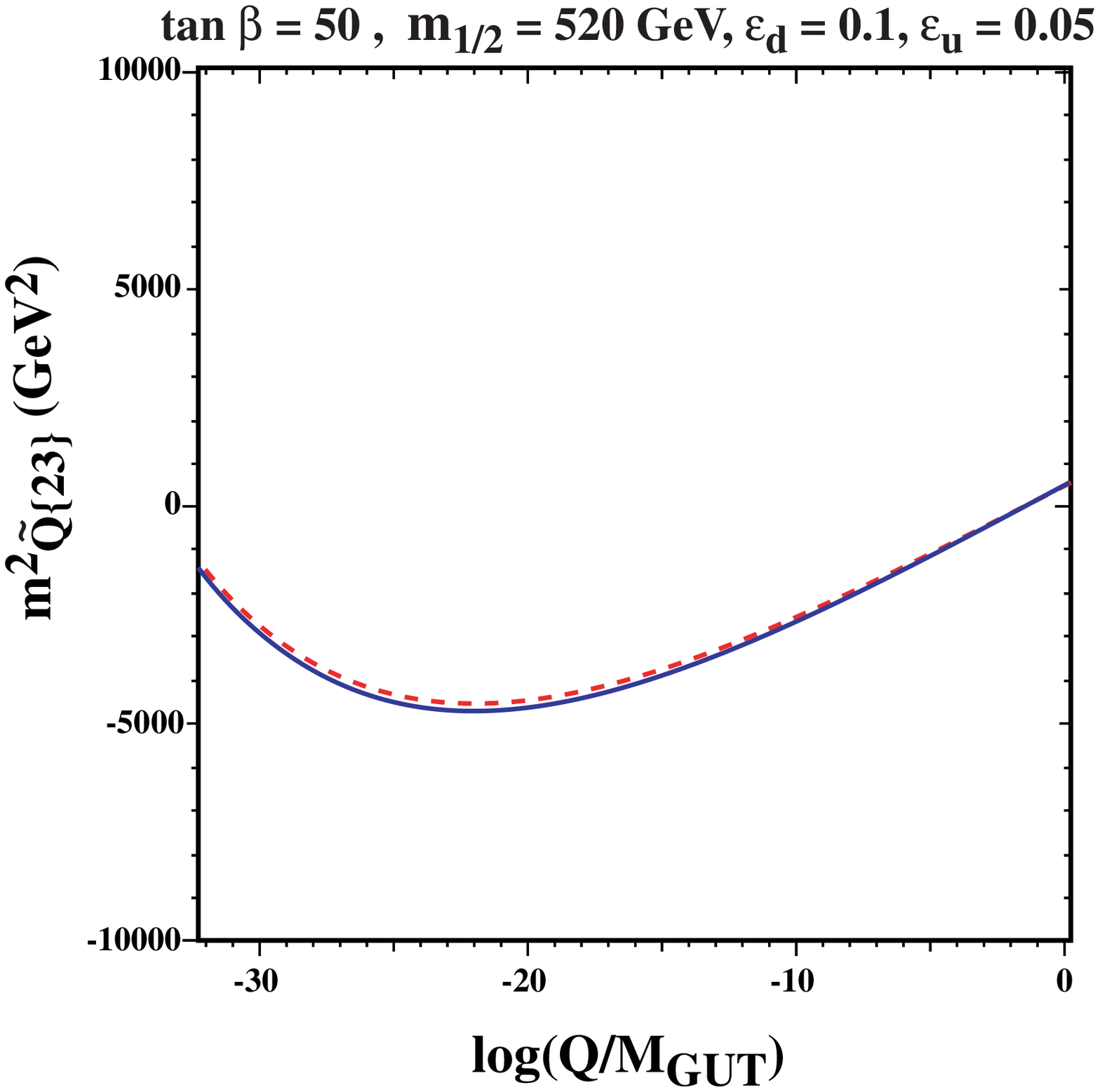}
\includegraphics[width=.45\textwidth]{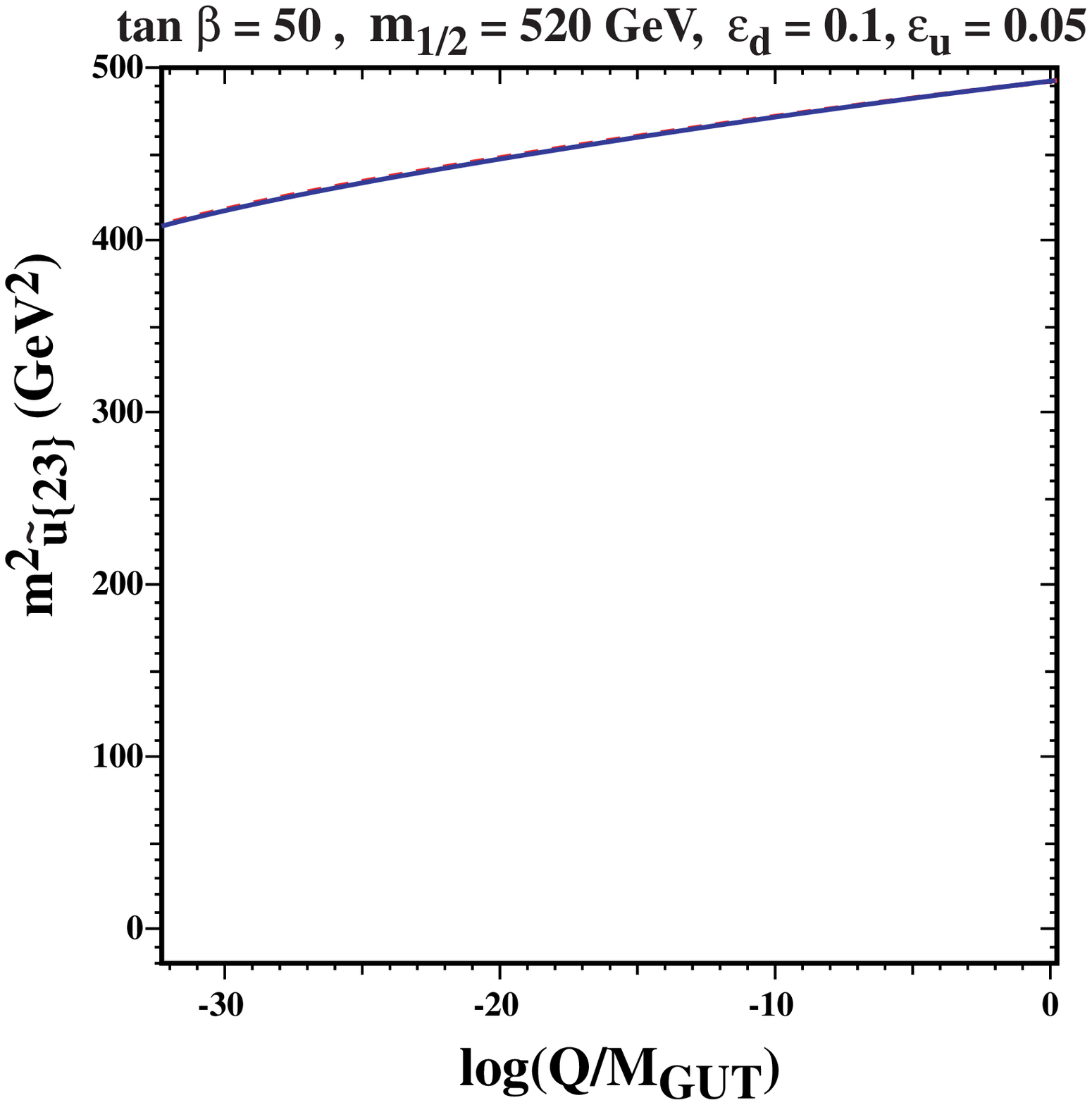}
\end{center}
\begin{center}
\includegraphics[width=.45\textwidth]{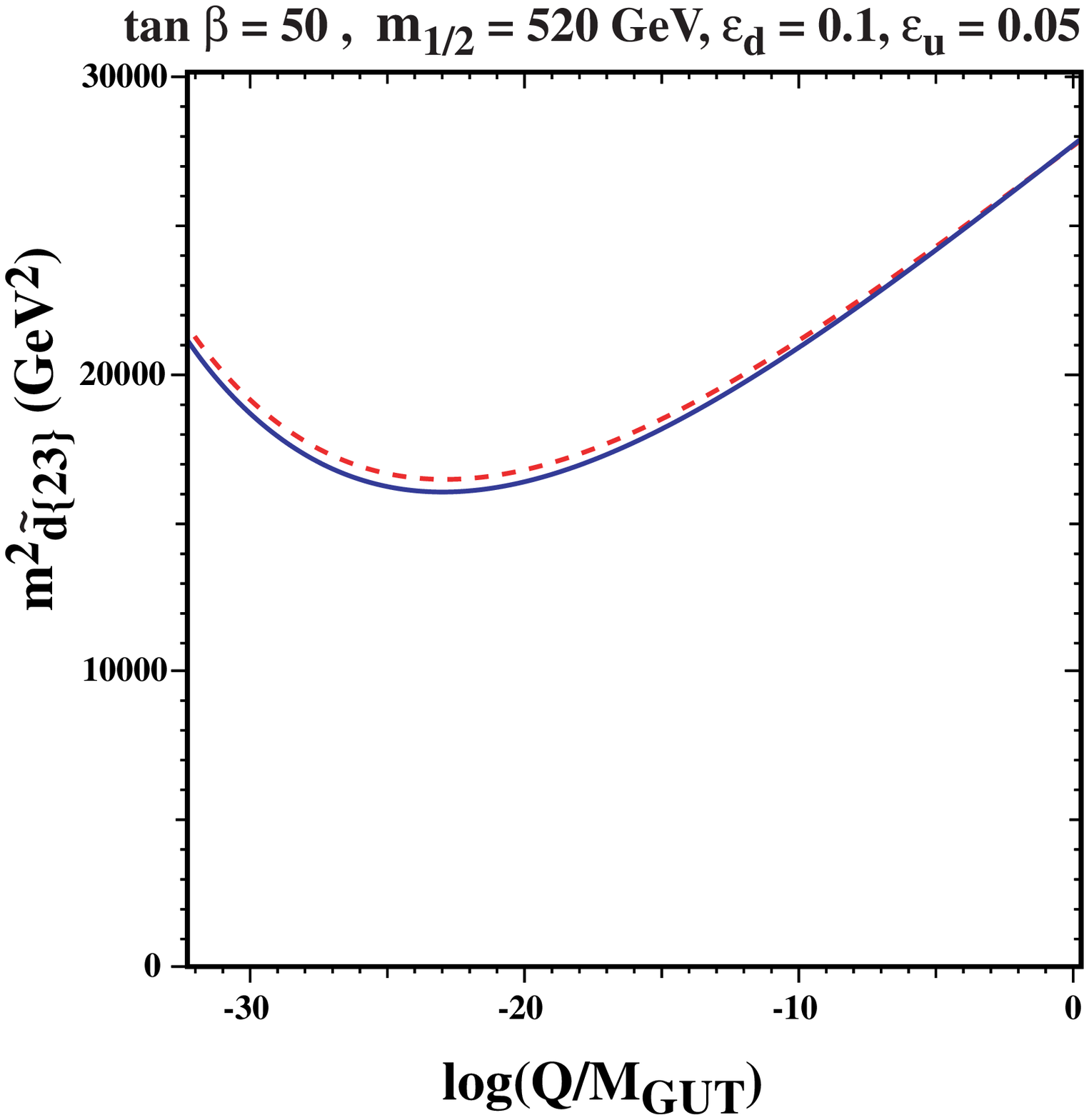}
\includegraphics[width=.45\textwidth]{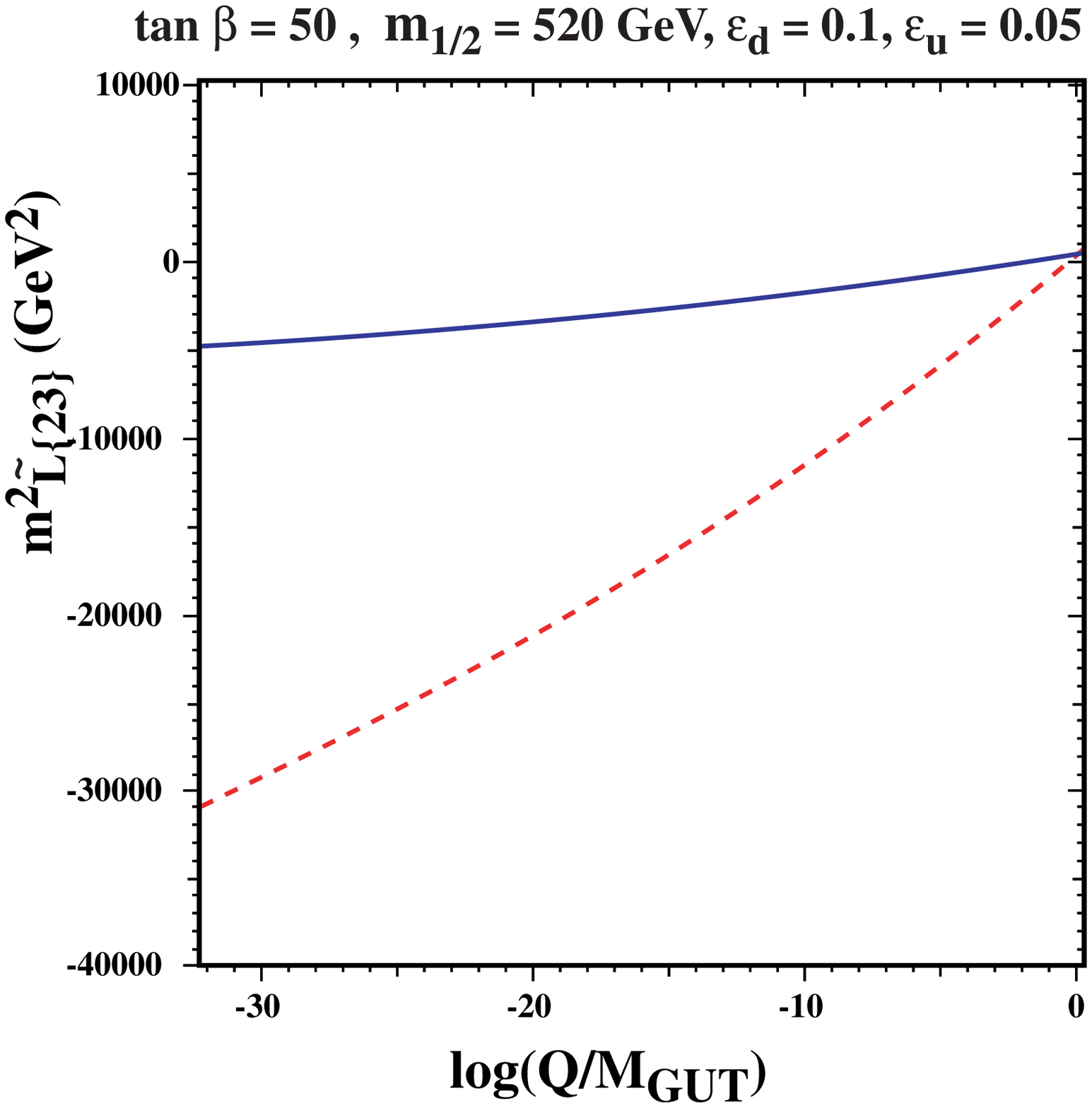}
\end{center}
\caption{\small As in Fig. \protect\ref{fig:ed_eu_softsqrd_comp_0.08_520_330_0} for case
I with $\ep_{Sd}=\ep_{Se}=0.45$,  $\ep_{Su}=0.06$ (red dashed lines), and for case
II with $\ep_{Sd}=0.45$, $\ep_{Se}=0.225$ and $\ep_{Su}=0.06.$ (blue solid lines). Here  $m_0=370$ GeV and $A_0 = -m_{1/2}$. }
\label{fig:eSf_ssqd_comp_520_370_-520}
\end{figure}

To this end, we first check once more the running of the
off-diagonal squared masses and trilinear couplings with respect to
the parameters $\ep_{Sf} > \ep_f$. For the case of $\ep_{Sf}=\ep_f$ we
have seen in the discussion of section \ref{sec:runn_off_soft_p}
(Fig. \ref{fig:ed_eu_softsqrd_comp_0.08_520_330_0}) that
$(M^2_{\tilde L})_{23}$ and $(M^2_{\tilde e})_{23}$ are more sensitive than the quark sector, other than $(M^2_{\tilde d})_{23}$. To be definite, we make the following choices for $\ep_{Sf}$:
\bea
& & I. ~~ \ep_{Sd} = \ep_{Se} =0.45, \quad \ep_{Su}=0.06, \nn\\
& & II.~  \ep_{Sd}=0.45, \quad 
\ep_{Se}=0.225, \quad  \ep_{Su}=0.06,
\label{eq:choices_epSf}
\eea
for both cases we keep $c^d_{22}=0.7$ and $\ep_d=0.1$,
$\ep_u=0.05$. In Figure \ref{fig:eSf_ssqd_comp_520_370_-520}, we plot
then the running of  $ M^2_{\tilde Q ~ 23}$,  $ M^2_{\tilde u ~ 23}$,
$ M^2_{\tilde d ~ 23}$ and $ M^2_{\tilde L ~ 23}$ for the two cases of
\eq{eq:choices_epSf}. These figures can be compared directly with those in Fig. 
\ref{fig:ed_eu_softsqrd_comp_0.08_520_330_0}. The only difference being the
increased values of the expansion parameters which are now distinct from those
in the Yukawa sector.

\begin{figure}[ht!]
\begin{center}
\includegraphics[width=.45\textwidth]{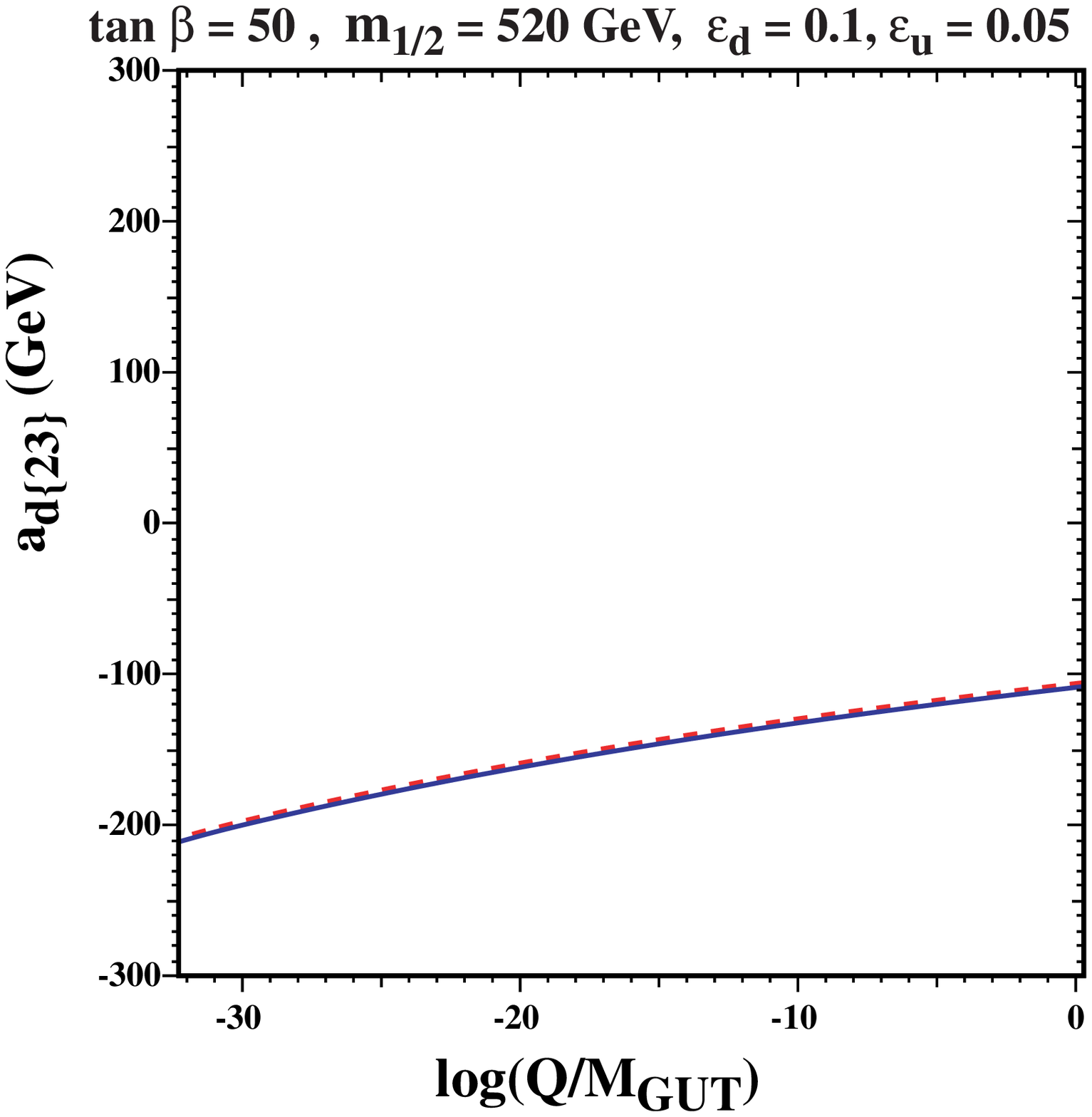}
\includegraphics[width=.45\textwidth]{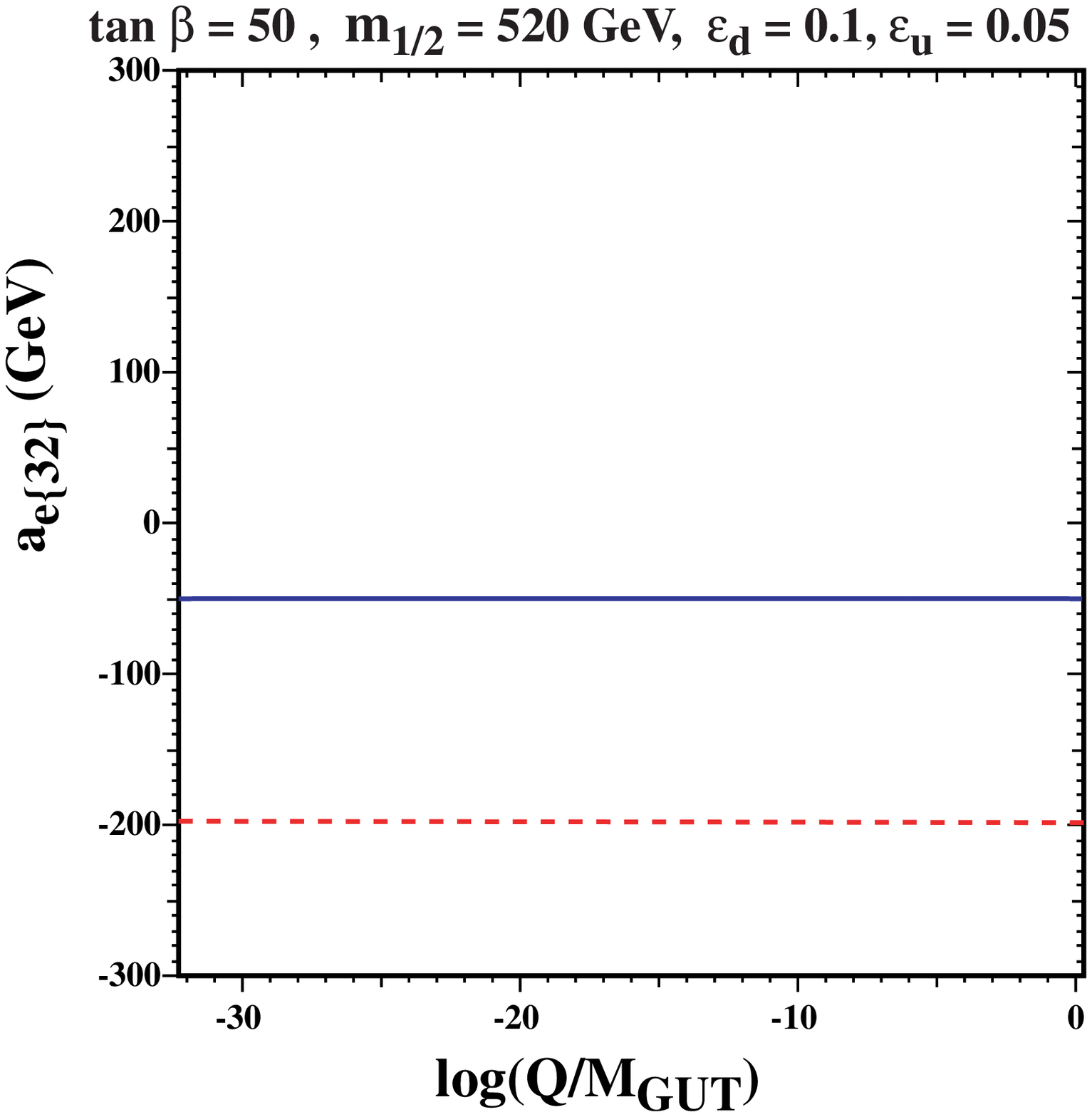}
\end{center}
\caption{\small As in Fig. \protect\ref{fig:eSf_ssqd_comp_520_370_-520}
for the running of a)  ${a_d}_{23}$ and b) ${a_e}_{32}$.}
\label{fig:eSf_eL_ssqd_comp_520_370_-520}
\end{figure}

The running of  $ M^2_{\tilde Q ~ 23}$ is significantly affected by the 
increase in $\ep_{Sd}$ as  one can see from the comparison between Figs. \ref{fig:ed_eu_softsqrd_comp_0.08_520_330_0}a
and \ref{fig:eSf_ssqd_comp_520_370_-520}a. This is largely due to the increased importance of the $a_d^\dagger a_d$ term in the beta function which is now competing with the Yukawa terms in Eq. \ref{eq:beta1l_soft}.
Not surprisingly the running of $ M^2_{\tilde u ~ 23}$ is quite close  to that of Figure \ref{fig:ed_eu_softsqrd_comp_0.08_520_330_0}b because we have kept $\ep_{Su}=0.06$ quite close to $\ep_{Su}=0.05$. In order to affect the branching ratio for $\bsgmm$, we are most interested in the effect on the $d$ sector. The initial value of $ M^2_{\tilde d ~ 23}$ is determined by $\ep_{Sd}$ (and $m_0$) and is increased relative to the case studied in Fig.  \ref{fig:ed_eu_softsqrd_comp_0.08_520_330_0}c by the the ratio $\ep_{Sd}/\ep_{d}=0.45^2/0.1^2=20.25$.
As a result, we can see in Figure \ref{fig:eSf_ssqd_comp_520_370_-520}c the enhanced running of  $ M^2_{\tilde d ~ 23}$. In panel d), we see the differences between cases I and II, as the running of $ M^2_{\tilde L ~ 23}$ is clearly sensitive to $\ep_{Se}$.
In Fig. \ref{fig:eSf_eL_ssqd_comp_520_370_-520}, we show the corresponding effect on the trilinear couplings which can be compared directly with Fig. \ref{fig:ed_eu_aterms_comp_0.08_520_330_0}.

\begin{figure}[ht!]
\begin{center}
\begin{center}
\includegraphics[width=.32\textwidth]{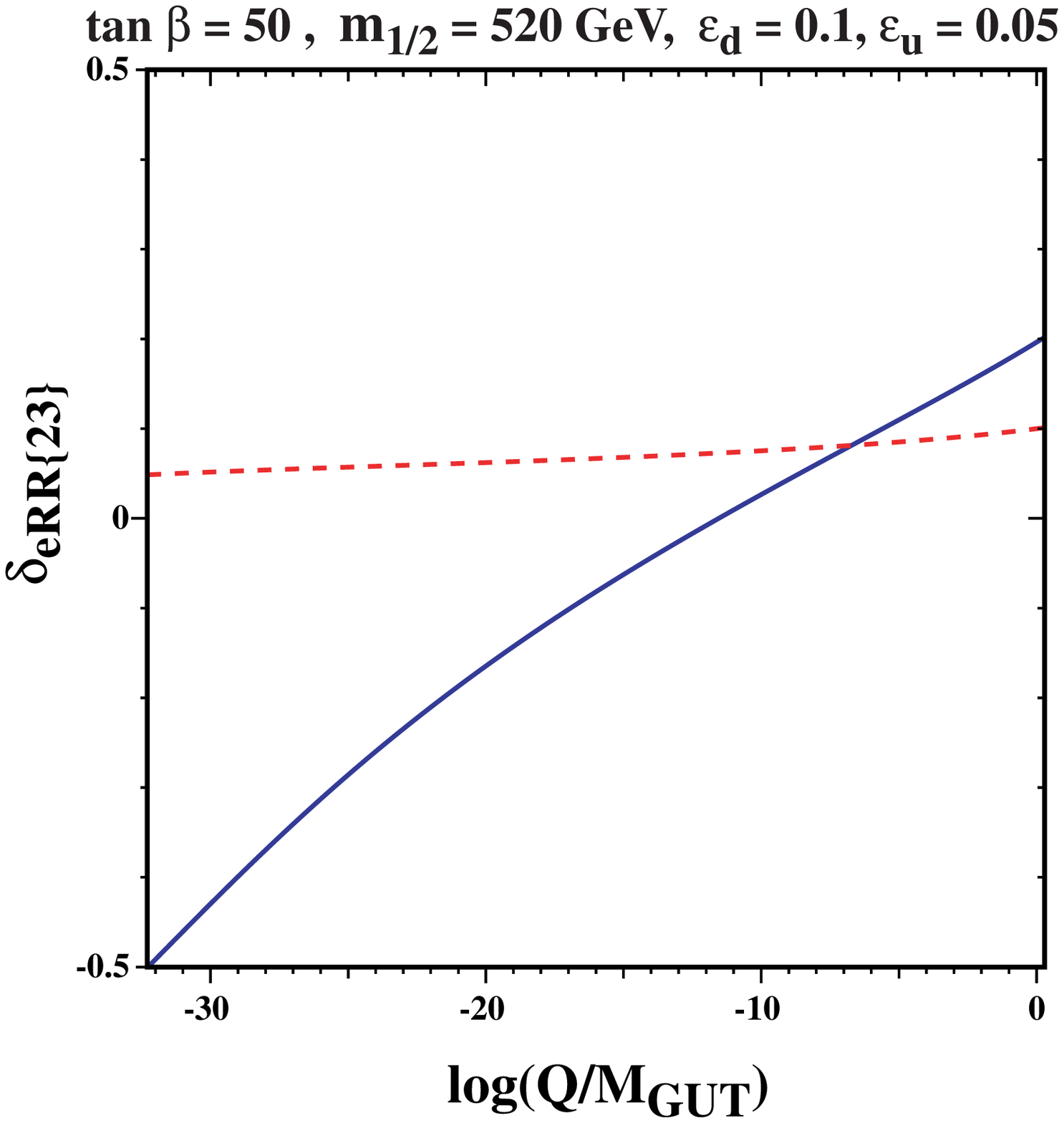}
\includegraphics[width=.32\textwidth]{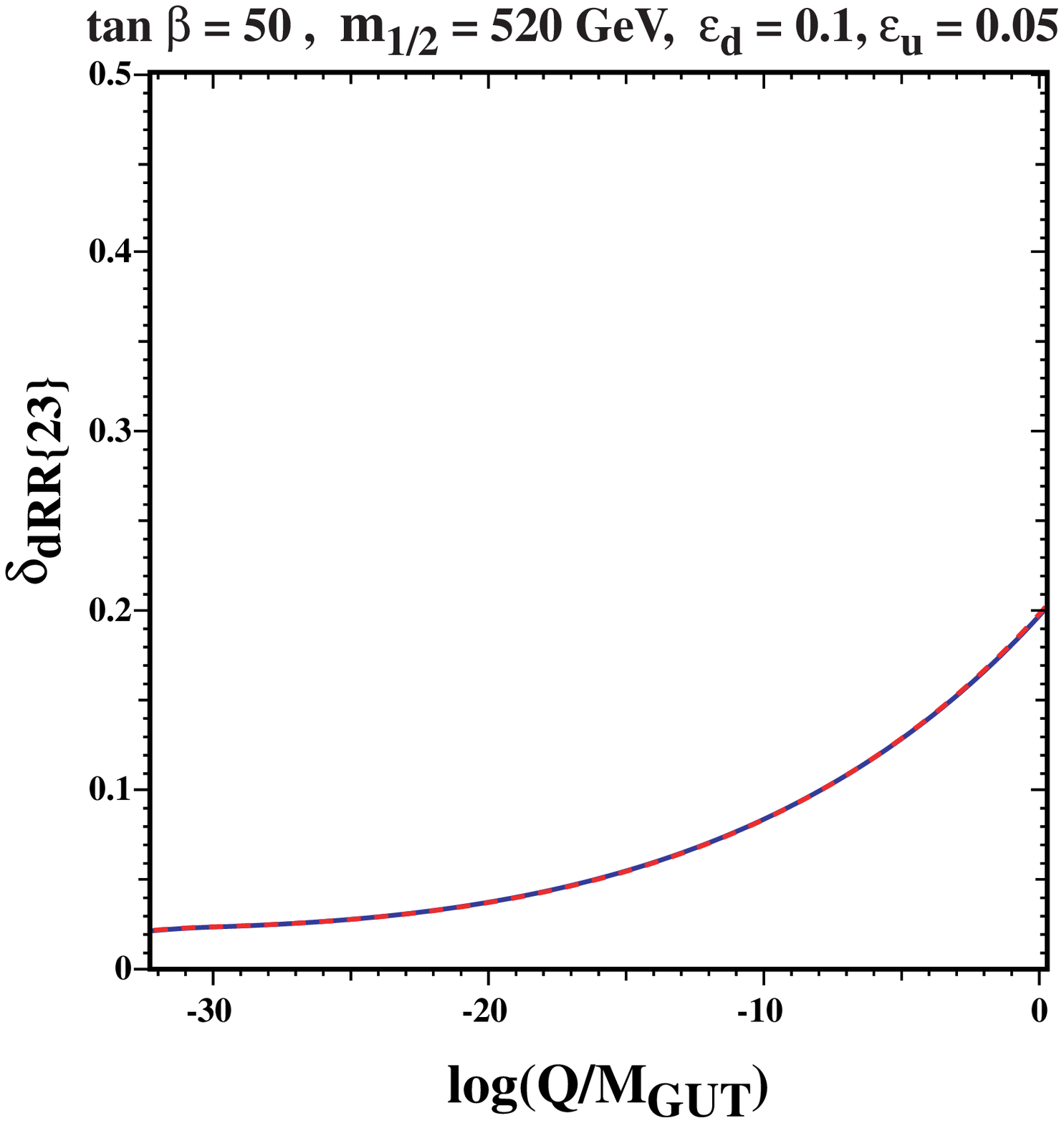}
\end{center}
\includegraphics[width=.32\textwidth]{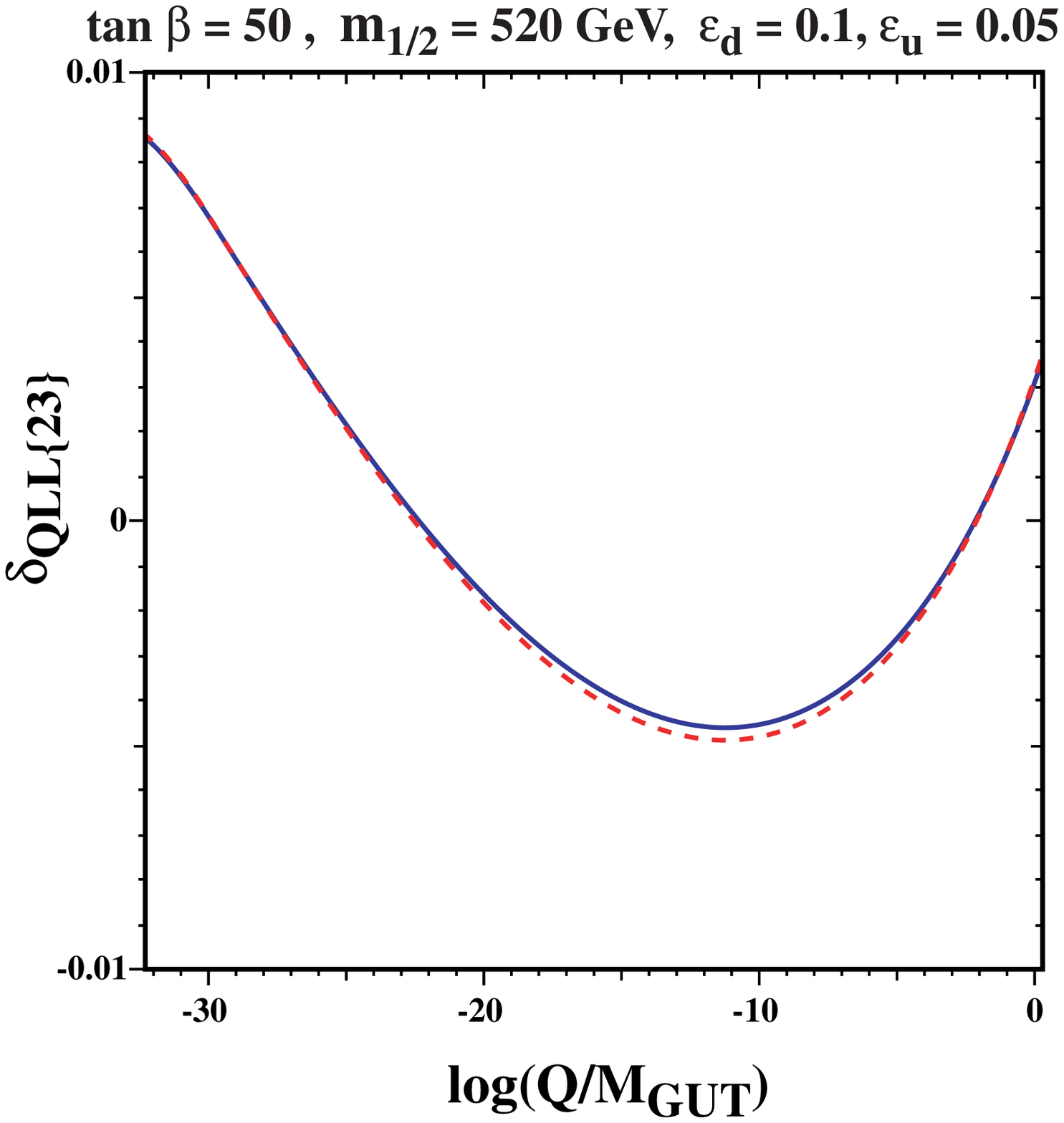}
\includegraphics[width=.32\textwidth]{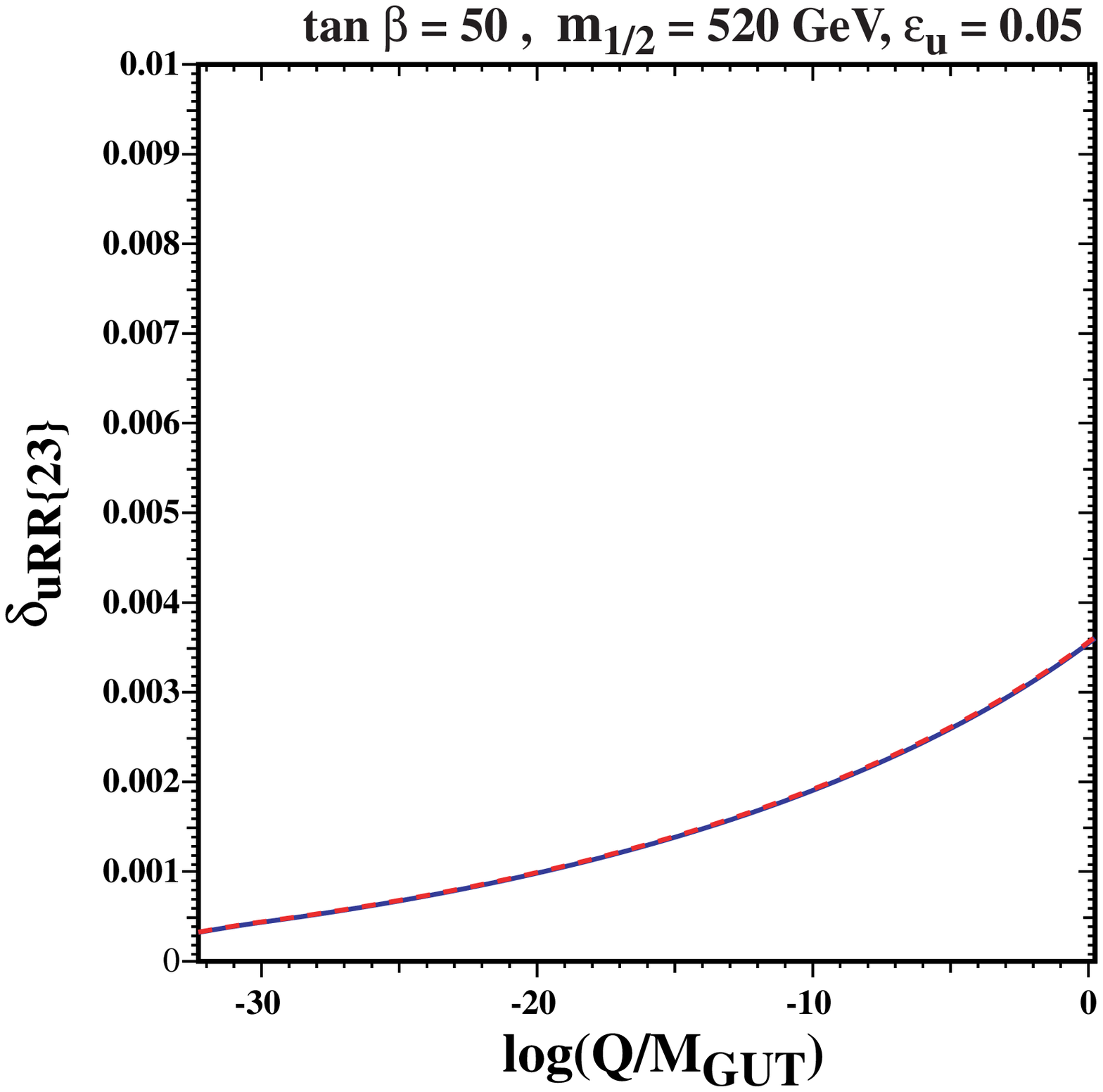}
\end{center}
\begin{center}
\includegraphics[width=.34\textwidth]{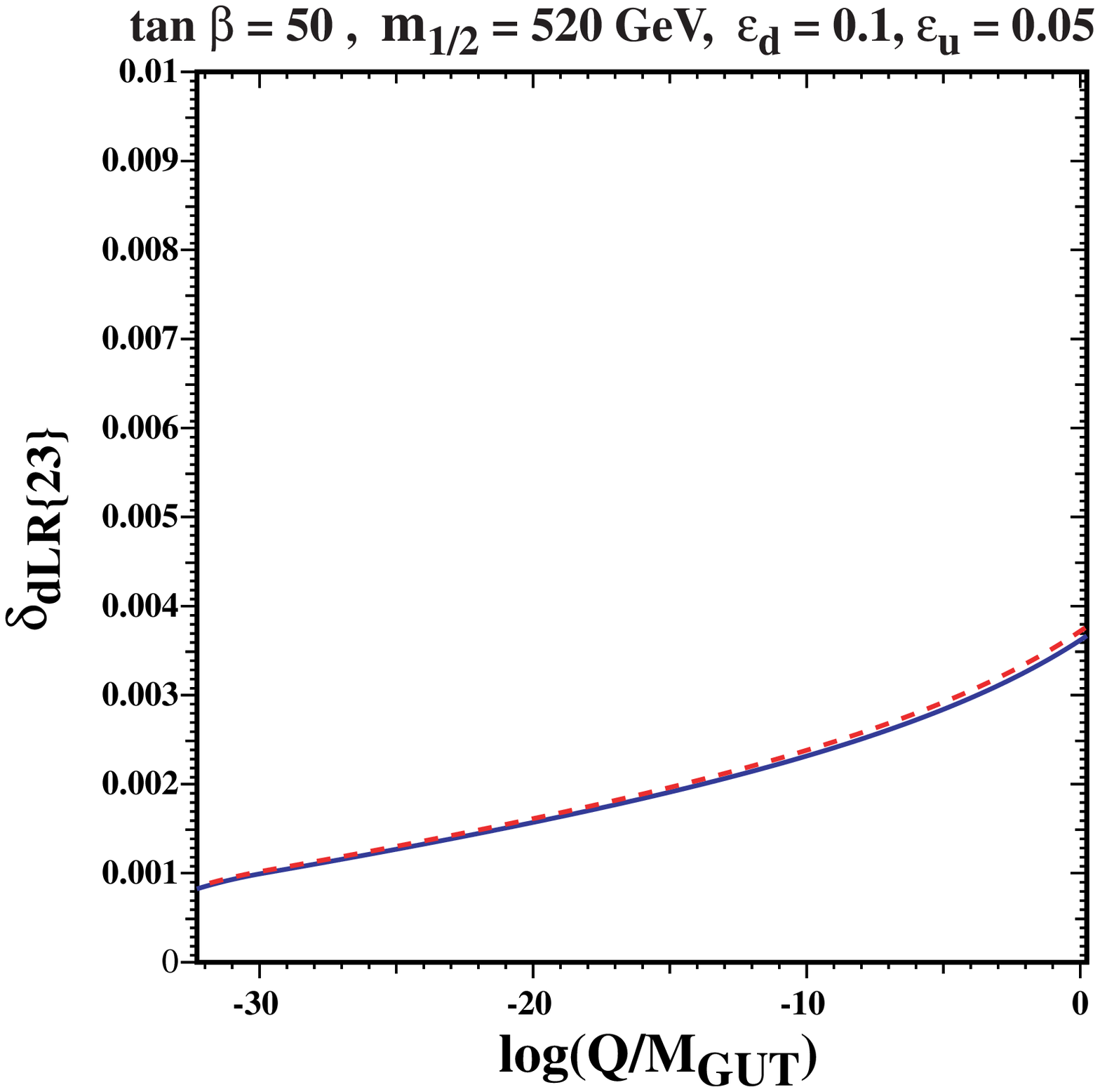}
\includegraphics[width=.32\textwidth]{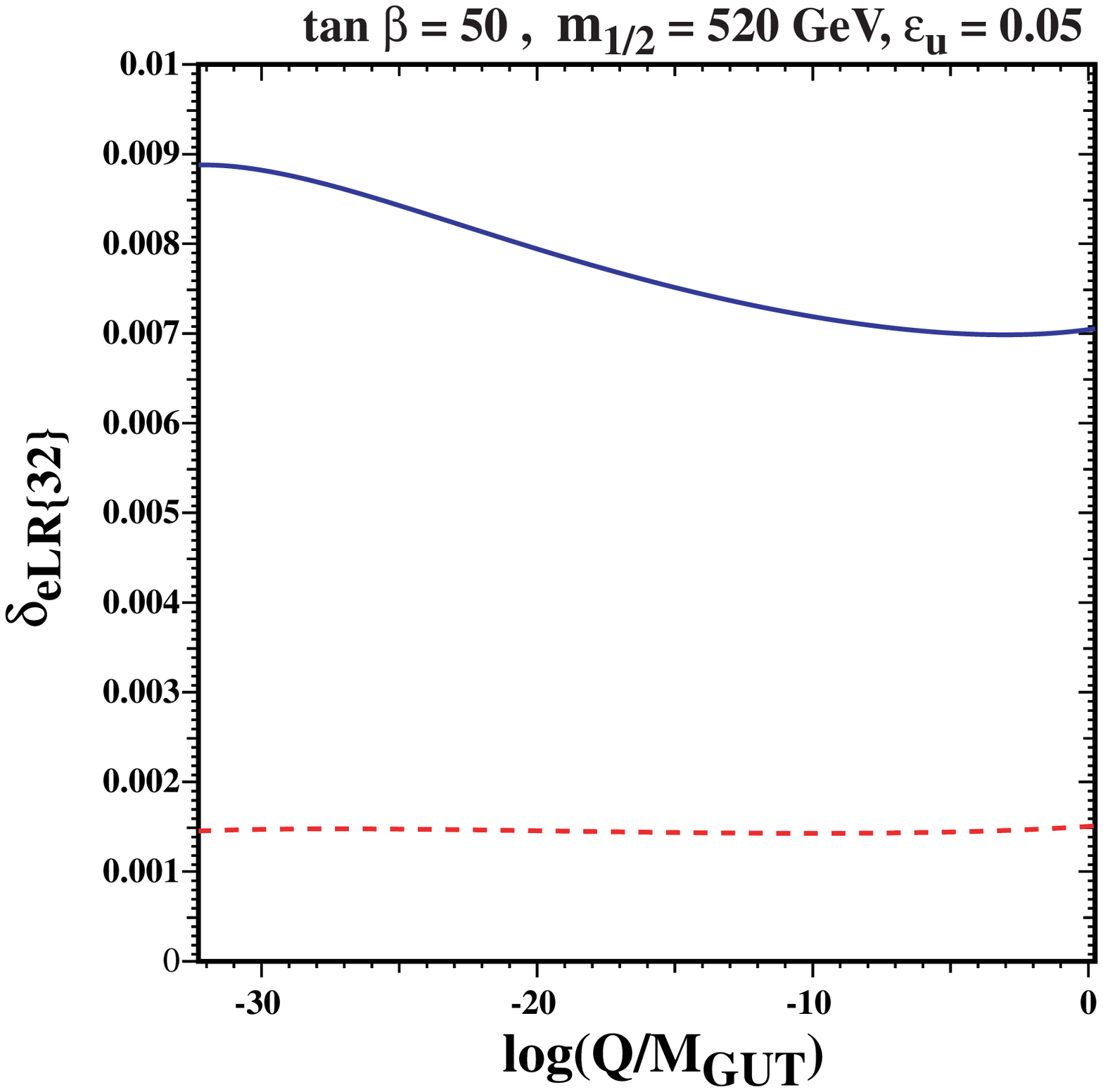}
\end{center}
\caption{\small As in Fig. \protect\ref{fig:eSf_ssqd_comp_520_370_-520}
for the running of $(\delta_{f,XY})_{ij}$ with $f,XY$ = a) $e,RR$ 
b) $d,RR$, c) $Q,LL$, d) $u,RR$, e) $d,LR$, f) $e,LR$.}
\label{fig:edS_euS_dlt_qXX_520_370_-520}
\end{figure}

In Fig. \ref{fig:edS_euS_dlt_qXX_520_370_-520},
we show the effect of the increased $\ep_S$ on the parameters
$(\delta_{f,XY})_{ij}$ defined in \eq{eq:del_f_XY}.
The increased effect on the running of these parameters is also
largely due to the increase in the initial values. For example, 
for $(\delta_{d,RR})_{23}$ the initial value is increased from  
$9\times 10^{-3}$ to  $0.021$, when
$\ep_d=\ep_{Sd}$ changes from 0.1 to $\ep_{Sd}=0.45$. The comparison
to the increase for the other $(\delta_{f,XY})_{23}$ is given in Table \ref{dLs_epSf}.

\begin{table}%+[!h]
\begin{center}
\begin{tabular}{|l | l l l |}
\hline
\multicolumn{4}{c}{$|(\delta_{f, XY})_{23}|$ for $(m_{1/2},m_0,A_0)=(520,370,-520)$ GeV }\\
\hline
                                 &                        &    I   &  II    \\
$|(\delta_{L_{,LL}})_{23}|$       & $ 0.0012 $               & $0.15$
& 0.014  \\
$|(\delta_{e_{,RR}})_{23}|$       & $ 0.0069 $               & $0.50$ 
& 0.05 \\
$|(\delta_{e_{,LR}})_{23,32}|$    & $ 6.0 \times 10^{-6} $   & $8.6
\times 10^{-3} $
& $1.5 \times 10^{-3} $ \\
$|(\delta_{e_{,RL}})_{23,32}|$    & $ 4.4 \times 10^{-6} $  & $7.2 \times 10^{-3} $ 
& $1.1 \times 10^{-3} $ \\
$|(\delta_{Q_{,LL}})_{23}|$       & $ 0.0069 $              & $0.009$ 
& $$0.009$ $ \\
$|(\delta_{d_{,RR}})_{23}|$       & $9\times 10^{-4}$       & $0.021$ 
& $0.021$     \\
$|(\delta_{d_{,LR}})_{23}|$       & $4.3 \times 10^{-6} $  &
  $8.0\times 10^{-4} $  
& $8.0\times 10^{-4} $ \\
\hline
\end{tabular}
\end{center}
\caption{\small Weak scale values for the $|(\delta_{f, XY})_{23}|$ (denoted in the 1st column) for
  $(m_{1/2},m_0,A_0)=520,370,-520)$ GeV. In the 2nd column, we assume 
  $\ep_{Sd}=\ep_d=0.1$, $\ep_{Su}=\ep_u=0.05$.  In column three, we take
$\ep_{Sd}=0.45$,  $\ep_{Su}=0.06$  and $\ep_{d}=0.1$,
  $\ep_{u}=0.05$ while in column four, we lower $\ep_{Se}$ to 0.225.}
\label{dLs_epSf}
\end{table}

If one would like to introduce right handed neutrinos for these cases, then the present analysis will be valid assuming the right handed neutrinos decouple at a higher scale than $M_{\rm{GUT}}$, or $M_{\nu_R}>M_{\rm{GUT}}$. For $M_{\nu_R}<M_{\rm{GUT}}$ the quark sector will not be largely affected but the lepton sector will be.  We will have then important changes to the running of $(M^2_{\tilde L})_{23}$ and $(M^2_{\tilde e})_{23}$. Since the goal in this paper is to study the constraints mainly in the quark sector, the introduction of $M_{\nu_R}<M_{\rm{GUT}}$ is beyond the scope of the present work. 
 For the present case we
 consider constraints from leptonic flavour violating process such as $\tau\rightarrow \mu \gamma$. For example in the CMSSM,  the bounds in terms of the flavour violating parameters \eq{eq:del_f_XY}, for $\tan\beta=50$, are \cite{Masina:2002mv}
\bea
|(\delta_{L, LL})_{23}| \lsim 0.1,\quad |(\delta_{e, RR})_{23}| \lsim 0.1,
\quad  |(\delta_{e, LR})_{23,32}|\lsim 0.02, 
\label{eq:msugra_dl_bds} 
\eea
using the most stringent current bounds. For the case $\ep_{Sf}=\ep_f$ the values obtained for $(\delta_{l,XX})_{23}$ are well below
these bounds:
\bea
&&|(\delta_{L, LL})_{23}| = 0.0012,\quad \quad  \ |(\delta_{e, RR})_{23}|=0.0069,\nn\\
&&|(\delta_{e, LR})_{23}| = 6.0 \times 10^{-6},\quad |(\delta_{e,RL})_{23}| =  4.4 \times 10^{-6}.
\eea
For case I of \eq{eq:choices_epSf} the bounds of \eq{eq:msugra_dl_bds} are already saturated while for case II, the parameters are still within the experimental limits.

\begin{figure}[ht!]
\begin{center}
\includegraphics[width=0.46\textwidth]{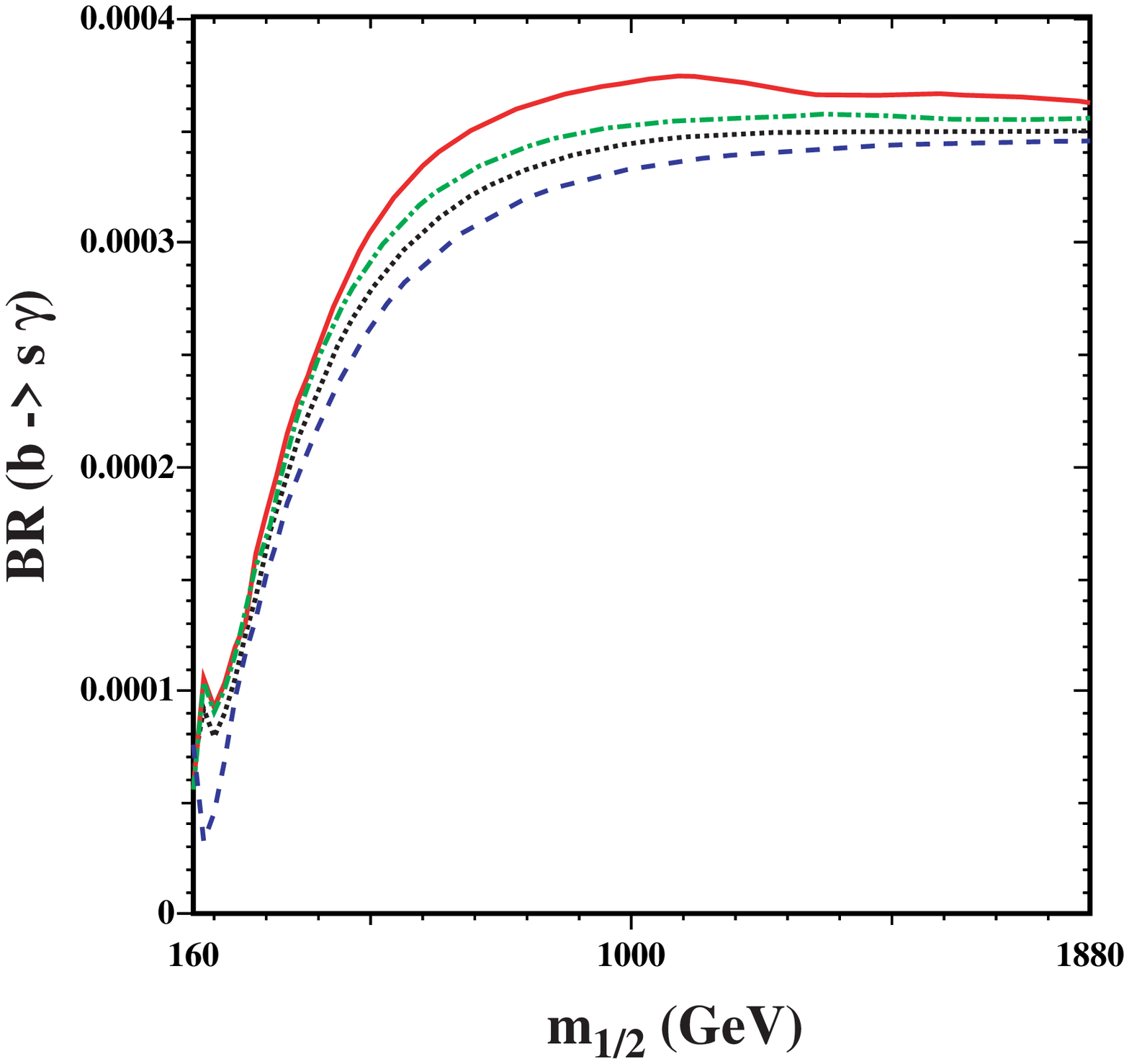} 
\includegraphics[width=0.47\textwidth]{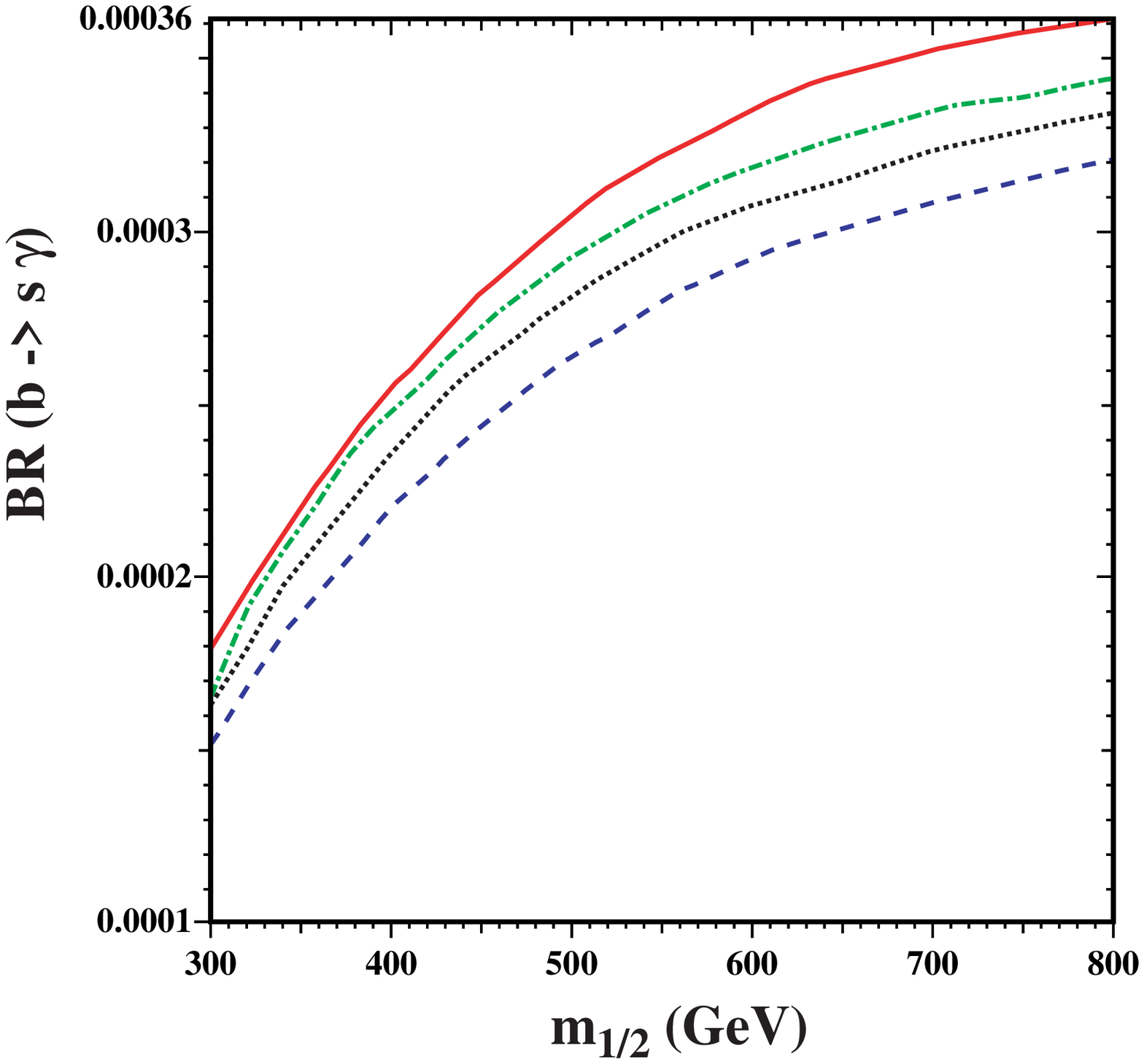} 
\end{center}
\caption{\small As in Fig. \protect\ref{fig:bsg_0.1_comp}. The curves from bottom to top are as follows: the single family limit is shown by the blue dashed curve; the choice of $\ep_d=\ep_{Sd}=0.1$ is shown by the black dotted curve; increased mixing with $\ep_{Sd}=0.45$, $\ep_{Se}=0.225$ is shown by the green dot-dashed curve; and $\ep_{Sd}=\ep_{Se}=0.45$ is given by the red solid upper line.} 
\label{fig:bsg_A0=520_efS}
\end{figure}

As can be expected,
the sensitivity of $\bsgmm$ on our expansion parameters 
for the  cases  $\ep_{dS}> \ep_{d}$ is greater
than that discussed in the previous section with $\ep_{dS}= \ep_{d}$. The value
for the branching ratio for $\bsgmm$ as a function of $m_{1/2}$ is shown in  Fig. \ref{fig:bsg_A0=520_efS}. The lower two curves correspond to the single family case (where
all mixings are ignored) and our nominal choice of $\ep_{Sd} = \ep_d$ and are taken directly 
from Fig. \ref{fig:bsg_0.1_comp}. The upper two curves show the effect of increasing
$\ep_{Sd}$ and $\ep_{Se}$ using values from case I (upper solid red curve) and case II
(green dot-dashed curve). Here we see more clearly that in regions with relatively low $m_{1/2}$,
the effect of mixings relaxes the constraint imposed by $\bsgmm$.
In these cases, the gluino
contributions can be up to $10\%$ for $m_{1/2}$ around 200 GeV but when
$m_{1/2}$ increases these contributions reduce considerably. Here the contribution
from gluinos does not vary too much from cases I and and II of
\eq{eq:choices_epSf}, as the running in the $\tilde{d}$ sector is not greatly
affected.

To better appreciate the effect of the mixing in the squark sector, we zoom in on the region between $m_{1/2}$ = 300 and 800 GeV as shown in panel b) of Fig. \ref{fig:bsg_A0=520_efS}. For $\tan \beta = 50$, this region is preferred when other observables such as $(g-2)_\mu$ are included \cite{ehoww,eos}. When theoretical and calculational errors are included, the 95\% exclusion value for BR($\bsgmm$) occurs at about 1.9 $\times 10^{-4}$, as discussed in the previous section and corresponds to a lower limit of 400 GeV on $m_{1/2}$ in the single family case most commonly treated in the CMSSM. This limit is substantially relaxed when two-family mixing is included. For $\ep_{Se} = \ep_{Sd} = 0.45$ as shown by the upper curve in the figure, we see that the limit is relaxed to $m_{1/2} \gsim 335$ GeV.
One should keep in mind that the exact numerical limit found here and in the CMSSM 
is subject to change when NNLO SM corrections are included. On the other hand, we expect the 
relative change in the limit when two-family mixing is included to be a robust result.

\begin{figure}[ht!]
\begin{center}
\includegraphics[width=0.46\textwidth]{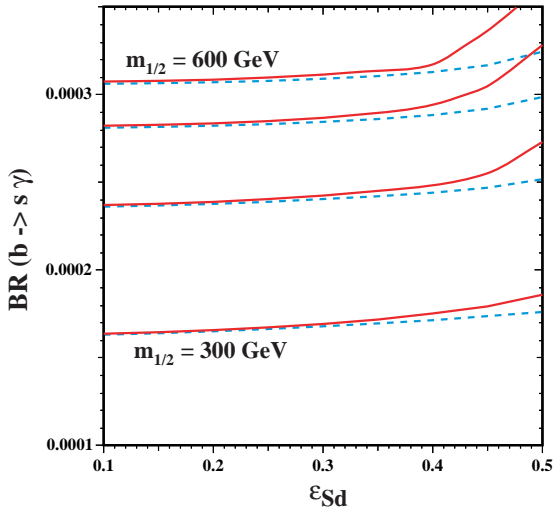}
\end{center}
\caption{\small Variation of $\BR(\bsgmm)$ with respect to
  $\ep_{Sd}$. From top to bottom the lines correspond to  $(m_{1/2}, m_0, A_0)=$
  $(600,410,-600)$ GeV,  $(500,360,-500)$ GeV,  $(400,340,-400)$ GeV
  and $(300,380,-300)$ GeV.} 
\label{fig:bsg_edS_A0=-400_-500}
\end{figure}

Finally, we show in Fig. \ref{fig:bsg_edS_A0=-400_-500} the behaviour of BR$(\bsgmm)$ on $\ep_{Sd}$ for several choices of $m_{1/2}$. As one can see, for low $m_{1/2}$, the branching ratio is lower than the experimental value when $\ep_{Sd}$ is small. When $\ep_{Sd} \gsim 0.4$, the branching ratio increases and the constraint on $m_{1/2}$ is relaxed.  

\subsection{Sensitivity on the determination of the MSSM spectra}

\begin{figure}[ht!]
\begin{center}
\includegraphics[width=0.3\textwidth]{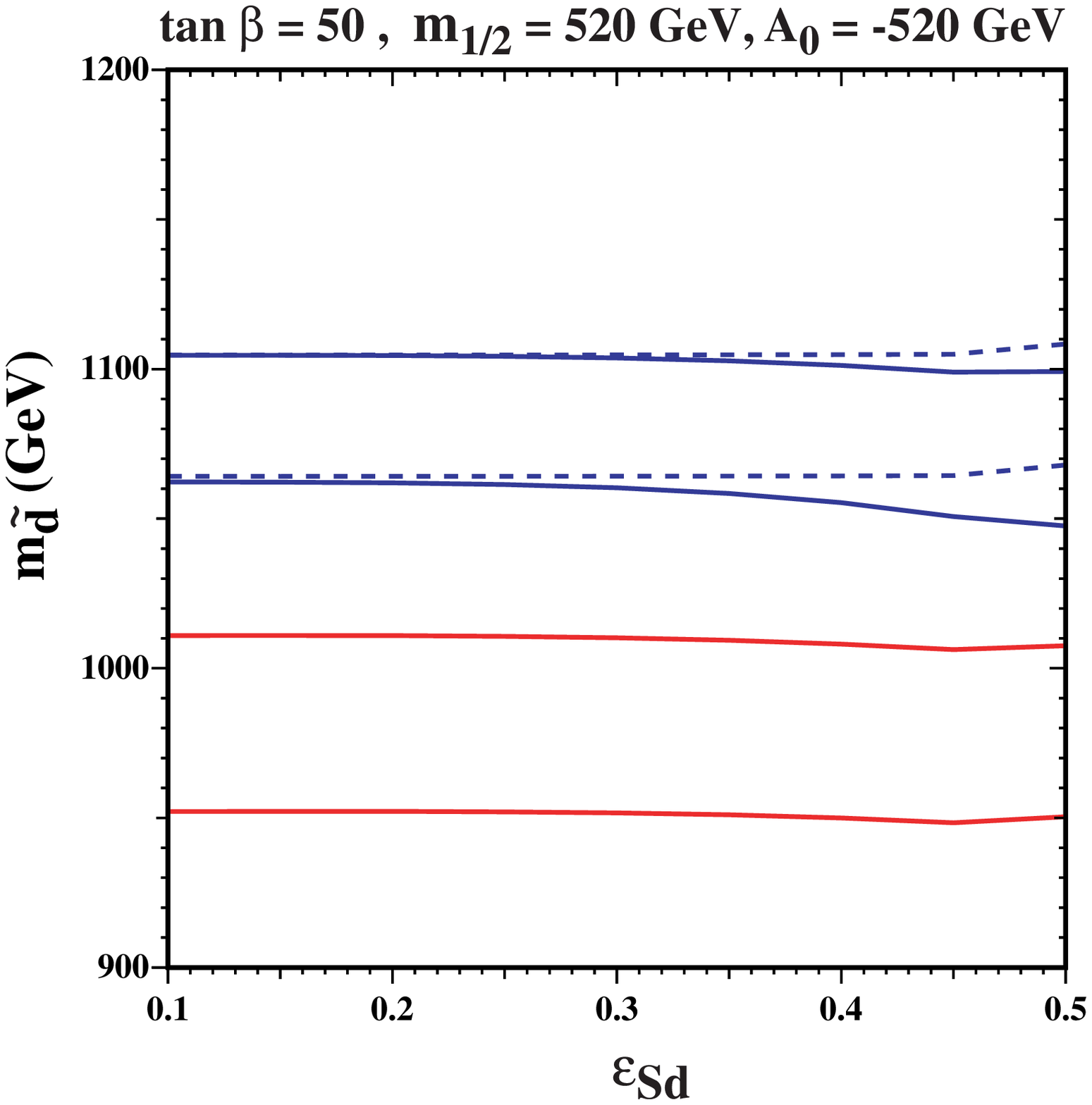} 
\includegraphics[width=0.3\textwidth]{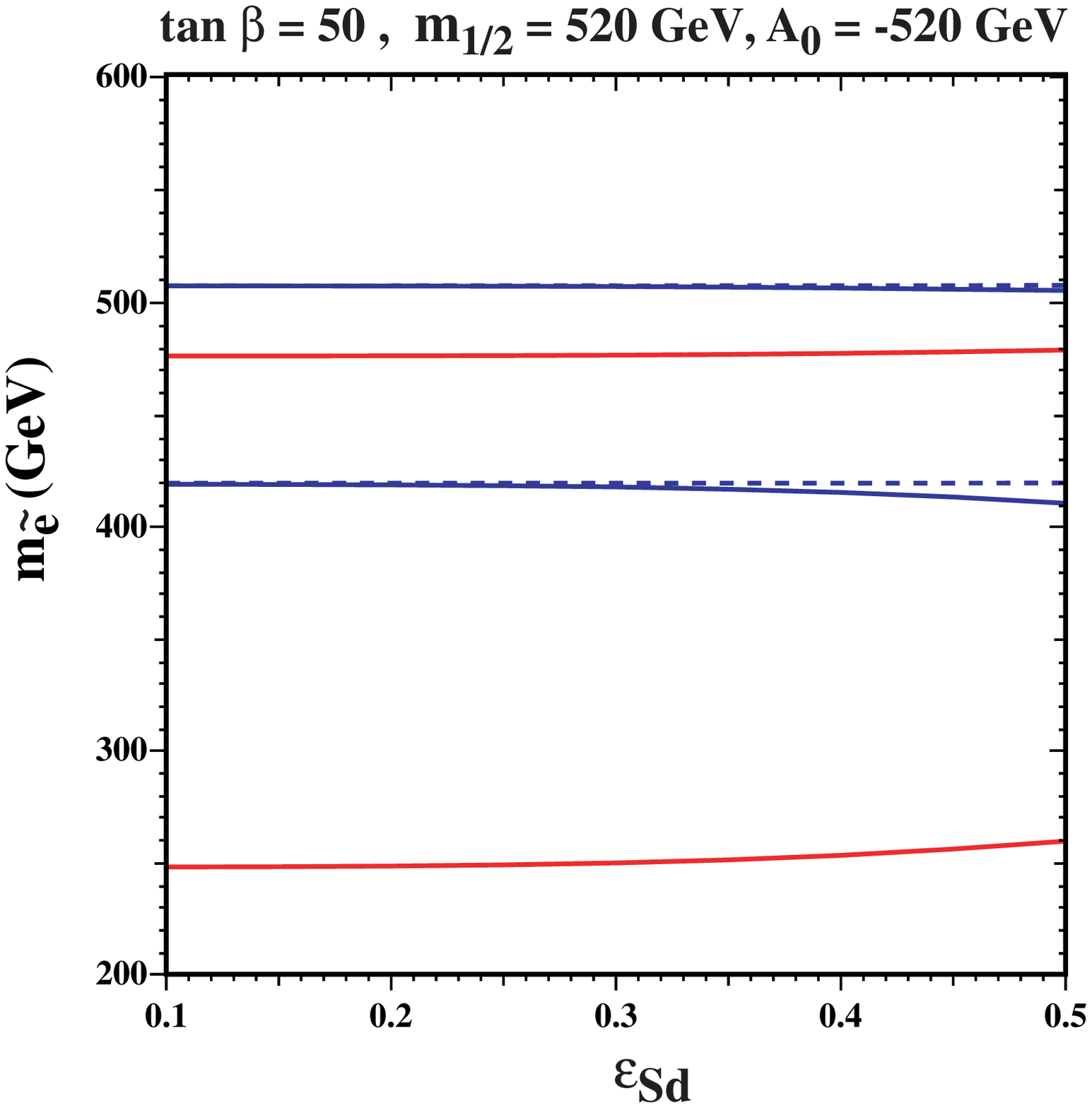} 
\includegraphics[width=0.3\textwidth]{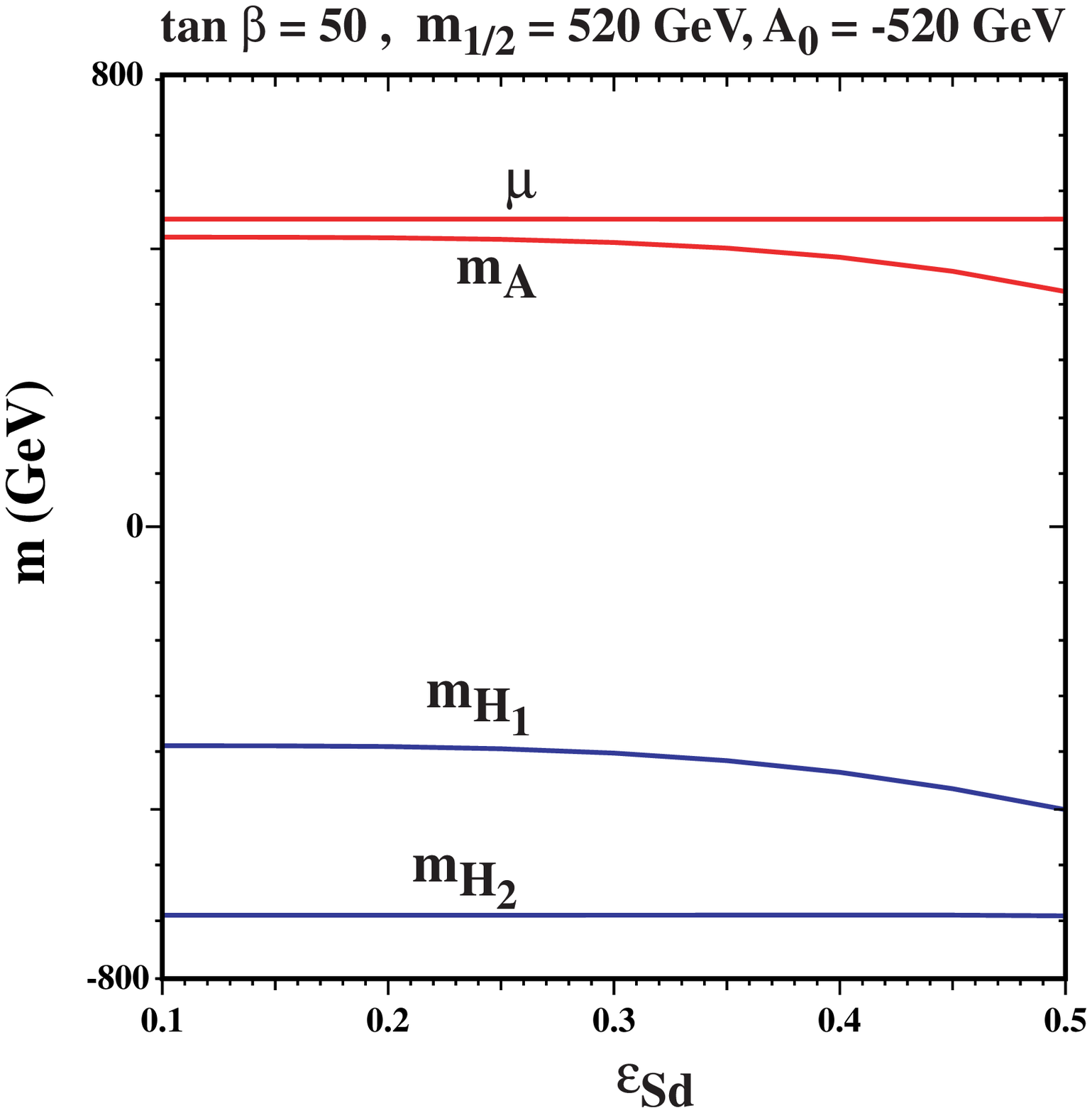} 
\end{center}
\caption{\small Masses for the sdowns (a), and charged sleptons (b) sectors as a function of
$\ep_{Sd}$ for the case $(m_{1/2},m_0,A_0)$
  $=(520,370,-520)$ GeV, $\ep_{Se}=\frac{\ep_{Sd}}{2}$,
  $\ep_{d}=0.1$, and  $\ep_{u}=0.05$.  In panel c, the soft Higgs masses, $\mu$ and $m_A$
  are shown as labeled.} 
\label{fig:mssm_spectra_520}
\end{figure}

In this last section, we would like to consider the effects of the
our flavour violating parameters on the resulting supersymmetric spectrum.
 Since the
parameters  $\ep_u$,  $\ep_d$, $\ep_{Su}$,  $\ep_{Se}$,  $\ep_{Sd}$ are small ($<1$)
and  their dependence on the Yukawa, trilinear and soft
squared matrices is quadratic, Eqs.~ (\ref{eq:sofmasses_offneq0}-\ref{eq:sofmasses_offneq0_ef_diff_efS}),
we do not expect a big effect on the 
spectra, except for sensitive parameters such as
\bea
\mu,\ m_{H_{1,2}},\  m_{A},
\eea
which are determined from the electroweak symmetry breaking
conditions. In Fig. \ref{fig:mssm_spectra_520}, we show the
dependence of the six masses of the sdown sector and of the
charged slepton sector with respect to $\ep_{Sd}$ for the case
\bea
\ep_{Se}=\frac{1}{2} \ep_{Sd},
\eea
with $\ep_{Su}=0.06$ which includes case II of  \eq{eq:choices_epSf}
for $\ep_{Sd}=0.45$. Here we have fixed $m_{1/2} = - A_0 =  520$ GeV, $m_0 = 370$ GeV,
and $\tan \beta = 50$.  
As one
can see, only when $\ep_{Sd}\gsim 0.35$, does the splitting in the
masses of the heaviest families (1st two generations) become relevant. 
The third generation (mostly sbottom) masses are relatively insensitive to 
$\ep_{Sd}$ and are close to their CMSSM (with no flavour violation) values
$(M_{\tilde d_1},M_{\tilde d_2})$ $=(956,1017)$ GeV as shown by the lower two (red)
curves in Fig. \ref{fig:mssm_spectra_520}a. where 1,2,...5,6 are ordered
lightest to heaviest.
Analogously for the 3rd generation sleptons we have $(M_{\tilde
  e_1},M_{\tilde e_4})$ $=(246,477)$ GeV, in the CMSSM as shown by 
  the two (red) curves in  panel b.
  Note that the heavier stau is heavier than the mostly right-handed
  selectron and smu.
  With no flavour violation, the masses of the 1st two generations
  are degenerate (due to the smallness of the associated fermion masses)
  in the CMSSM.  Here we see the degree of splitting induced when $\ep_{Sd} \ne 0$. 
  We do not show the dependence of the of the sups with respect to
$\ep_{Sd}$ or $\ep_{Su}$ because their change (with respect the
running of just the third family) is less than the $0.4\%$. 
We note, however,  a more important difference in the determination of
the running Higgs sector parameters which is displayed in panel c of Fig. \ref{fig:mssm_spectra_520}.
When $\ep_{Sd} = 0$, we have
$(\mu,m_{H_1},m_{H_2},m_{A})=(568,-403,-708,516)$ GeV
at the electroweak scale, 
where $m_{H_i}$ are the two soft Higgs masses, and the minus
refers to the sign of the squared mass.  For $\ep_{Sd}=0.45$ these become $(\mu,m_{H_1},m_{H_2},m_{A})=(545,-465,-688,452)$ GeV.
Since the most important supersymmetric contribution to BR$(B_s\rightarrow \mu^+ \mu^-)$ scales as  $f^2_{B_s}/m^4_A$ and in the analysis that we have done the inclusion of the second family tends to lower $m_A$, we expect that in general BR$(B_s\rightarrow \mu^+ \mu^-)$ increases from the case $\ep_{Sd}=0$ to $\ep_{Sd}=0.45$, but remains within the experimental limit.

\section{Summary}

Flavour symmetries (FS) are often constructed using supersymmetry.
Departures from MFV will affect the supersymmetric spectra and
constraints at the electroweak scale can be employed to 
determine the profile of the Yukawa matrices at the GUT
scale. Flavour violating processes then can be used to further
constrain the shape of such Yukawa matrices and the parameters associated with
the FS.

Clearly,  a completely general analysis is not practical for
computational purposes, as it involves many new supersymmetric parameters in
addition to the Yukawa matrices. To
constrain the parameters of the latter we can rely on the
determination of the Yukawa matrices at $M_{\rm{EW}}$ from flavour
violating parameters in the SM,  assuming a hierarchical symmetric
form, \eq{eq:yuksymm} (although in this context a non-symmetric form
is equally plausible, it requires more parameters) 
such that the expansion parameter in the up
sector is smaller than in the down sector, motivated by the different
hierarchies between up and down quarks respectively, $\ep_u\sim
\ep_d^2$. For the latter case we can simplify the usual relations
obtained in FS between the expansion parameters of the Yukawa matrices
and the off-diagonal parameters of the trilinear couplings and soft squared mass matrices at 
the GUT scale. 

In our analysis, we used an iterative procedure as is common in the analysis of the
CMSSM  \cite{cmssm,cmssmwmap}, with the inclusion of
three more parameters describing the Yukawa matrices ($\ep_u$,
$\ep_d$ and $c^d_{22}$), at the GUT scale. Other necessary inputs are the values of
$m_s(2{\rm{GeV}})$, $m_c(2{\rm{GeV}})$, $m_\mu(M_{\rm{EW}})$, and the
values of the CKM matrix relevant to the two heaviest families. These are used to
determine seven more parameters in the Yukawa sector at the GUT
scale, as in \eq{eq:yuk_o_init}.

The important results of this analysis in the fermion sector are as
follows. First, using the supersymmetric corrections
to the fermion masses of the two heaviest families,
at the EW scale,  we can accurately
determine the profile of the Yukawa matrices at the GUT scale. 
Particularly important is the ratio 
$(Y_d)_{23}(M_{\rm{GUT}})/(Y_d)_{22}(M_{\rm{GUT}})$.
Whereas in an analysis where the running of supersymmetric parameters
is neglected, this ratio is found to be 1.3, when supersymmetric 
corrections are taken into account, this ratio is enhanced to 1.8, cf.,  \eq{eq:ydk22_o_yd23}.
Second, the value of the expansion parameter, $\ep_d$,
is effectively constrained by the appropriate determination of
$m_s(M_{\rm{EW}})$, with its corresponding supersymmetric corrections. Of
course the exact value depends on the values for the coefficients used in the
expression of the Yukawa matrices, for example
$(Y_d)_{22}=c^d_{22}\ep^2_d$. Assuming that the
coefficients $c^d_{ij}$ should be of order one, we find for example that for
$c^d_{22}=0.7$,  $\ep_d\in[0.085,0.10]$, otherwise $m_s$ is
predicted to be either too small or too big, respectively.
This determination is important when constructing FS.  The
exact values of $(Y_f)_{ij}(M_{\rm{GUT}})$ are used to justify different choices of
family groups  (e.g. $SU(3)$, $SO(3)$). We have only considered the running
of the two heaviest families and hence the structure of the
lightest family cannot be determined at the GUT scale by this analysis. However, the
inclusion of the lightest family will not have a strong impact on the
results obtained here.

Our results are also relevant  for the supersymmetric sector.
When off-diagonal terms are introduced in the sfermion mass matrices
or trilinear terms at the GUT scale, departures from MFV are induced at the EW
scale. In the most constrained models, ie., when no new parameters are introduced
as in  \eq{eq:sofmasses_offneq0}, this departure is small and 
the resulting sensitivity of $\BR(\bsgmm)$
is not significant. In contrast, when the expansion parameters, $\ep_{Sf}$ , 
for the supersymmetric sector differ from that of the fermion sector,  $\ep_{f}$,
we may obtain more sizable departures from MFV. 
We have compared the running of
the off-diagonal trilinear couplings and soft squared mass matrices
for these two possibilities.

Our calculations were performed using the values of the
supersymmetric parameters as obtained when running the full MSSM, with
the simplified structure for the trilinear couplings and
soft squared masses at two loops as described above.
Hence we do not rely on
estimating the value of the flavour violating parameters
$(\delta_{f_{XY}})$, \eq{eq:del_f_XY}, at  the EW scale.
Indeed, these are calculated directly in terms of the parameters
introduced at the GUT scale as an extension of the CMSSM.
These parameters were introduced in  Section \ref{sec:runn_off_soft_p} to emphasize their running
from $M_{\rm{GUT}}$ to $M_{\rm{EW}}$.
Current limits on the charged slepton sector
could be used to further constrain the choice of the expansion
parameters $\ep_{f}$ and $\ep_{Sf}$.

We have seen that the running of $(\delta_{f_{XY}})_{23}$ is quite limited
when initial conditions at the GUT scale are constrained. This is true for both the case of
MFV at the GUT scale as well as the case when the expansion parameters of
the supersymmetric sector differ from those of the fermion
sector, $\ep_{Sf} \ne \ep_{f}$, as we saw from Table
\ref{dLs_epSf}. In column 2 of that table, we have assumed MFV at the GUT scale
and the departures from MFV at the weak scale are small as seen by the small values
of the $\delta$-parameters. 
In this case, the bounds on the parameters  $(\delta_{{(e,L)}_{XY}})_{23}$
are well below the bounds imposed by $\BR(\tau \rightarrow \mu\gamma)$ in 
\eq{eq:msugra_dl_bds}.
However, if we take for example $\ep_{Sd}=\ep_{Se}=0.45$, we can see that
the values obtained for $(\delta_{{(e,L)}_{XY}})_{23}$ in column 3 
already saturate the bounds imposed by
$\BR(\tau \rightarrow \mu\gamma)$. 
In the last column, we showed the $\delta$'s when
$\ep_{Sd}=0.45$ and $\ep_{Se}=0.225$ where the bounds imposed by $\BR(\tau \rightarrow \mu\gamma)$ are satisfied.

We have linked the down-type quarks to the charged leptons by 
generating the off-diagonal terms of the Yukawa matrices of both 
with $V_{\rm{CKM}}$ at EW scale as in \eq{eq:rel_yd_ye}. This is
motivated by GUT theories where often charged leptons and down-type
quarks acquire Yukawa matrices of the same form. Extending the
relation to the supersymmetric sector, by taking the expansion parameters 
$\ep_{Sd}$ and $\ep_{Se}$ to be of the same order, we naturally find that
the parameters $|(\delta_{{(Q,d)} XY})_{23}|$ and  $|(\delta_{{(L,e)} XY})_{23}|$
are correlated at EW scale. From Table \ref{dLs_epSf}, we see that
for $\ep_{Sd}=\ep_{Se}$ we have in fact  $|(\delta_{{(Q,d)} XX})_{23}| =O\left(
|(\delta_{{(L,e)} XX})_{23}|\right)$. Applying the bounds on
$|(\delta_{{(L,e)} XX})_{23}|$ imposed by $\BR(\tau \rightarrow \mu\gamma)$,
we have taken $\ep_{Se}=0.5\ep_{Sd}=0.225$ and this effectively decreases
 $|(\delta_{{(L,e)} XX})_{23}|$ by an order of magnitude with respect to
 $|(\delta_{{(Q,d)} XX})_{23}|$.

Finally, we have calculated $\BR(\bsgmm)$ at NLO for the charged
Higgs, chargino and neutralino contributions and at LO for the gluino
contributions (since a NLO computation is not yet fully
available for this case).  BLO corrections for the gluino contributions were
considered in \cite{Foster:2005wb}, but given the values of $(\delta_{f_{,XY}})_{23}$
that we have derived, these corrections will have little impact 
on the value $\BR(\bsgmm)$ that we obtain. 

Given the SM NNLO computation and the continuously improved
experimental value, it is important to determine $\BR(\bsgmm)$  as precise as possible
in a given supersymmetric model. We have seen that the
structures for Yukawa matrices, trilinear terms and soft-matrices of
the form of Eqs. (\ref{eq:sofmasses_offneq0}-\ref{eq:sofmasses_offneq0_ef_diff_efS})
yield a positive contribution to the calculation of $\BR(\bsgmm)$  for $\mu>0$ bringing the branching
ratio closer to its experimental value than what would have been obtained in the
absence of any FV. 
In fact when $\ep_{Sf}\neq\ep_{f} $, such that $\ep_{Sd}> \ep_{d}>\ep_{Su} > \ep_{u}$,
it is possible to relax the lower limit of $\sim 400$ GeV on $m_{1/2}$ found 
in the single family case most commonly treated in the CMSSM. For $\ep_{Se} = \ep_{Sd} = 0.45$
this limit is relaxed to  $m_{1/2} \gsim 335$ GeV, as we have
discussed in Section \ref{sec:const_efS}.

%
%%%%%%%%%%%%%%%%%%%%%%%%%%%%%%%%%%%%%%%%%%%%%%%%%%%%%%%%%%%%%%%%%%%%%%%
%
\appendix
%
%%%%%%%%%%%%%%%%%%%%%%%%%%%%%%%%%%%%%%%%%%%%%%%%%%%%%%%%%%%%%%%%%%%%%%%
%

\section{${\mathbf{b}}\rightarrow {\mathbf{s \gamma}}$ beyond minimal flavour violation  \label{sec:bsg_bmfv}}
  At LO the renormalization evolution of the gluino contribution 
to the effective Hamiltonian, ${\mathcal{H}}_{\rm eff}$, from
$M_{\rm{EW}}$  down to $\mu_b$ is different than the
contribution from $W^-$, $H^-$, charginos and neutralinos
\cite{Borzumati:1999qt}. Hence,  ${\mathcal{H}}_{\rm eff}$ can be
  written as:
\bea
{\mathcal{H}}_{\rm eff}={\mathcal{H}}^{W^-, \ H^-,\ \chi}_{\rm eff} +
{\mathcal{H}}^{\tilde g}_{\rm eff}.
\eea
``Mixed'' diagrams containing the W boson, gluinos and squarks give
rise to $\alpha_s$ corrections to the Wilson coefficients in ${\mathcal{H}}^{W^-, \ H^-,\ \chi}$
at the matching scale, and they are taken into account at the NLO
level. We  include the contributions of  gluinos at LO and those from $H^-$, 
charginos and neutralinos at NLO.

We recall that the starting point in the calculation of inclusive $B$ decay rates is
the low-energy effective Hamiltonian \cite{Buchalla:1995vs}
\begin{equation}
  {\mathcal{H}}^{W^-, \ H^-,\ \chi} _{\rm eff} = -\frac{4 G_F}{\sqrt2}\,V_{ts}^* V_{tb}
   \sum_i C_i(\mu_b) O_i(\mu_b) \,.
\label{Heff}
\end{equation}
In the massless strange quark limit, the operators relevant to our discussion are
\begin{eqnarray}
   O_2 &=& \bar s_L\gamma_\mu c_L\bar c_L\gamma^\mu b_L \,, \nonumber\\
   O_7 &=& \frac{e\,m_b}{16\pi^2}\,\bar s_L\sigma_{\mu\nu}
    F^{\mu\nu} b_R \,, \nonumber\\
   O_8 &=& \frac{g_s m_b}{16\pi^2}\,\bar s_L\sigma_{\mu\nu}
    G_a^{\mu\nu} t_a b_R \,.\label{eq:basic_op}
\end{eqnarray}
To an excellent approximation, the contributions of other operators can
be neglected. The renormalization scale $\mu_b$ in (\ref{Heff}) is
conveniently chosen of order $m_b$, so that all large logarithms reside
in the Wilson coefficient functions. However when the mass of the strange quark is taken into account we need to consider the operators
\begin{eqnarray}
    O'_7 &=& \frac{e\,m_b}{16\pi^2}\,\bar s_R\sigma_{\mu\nu}
    F^{\mu\nu} b_L \,, \nonumber\\
    O'_8 &=& \frac{g_s m_b}{16\pi^2}\,\bar s_R\sigma_{\mu\nu}
    G_a^{\mu\nu} t_a b_L \,.\label{eq:sr_op}
\end{eqnarray}
For these operators the corresponding Wilson coefficients
$\tilde C^{H^-}\propto m_s m_b/m^2_t\tan^2\beta$ and $\tilde
C^{\chi^\pm}$ are quite small. 
For ${\mathcal{H}}^{\tilde g}_{\rm eff}$ we have
\bea
{\mathcal{H}}^{\tilde g}_{\rm eff}=\sum_i C_{i,\tilde g}(\mu_b)
O_{i\tilde g}(\mu_b) + \sum_i C'_{i,\tilde g}(\mu_b)
O'_{i\tilde g}(\mu_b) + \sum_i\sum_q C^q_{i,\tilde g}(\mu_b)
O^q_{i\tilde g}(\mu_b),
\label{Heff_gln}
\eea
where $q$ runs over the light quarks $q=u, d, c, s, b$.  The operators  $O_{i,\tilde g}(\mu_b)$ can be of two types, one for $O_{j\ b, \tilde g}(\mu_b)$ and the other for  $O_{j\  \tilde g, \tilde g}(\mu_b)$,  $j=7,8$. The former can be found from Eq. (\ref{eq:basic_op}) with the substitution of $(1/16\pi^2)$ with  $g_s^2$ and the latter with the substitution of $(m_b/16\pi^2)$ with $g_s^2$.
The primed operators in \eq{Heff_gln} come from the interchange of $L\leftrightarrow R $. The operators $O^q_{i\tilde g}(\mu_b)$ are suppressed with respect to the magnetic operators by the mass of the corresponding quark or by an extra power of $g_s$. We neglect them here since their contribution is quite tiny with respect of the operators in the first two terms of \eq{Heff_gln}.

In the SM, the  complete theoretical prediction for the $B\to X_s\gamma$ decay rate at NLO  was presented for the first time by Chetyrkin et al.\ \cite{Chetyrkin:1996vx}. It depends on a parameter $\delta$ defined by the condition that the photon energy be above a threshold given by $E_\gamma>(1-\delta) E_\gamma^{\rm max}$, where $E_\gamma^{\rm max}=m_b/2$ is the maximum photon energy attainable in the parton model. Thus, $\delta=1-2E_\gamma^{\rm max}/m_b$, and we use $E_\gamma^{\rm max}= 1.6$ GeV, as it is the value at which the HFAG collaboration has presented its latest results  \cite{bsgexp,hfag}.  

The prediction for the $B\to X_s\gamma$ branching ratio is usually obtained by normalizing the result for the corresponding decay rate to that for the semileptonic decay rate, thereby eliminating the strong dependence on the $b$-quark mass:
\bea
   {\rm BR}(B\to X_s\gamma)\big|\!\!\!\!\!\!&&\!\!\!\!\! _{E_\gamma>(1-\delta)E_\gamma^{\rm max}}
   ={\rm BR}(B\rightarrow X_c e \bar\nu)_{\rm{exp}} \frac{\Gamma(B\rightarrow X_s \gamma)|_{E_\gamma>(1-\delta)E_\gamma^{\rm max}}}{\Gamma(B\rightarrow X_c e \bar\nu) }.
\eea
Following closely the formalism of  \cite{Gambino:2001ew,Hurth:2003dk}, we write
\bea
  {\rm BR}(B\to X_s\gamma)\big|\!\!\!\!\!\!&&\!\!\!\!\! _{E_\gamma>E_0}
   ={\rm BR}(B\rightarrow X_c e \bar\nu)_{\rm{exp}} \!\!\left|\frac{V^*_{ts}V_{tb}}{V_{cb}}\right|^2\! \frac{6\alpha_{\rm{em}}}{\pi}\! \left[P(E_0)+N(E_0)\right]\! \frac{1}{r(\Gamma_u/\Gamma_c)}
\eea
where perturbative, $P(E_0)$, and non-perturbative, $N(E_0)$, contributions are treated separately,
and $E_0=(1-\delta)E_\gamma^{\rm max}$. 
The function $N(E_0)$ is a correction which does not depend on the renormalization of the running mass $m_b$ from $\mu_{\rm{EW}}$ to $m_b$,  it contains  power corrections that arise from the interference between the current-current operator  and the magnetic dipole operator, hence called the ``non-perturbative'' contribution. The precise definition is given in Eq. 3.10, of ref. \cite{Gambino:2001ew}.
In the SM this is \cite{Gambino:2001ew}, $N(E_0=1.6\ {\mathrm{GeV}})= 0.0036 \pm 0.0006$. The ratio $r(\Gamma_u/\Gamma_c)$ is a semileptonic phase space factor which is independent of the calculational problem of convergence of the perturbation series in $b\rightarrow X_s \gamma$. This has been calculated accurately \cite{Gambino:2001ew} up to NNLO in the SM.  
 
Since $\BR(B\rightarrow X_c e \bar\nu)$ is an experimental input, only the function $P(E_0)$ has contributions from processes beyond the SM.
Using the approach of Gambino  et. al. \cite{Gambino:2001ew}, rather than writing the terms $P(E_0)+N(E_0)$ as functions of the standard Wilson coefficients, \eq{Heff}, at the decay scale, $\mu_b$, these functions can be written in terms of functions involving the renormalization effects from $\mu_0$ ($m_t$ or $\mu_{\rm{EW}}$ ) to $\mu_b$ and the Wilson coefficients at $\mu_0$. The contributions from the
charm quark do not get re-normalized from $M_{\rm{EW}}$ to $m_b$, while the contributions from $t$ do. 
Hence instead of the familiar contribution of $|C^{\rm{eff}}_7(\mu_{\rm{EW}})|^2$ to ${\rm BR}(B \rightarrow X_s \gamma)$, we have 
\bea
|C^{\rm{eff}}_7(\mu_b)|^2 &\rightarrow& \hat K^2_{7\rm{eff}} \equiv P(E_0) + N(E_0)\nn\\
P(E_0) &=&\left| X_c+ X_t  + \epsilon_{\rm{EW}} \right|^2 +B(E_0),
\label{eq:pertb_conts_bsg}
\eea
where $X_c$ and $X_t$ represent the $c$ and $t$ contributions respectively, $\epsilon_{\rm{EW}}$ is an electroweak correction and $B(E_0)$ is the bremsstrahlung correction  from $b\rightarrow s\gamma g$ and $b\rightarrow s \gamma q\bar q$ with $q=u,\ d,\ s$ \cite{Hurth:2003dk}. 

Analogous to the usual Wilson coefficients, we can decompose
\bea
X_f(\mu_b)=X^{(0)}_f(\mu_b) + \frac{\alpha_s(\mu_b)}{4\pi} X^{(1)}_f(\mu_b),\ f=c,t.
\eea
The effects of physics beyond the SM can be incorporated \cite{Hurth:2003dk} as:
\bea
C^{(0)\rm{TOT}}_i(\mu_{\rm{EW}})=C^{(0)\rm{SM}}_i(\mu_{\rm{EW}})+C^{(0)\rm{BSM}}_i(\mu_{\rm{EW}}) ,\ i=7,8,
\label{eq:totwicoeff}
\eea
where $C^{(0)\rm{SM}}_i(\mu_{\rm{EW}})=C^{(0)\rm{eff}}(\mu_{\rm{EW}})_i,\ i=7,8$ correspond to the coefficients of Eq. (12) of \cite{Chetyrkin:1996vx}, i.e. the scheme independent Wilson coefficients, the coefficients $C^{(0)\rm{BSM}}_i(\mu_{\rm{EW}})$ of course refer to processes beyond the SM, in this case the MSSM. From here on, we will refer to the following coefficients
%
%\bea
$\tilde C^{(0)}_7({\mu_{\rm{EW}}}) =   C^{(0){\rm{TOT}}}_{7}({\mu_{\rm{EW}}}) +  \frac{23}{36}~$%,\nn\\
and $\tilde C^{(0)}_8({\mu_{\rm{EW}}}) =  C^{(0){\rm{TOT}}}_{8}({\mu_{\rm{EW}}}) +  \frac{1}{3}$.
%\eea
%
For the MSSM, the general form for $C^{(0){\rm{MSSM}}}_{i}(\mu_{\rm{EW}})$ is known, although the contribution from gluinos does not follow the same pattern of renormalization from $M_{\rm{EW}}$ to $m_b$ and it should be implemented in a different fashion. The contribution from charginos, neutralinos and charged Higgs bosons can be renormalized as in the SM, although the complete NLO corrections to  $C^{{\rm{MSSM}}}_{i=7,8}(\mu_b)$ are not known.

Thus, if we have the following hierarchy of masses
\bea
\mu_{\rm{SUSY}}\sim O(m_{\tilde g}, m_{\tilde q}, m_{\tilde t_1}, m_{\tilde t_2}) \gg
{\mu_{\rm{EW}}}\sim O(M_{\rm{EW}},m_t,m_{H^{\pm}})\gg \mu_b,
\eea
we expect that the following assumptions represent a first good approximation for taking into account the contribution from charginos, neutralinos and charged Higgs bosons:
\bea
\tilde C^{(0)}_{i} &=& \tilde C^{(0)SM}_{i} + C^{(0)\chi, H^{\pm}}_{i} ,\ i=7,8.
%\tilde C^{(0)}_{8} &=& \tilde C^{(0)SM}_{7} + C^{(0)\chi, H^{\pm}}_{8}
\eea
We note that $C^{(0)\tilde g}_7 (\mu_b)$ must be added on directly to
$X^{(0)}_t(\mu_b)$ because gluinos undergo a different renormalization
from $M_{\rm{EW}}$ to $\mu_b$. The expressions that we use for $C^{(0)\chi,
  H^{\pm}}_{7,8}$ and $C^{(0)}_{\tilde g}$ are given in  
  Eqs. A16-A20 of \cite{Besmer:2001cj}. At NLO for $\tilde C^{X H^{\pm}}_{7,8}$,
we follow the approach of \cite{Degrassi:2000qf}.

Thus at LO,  \eq{eq:pertb_conts_bsg} would receive the
  following contributions from gluinos at $\mu_b$:
\bea
P(E_0)_{\tilde g} &=&\left|X^{(0)}_{\tilde g}(\mu_b) +  X_c(\mu_b)+ X_t(\mu_b)
  + \epsilon_{\rm{EW}} \right|^2 +  \left|X^{(0)'}_{\tilde g}(\mu_b) \right|^2
+  B(E_0),
\label{eq:pertb_conts_bsg_with_gl}
\eea
where 
\bea
\label{eq:cont_bsg_fgluinos}
X^{(0)}_{\tilde g}(\mu_b) = -\frac{16\sqrt{2}\pi^3 \alpha_s(\mu_b)}{G_F V_{tb}V^*_{ts}}\left[C^{(0)}_{7 b, \tilde g}(\mu_b) + \frac{1}{m_b} C^{(0)}_{7 \tilde g, \tilde g}(\mu_b) + \frac{m_c}{m_b}  C^{(0)}_{7 c, \tilde g}(\mu_b) \right],
\eea
and $X^{(0)'}_{\tilde g}(\mu_b)$ is the analogous contribution from
the primed operators (of flipped chirality $L\leftrightarrow R$ ) of \eq{Heff_gln}.
The bremsstrahlung function $B(E_0)$, also receives gluino
contributions \cite{{Hurth:2003dk}}, which are added to the Wilson coefficients at $\mu_b$
\bea
C_i^{(0){\rm  TOT}}(\mu_b)= C^{(0) SM}_i(\mu_b) + C^{(0)\chi, H^\pm} (\mu_b) + X^{(0)}_{\tilde g}(\mu_b), i=7,8,
\eea 
the contribution from BSM processes to the rest of the Wilson
coefficients, $i=1,..,6$, is negligible. We will denote  by $
K^2_{7\rm{eff}\tilde g}$ the contribution to $\BR(\bsgmm)$ containing the Wilson
operators from gluinos:
\bea
\hat K^2_{7\rm{eff}\tilde g} \equiv P(E_0)_{\tilde g} + N(E_0).
\label{eq:Kgluin}
\eea

As stated before we have implemented the calculation of $\BR(b\rightarrow s \gamma)$  as introduced in \cite{Gambino:2001ew}, and adopted in \cite{Hurth:2003dk} for the contribution of beyond SM effects. The main difference in the way $\BR(b\rightarrow s \gamma)$ is calculated in the SM as introduced in \cite{Gambino:2001ew}, with respect to previous analyses is that $m^{pole}_c/m^{pole}_b$ is replaced by ${m^{\bar{MS}}_c(m_c)}/{m^{pole}_b}$ in the element $\langle X_s\gamma | (\bar{s}c)_{V-A} (\bar{c}b)_{V-A}|b \rangle$. In addition, the running of the charm and top contributions from the EW scale to the decay scale $\mu_b$ is different. In Table \ref{tbl:input_par}, we compare the values obtained in our calculation for the different choices of  $m^{pole}_c/m^{pole}_b=0.22$ and ${m^{\bar{MS}}_c(m_c)}/{m^{pole}_b}=0.3003$. Lines 1 to 5 correspond to the calculation of $\BR(\bsgmm)$ implemented here following
\cite{Gambino:2001ew}.
Line 5 is the SM value that we obtain in the decoupling limit. Line 4 is what we obtain when using the top mass assumed in \cite{Gambino:2001ew}, and is in perfect agreement with the value obtained there (Eq. 4.14 of that reference). Line 3 corresponds to the SM when   $\delta=0.9$ shown for comparison to other works. Lines 1 and 2 assume  $m^{pole}_c/m^{pole}_b=0.22$ for two values of $m_t$. The remaining parameters for the calculation of $\BR(\bsgmm)$ that we use and that were not specified before %in Section 2.1 
are $\BR(B\rightarrow X_c e \bar\nu)_{\rm{exp}}$ $=0.1059$, $\left|\frac{V^*_{ts}V_{tb}}{V_{cb}}\right|^2$ $=0.97$ and $\mu_b=5$ GeV.

\begin{table}%[!h]
\begin{center}
\begin{tabular}{|l l l  |l|}
\hline
\multicolumn{4}{c}{$\BR(\bsgmm)$ for the SM}\\
\hline
 $m_t$ & $E_{\gamma}$ &  $\frac{m_c}{m_b}$ & ${\BR(b\rightarrow s \gamma)}{[ 10^{-4}]}$\\
\hline
%$\hat K_{7\rm{eff}}(\mu_b)=P(E_0)+B(E_0)$: &       &        &        \\
\hline
 $174.3$ GeV & $1.6$ GeV    &  $0.3003$& $3.32\pm 0.19$ \\
 $171.4$ GeV & $1.6$ GeV    &  $0.3003$& $3.27\pm 0.19$ \\
\hline
$171.4$ GeV & $0.2345$ GeV &  $0.22$& $3.71\pm 0.20$ \\
$174.3$ GeV & $1.6$ GeV &  $0.22$& $3.61\pm 0.20$ \\
$171.4$ GeV & $1.6$ GeV &  $0.22$& $3.59\pm 0.20$ \\
\hline
%$|C^{\rm{eff}}_7(\mu_b)|^2$: &       &        &        \\
%\hline
% $171.4$ GeV & $1.6$ GeV & $0.22$ & $3.68$ \\
\hline
\end{tabular}
\end{center}
\caption{\small Input values for the evaluation of $\BR(b\rightarrow s
  \gamma)$. For these cases, we have used $\alpha_s(m_Z)=0.1185$. The
  value of $E_{\gamma}=0.2345$ GeV corresponds to $\delta=0.9$. The
  error quoted is just the parametric error.} \label{tbl:input_par}
\end{table}

%%%%%%%%%%%%%%%%%%%%%%%%%%%%%%%%%%%%%%%%%%%%%%%%%%%%%%%%%%%%%%%%%%%%%%%%%%%%%%%%
\section{Interactions of fermion, sfermions and sparticles {\label{app:inter}}}
%
%%%%%%%%%%%%%%%%%%%%%%%%%%%%%%%%%%%%%%%%%%%%%%%%%%%%%%%%%%%%%%%%%
%
\subsection{Quarks-squarks-gluino interactions}
The interaction between quarks-squarks and gluinos is described by the Lagrangian:
\bea
-{\mathcal{L}}_{q-\tilde{q}-\tilde{g}}=\sqrt{2}g_s T^a_{\alpha\beta}\left[-\bar{q^\prime}^{\alpha}_i{\mathcal{P}}_L \tilde{g}_a \tilde{q^\prime}^\beta_{R i}+ \bar{q}^\alpha_i{\mathcal{P}}_R \tilde{g}_a\tilde{q}^{\beta}_{L i}+h.c
\right],
\eea
in the electroweak-color basis, where ${\mathcal{P}}_{R,L}=(1\pm \gamma_5)/2$, $a$ is the gluino color index, $\alpha,\beta$ are the quark-squark color indices and $i$ is the generation index.  In the gauge basis, the couplings at the quark-squark-gluino vertex are given by, for {\it incoming} gluinos to sfermion-fermion, 
$
C_{\tilde g\tilde
  f_{(2l-1)}(f^\prime_L)_l}=-ig_3\sqrt{2}T=-i\frac{2}{\sqrt{3}}g_3$. In this basis the effective mass Lagrangian that we construct from the soft Lagrangian of \eq{eq:sftLg} is
\bea
\label{eq:eff_soft}
{\cal{L}}^{\rm{eff}}_{m_{\tilde q}} = -(\tilde q'_{L}, \tilde q'_{R})_i ({\mathcal{M}}^2_{\tilde q'})_{ij} \left(\begin{array}{c} \tilde q^{'*}_{L}\\ \tilde q^{'*}_{R} \end{array}\right)_j, 
\eea
where
\bea
\label{eq:effsfmssterms}
({\mathcal{M}}^2_{\tilde f})_{ij}&=&\left[
\begin{array}{cc}
M^2_{LL}          & M^{2\dagger}_{LR}\\
M^{2}_{LR} & M^2_{RR}
\end{array}
\right]_{ij}\nn\\
&=&
\left[\begin{array}{cc}
(M^2_{\tilde Q})_{ij}+(M^{2}_f)_{ij}+D^f_{L} & 
-({a_f}_{ij}v_f+\mu \tan^p\beta (M_f)_{ij}) \\
 -({a^{*}_f}_{ij}v_f+\mu \tan^p\beta (M^{*}_f)_{ij})&
 (M^2_{\tilde f_R})_{ij}+(M^{2}_f)_{ij}+D^f_{R}
\end{array}\right]
\nn\\
D^f_{L,R} &=& \cos 2\beta M^2_Z(T^3_f-{\mathcal{Q}}_{f_{L,R}} \sin^2\theta_W), \quad
p\ =\  \left\{\begin{array}{c} 1,\ f=d\\ -1,\ f=u. \end{array}\right.
\eea
In \eq{eq:effsfmssterms} $i, j=1,2,3$ are the family indices, and $(M_f)_{ij}$ are the non-diagonalized fermion mass matrices.  We recall that in the eigenmass formalism all the interactions are computed with the particles that are mass eigenstates. Hence the effective soft mass matrix of \eq{eq:eff_soft}, with the corresponding family elements \eq{eq:effsfmssterms}, needs to be rotated where the fermions are mass-eigenstates, i.e. to the so-called super CKM  (SCKM) basis. For the diagonalization of fermions we adopt the convention
\bea
(u_{L,R})_i = (V^{u\dagger}_{L,R})_{ij} (u'_{L,R})_j,\quad
(d_{L,R})_i = (V^{d\dagger}_{L,R})_{ij} (d'_{L,R})_j,\quad \rightarrow \quad
V_{\rm{CKM}}= V^{u\dagger}_L V^d_L,
\label{eq:fer_eig-mass}
\eea
and thus the sfermions must be rotated similarly:
\bea
(\tilde u_{L,R})_i = (V^{ u\dagger}_{L,R})_{ij} (\tilde u'_{L,R})_j,\quad
(\tilde d_{L,R})_i = (V^{ d\dagger}_{L,R})_{ij} (\tilde d'_{L,R})_j.
\label{eq:sfer_eig-mass}
\eea
Then the soft mass matrix \eq{eq:effsfmssterms} becomes
\bea
({\mathcal{M}}^{\rm{SCKM} }_{\tilde f})^2_{ij}\!\!\!&=&\!\!\!\! \left[
\begin{array}{cc}
{M^{\rm{SCKM}}}^2_{LL} & {M^{\rm{SCKM}\dagger}}^2_{LR}\\
{M^{\rm{SCKM}}}^2_{LR} & {M^{\rm{SCKM}}}^2_{RR}
\end{array}
\right]_{ij} \equiv  (\widehat{\mathcal{M}}^{2}_{\tilde f})_{ij} \nn\\
&=& \!\!\!\! \left[\begin{array}{cc}
(V^{fT}_{L}M^2_{\tilde Q}V^{f*}_L)_{ij}+\hat M^2_{f_i}+D^f_{L} & 
-((V^{fT}_L a_{f} V^{f*}_R)_{ij} v_f+\mu^* \tan^p\beta \hat M_{f_i}) \\
 -( (V^{f\dagger}_L (a^{*}_f)_{ij} V^{f}_{R})v_f+\mu \tan^p\beta \hat M_{f_i})&
 (V^{fT}_{R} M^2_{\tilde f_R} V^{f*}_R)_{ij}+ \hat M^{2}_{f_i}+D^f_{R}
\label{eq:eff_rotatedmassm}
\end{array}\right]\! \!,\nn\\
\eea
where $D^f_{L,R}$ remain diagonal, and $\hat M_f$  is the diagonal matrix of the $f$ type fermions. The sources of flavour violation then come from 
\bea
\left({M^{\rm{SCKM}}}^2_{\tilde Q}\right)_{ij}    & = & (V^{fT}_{L}M^2_{\tilde Q}V^{f*}_L)_{ij}\nn\\
\left(a^{\rm{SCKM}}_{f}\right)_{ij}               & = &  -(V^{fT}_L a_{f} V^{f*}_R)_{ij}\nn\\
\left({M^{\rm{SCKM}}}^2_{\tilde f_R}\right)_{ij}  & = &  (V^{fT}_{R} M^2_{\tilde f_R} V^{f*}_R)_{ij},
\eea
which are not generically diagonal in the SCKM basis.  For the case of 1 family the matrix $({\mathcal{M}}^{\rm{SCKM}}_{\tilde f})^2_{ij}$ is the usual  $2 \times 2 $ soft mass matrix whose diagonalization is obtained through
\bea
\left[\tilde f_L, \tilde f_R \right]
\left[
\begin{array}{cc}
{M^{\rm{SCKM}}}^2_{LL} & {M^{\rm{SCKM} \dagger}}^2_{LR}\\
{M^{\rm{SCKM}}}^2_{LR} & {M^{\rm{SCKM}}}^2_{RR}
\end{array}
\right]
\left[
\begin{array}{c}
\tilde{f}^*_L\\
\tilde{f}^*_R
\end{array}
\right]=
\left[\tilde f_L, \tilde f_R \right]K^{\dagger}
\left[
\begin{array}{cc}
M^2_{\tilde f_1} &    \\
      & M^2_{\tilde f_2}     \\
\end{array}
\right]K
\left[
\begin{array}{c}
\tilde{f}^*_L\\
\tilde{f}^*_R
\end{array}
\right],
\label{eq:diagM2by2}
\eea
where then the sfermion mass eigenstates, $\tilde{q}_i$ are defined by 
\bea
\left[
\begin{array}{c}
\tilde{f}_L\\
\tilde{f}_R
\end{array}
\right]=K^T
\left[
\begin{array}{c}
\tilde{f}_1\\
\tilde{f}_2
\end{array}
\right].
\label{eq:sf_gaug-mass}
\eea
Analogously for three families, we have
\bea
\label{eq:softmass6by6}
\widehat {\mathcal{M}}^{2\dagger}_{\tilde f}=\left[
\begin{array}{ccc}
(\widehat {\mathcal{M}}^2_{\tilde f})_{11}           & (\widehat {\mathcal{M}}^2_{\tilde f})_{12} &
(\widehat {\mathcal{M}}^2_{\tilde f})_{13} \\
(\widehat {\mathcal{M}}^2_{\tilde f})^{\dagger}_{12} & (\widehat {\mathcal{M}}^2_{\tilde f})_{22} &
(\widehat {\mathcal{M}}^2_{\tilde f})_{23} \\
(\widehat {\mathcal{M}}^2_{\tilde f})^{\dagger}_{13} & (\widehat {\mathcal{M}}^2_{\tilde f})^{\dagger}_{23}& (\widehat {\mathcal{M}}^2_{\tilde f})_{33}\\
\end{array}
\right], \quad
\tilde u^\prime=\left[
\begin{array}{c}
\tilde u_L\\
\tilde u_R\\
\tilde c_L\\
\tilde c_R\\
\tilde t_L\\
\tilde t_R
\end{array}
\right],\quad
\tilde d^\prime=\left[
\begin{array}{c}
\tilde d_L\\
\tilde d_R\\
\tilde s_L\\
\tilde s_R\\
\tilde b_L\\
\tilde b_R
\end{array}
\right].
\eea
Hence the effective soft mass matrix $(\widehat {\mathcal{M}}^2_{\tilde f})_{ij}$ is a $6 \times 6$ matrix, and the electroweak eigenstates and mass eigenstates are related through
\bea
\tilde f'=(K^f)^T\tilde f,
\label{eq:f_gaug-mass}
\eea
$f'$ and $f$ here are column vectors. The primed states are the eigenstates of the Lagrangian in the gauge basis and the unprimed in the mass eigenbasis. 
Using the above definition of \eq{eq:sf_gaug-mass} and \eq{eq:f_gaug-mass}  the couplings of the mass eigenstates involved in gaugino/ sfermion vertex in terms of the gauge eigenstates are given as follows
\bea
C_{\tilde g \tilde f^*_j (f_L)_k}  =   K^{f*}_{j(2k-1)} C_{\tilde g \tilde f^{'*}_{(2k-1)} (f^{'}_L)_k},\quad
C_{\tilde g \tilde f^*_j (f_R)_k}  =   K^{f*}_{j(2k)} C_{\tilde g \tilde f^{'*}_{(2k)} (f^{'}_R)_k}.
\eea
We are working in the basis where  $Y_u$ is diagonal at the GUT scale.
At the EW scale  two loop corrections will make it non-diagonal but due to
the smallness of the expansion parameter $\ep_u$, we are neglecting
this contribution. Then $V_{\rm{CKM}}=V^d_L$, contrary to the convention used in 
many other papers (e.g. \cite{Borzumati:1999qt,Foster:2005wb}) where
$Y_d$ is taken to be diagonal at the EW scale. Hence the couplings of fermions and sfermions to other supersymmetric
particles must be defined accordingly. These couplings are given in the next two appendices.

%%%%%%%%%%%%%%%%%%%%%%%%%%%%%%%%%%%%%%%%%%%%%%%%%%%%%%%%%%%%%%%%%
%
\subsection{Quarks-squarks-neutralino interactions}
%
%%%%%%%%%%%%%%%%%%%%%%%%%%%%%%%%%%%%%%%%%%%%%%%%%%%%%%%%%%%%%%%%%
%
The interaction between quark-squark and neutralinos it is described
by the Lagrangian, in the mass eigenstate basis, by: 
\bea
-{\mathcal{L}}_{q-\tilde{q}-\tilde{\chi}^0} &=& \tilde{d}_j \bar{d}_k\left[C_{\tilde \chi_i^o \tilde d_j (d^\dagger_L)_k}{\mathcal{P}}_L + C_{\tilde \chi_i^o \tilde d_j (d^\dagger_R)_k} {\mathcal{P}}_R \right]\chi^o_i\nn\\
&+& \tilde{d}^\dagger_k \overline{\chi^o}_i\left[C^\dagger_{\tilde
    \chi_i^o \tilde d_j (d^\dagger_L)_k} {\mathcal{P}}_R +
  C^\dagger_{\tilde \chi_i^o \tilde d_j
    (d^\dagger_R)_k}{\mathcal{P}}_L \right]d_j,\quad  \chi^o_i=\left[\tilde\chi^o_i, \ \overline{\tilde\chi^o_i}\right].
\label{eq:lagr_int_neut}
\eea
The mass Lagrangian is given by
${\mathcal{L}}=-\frac{1}{2}\tilde\psi^0 N N^\dagger
{\mathcal{M}}_{\tilde\psi^0} N^* N^T (\tilde \psi^0)^T \ + \
{\mathrm{h.c.}}$, 
where the gauge eigenstates are $\tilde\psi^o = [-i\tilde b, -i\tilde w,\ \tilde h_d, \ \tilde h_u],$
for $\tilde h_{1,2}=\tilde h_{d,u}$. The mass matrix of neutralinos is
diagonalized by  $M_{\chi^0}=N^\dagger  {\cal M}_{\tilde\psi^0} N^*$,
and hence the mass eigenstates, $\chi^0_i$, are given by $\tilde \chi^0_i=(\tilde\psi^0 N)_{i}$.

The couplings of {\it incoming} neutralinos to sfermion-fermion in the
gauge basis, \eq{eq:lagr_int_neut}, can be identified from those of
Eq. D20 in \cite{Pierce:1996zz} as $C_{\tilde\psi^o_i \tilde
  f'_{(p)} (f'_{R})_m}=a_{\tilde\psi^o_i \tilde f'_{(p)}(f'_{R})_m}$,
and  $C_{\tilde\psi^o_i \tilde
  f'_{(p)} (f'_{L})_m}=b_{\tilde\psi^o_i \tilde f'_{(p)}(f'_{L})_m}$,
for $p=L,R$. Then the coupling at each vertex in the basis of mass
eigenstates can be written as $-i(C_{u_R}{\mathcal{P}}_L+C_{u_L}{\mathcal{P}}_R)$ and
$-i{\mathcal{C}}^{-1}(C_{d_R}{\mathcal{P}}_L+C_{d_L}{\mathcal{P}}_R)$,
for up and down quarks respectively. $C_{f_R}$ is any coupling involving a right-handed fermion state, for example $C_{\tilde\chi^o_i \tilde f_{j} (f_{R})_m}$. Then the couplings of the mass eigenstates involved in neutralino or chargino/ sfermion loops in terms of the gauge eigenstates are given as 
\bea
{\mathbf{C}}^u_{ijk} = C_{\tilde\chi^o_i \tilde u^\dagger_{j} (u_{R})_k} &=& \left[N^*_{1i}K^u_{j(2k)}\left(\frac{g_1}{\sqrt{2}}\right){\mathcal{Y}}_{u_R} +N^*_{4i}K^u_{j(2k-1)}\hat{Y}^u_{kk}\right]\nn\\
{\mathbf{D}}^u_{ijk} = C_{\tilde\chi^{o*}_i \tilde u^\dagger_{j} (u_{L})_k} &=& \left[N_{1i}K^{u}_{j(2k-1)}\left(\frac{g_1}{\sqrt{2}}\right){\mathcal{Y}}_{u_L} + N_{2i}K^{u}_{j(2k-1)} (\sqrt{2}g I^{u_L}_3)  + \  N_{4i}K^{u}_{j(2k)}\hat{Y}^u_{kk}\right]\nn\\
{\mathbf{C}}^d_{ijk}=C_{\tilde{\chi}^o_i \tilde{d}_j (d^\dagger_R)_k  }&=& \left[N^*_{1i}K^d_{j(2k)}\frac{g_1}{\sqrt{2}}{\mathcal{Y}}_{d_R} + N^*_{3i} K^d_{j(2k-1)} \hat{Y}^d_{kk} \right]\nn\\
{\mathbf{D}}^d_{ijk}=C_{\tilde{\chi}^{o*}_i \tilde{d}_j
  (d^\dagger_L)_k}&=&\left[N_{1i}K^d_{j(2k-1)} \frac{g_1}{\sqrt{2}}
  {\mathcal{Y}}_{d_L} +N_{2i} K^d_{j(2k-1)} \sqrt{2} g I^{d_L}_3 +
  N_{3i} \hat{Y}^d_{kk} K^d_{j 2k} \right]
\label{eq:neutcharg_cps}
\eea

%%%%%%%%%%%%%%%%%%%%%%%%%%%%%%%%%%%%%%%%%%%%%%%%%%%%%%%%%%%%%%%%%
%
\subsection{Quarks-squarks-chargino interactions}
%
%%%%%%%%%%%%%%%%%%%%%%%%%%%%%%%%%%%%%%%%%%%%%%%%%%%%%%%%%%%%%%%%%
%
The quarks-squarks-chargino interactions are given by
\bea
-{\mathcal{L}}_{q-\tilde{q}-\tilde{\psi}^{\pm}} &=&
\tilde u_j \bar{d}_k\left[C_{\tilde{\chi}^{- c}_i \tilde{u}_j (d^\dagger_L)_k} {\mathcal{P}}_L  +  C_{\tilde{\chi}^{+ c}_i \tilde{u}_j (d^\dagger_R)_k} {\mathcal{P}}_R \right]\chi^c_i\nn\\
&+& \tilde u^{\dagger}_k \overline{\chi^c}_i \left[C^\dagger_{\tilde{\chi}^{- c}_i \tilde{u}_j (d^\dagger_L)_k} {\mathcal{P}}_R  +  C^\dagger_{\tilde{\chi}^{+ c}_i \tilde{u}_j (d^\dagger_R)_k} {\mathcal{P}}_L \right] d_i\nn\\
&+& \tilde d_j \bar{u}_k\left[C_{\tilde{\chi}^{- c}_i \tilde{d}_j (u^\dagger_L)_k}  {\mathcal{P}}_L + C_{\tilde{\chi}^{+ c}_i \tilde{d}_j (u^\dagger_R)_k}  {\mathcal{P}}_R\right]\chi^c_i \nn\\
&+& \tilde{d}^\dagger_k \overline{\chi^c}_i
\left[C^\dagger_{\tilde{\chi}^{- c}_i \tilde{d}_j (u^\dagger_L)_k}
  {\mathcal{P}}_R + C^\dagger_{\tilde{\chi}^{+ c}_i \tilde{d}_j
    (u^\dagger_R)_k}  {\mathcal{P}}_L\right] u_j,\quad \chi_i=\left[\chi^+_i,\ \overline{\chi^-_i} \right].
\label{eq:lagr_ch_int}
 \eea
%
%%%%%%%%%%%%%%%%%
%
The mass Lagrangian is given by 
${\mathcal{L}}  = -\frac{1}{2}\left[ \tilde \psi^+ V V^+ {\cal{M}}_{\tilde
    \psi^+} U^* U^T (\tilde \psi^-)^T +  \tilde \psi^-
  {\cal{M}}^T_{\tilde \psi^+}(\tilde \psi^+)^T  \right]$, where
$\tilde\psi^+  = [-iw^+,\ \tilde h^+_u]$, $\tilde\psi^-   =
[-iw^-,\ \tilde h^-_d], \ \tilde h_{1,2}=\tilde h_{d,u}$. 
The mass matrix is diagonalized by $M_{\tilde \chi^{\pm}} = V^+  {\cal M}_{\tilde\psi^+} U^*=
U^+  {\cal M}^T_{\tilde\psi^+} V^*$, hence the rotation to chargino mass eigenstates is
$\tilde \chi^{+}=(\tilde\psi^+_1,\ \tilde\psi^+_2) V$, $\tilde \chi^-=(\tilde\psi^-_1 ,\ \tilde\psi^-_2) U$.

The couplings of {\it incoming} charginos to sfermion-fermion in the
gauge basis can also be identified to those of Eq. 20 of
\cite{Pierce:1996zz}. Using the transformations described in this
section to go to the eigenmass basis we have
\bea
\label{eq:coups_for_mag_wilcoeff}
{\mathbf{A}}^u_{ijk} = C_{\tilde\chi^+_i \tilde d^*_{j} (u_{L})_k} &=& V^*_{2i} \sum^3_{n=1}\left[K^{d*}_{j(2n-1)} (V^\dagger_{\rm{CKM}})_{nk}(-\hat{Y}^{u}_{kk}) \right]\nn\\
{\mathbf{B}}^u_{ijk} = C_{\tilde\chi^{- c}_i \tilde d^*_{j} (u_{R})_k} &=& \sum^3_{n=1}\left[U_{1i}K^{d*}_{j(2n-1)}  (V^\dagger_{\rm{CKM}})_{nk} g\ + U_{2i} K^{d*}_{j(2n)}  (-\hat{Y}^d_{nn}) (V^\dagger_{\rm{CKM}})_{nk}\right]\nn\\
{\mathbf{B}}^d_{ijk}=   C_{\tilde{\chi}^{+c}_i \tilde{u}_j (d^\dagger_R)_k }&=&\sum^3_{n=1} \left[V_{2i} K^u_{j (2n)} (V^{\dagger}_{\rm{CKM}})_{kn}(-\hat{Y}^{u}_{nn}) + g V_{1i} K^u_{j(2n-1)} (V^{\dagger}_{\rm{CKM}})_{kn}\right]\nn\\
{\mathbf{A}}^d_{ijk}=  C_{\tilde{\chi}^{-}_i \tilde{u}_j (d^\dagger_L)_k }&=& U^*_{2i} \sum^3_{n=1} \left[K^u_{j(2n-1)} (-\hat{Y}^{d}_{kk})(V^{\dagger}_{\rm{CKM}})_{kn}\right],
\eea
where it is useful to remember than in this notation the components $K^f_{j(2n-1)}$ are related to left-handed transformations and $K^f_{j2n}$ with right-handed ones. 

%
%%%%%%%%%%%%%%%%%%%%%%%%%%%%%%%%%%%%%%%%%%%%%%%%%%%%%%%%%%%%%%%%%%%%%%%%%%%%%%%%%%
%
\section*{Acknowledgments}

We would like to thank P. Gambino, G. Ganis, P. Ko, and A. Vainshtein for
useful discussions and/or correspondence. We also thank the authors
of \cite{ex_mvf_gut} for helpful comments on the original version of
the paper. This work  was supported in part 
by DOE grant DE-FG02-94ER.

%
%%%%%%%%%%%%%%%%%%%%%%%%%%%%%%%%%%%%%%%%%%%%%%%%%%%%%%%%%%%%%%%%%%%%%%%
%

\end{document}